%% file: 150NewGammaRaySources.tex
\documentclass[structabstract]{aa}
\usepackage{graphicx}
\usepackage{amssymb}
\usepackage{aalongtable}
\usepackage{lscape}
\usepackage{natbib}

\newcommand{\nufnupeak}{$\nu_{\rm peak}$f$_{\nu_{\rm peak}}$}

\newcommand{\lsim}{{\lower.5ex\hbox{$\; \buildrel < \over \sim \;$}}}
\newcommand{\gsim}{{\lower.5ex\hbox{$\; \buildrel > \over \sim \;$}}}

\begin{document}

   \title{Searching for $\gamma$-ray signature in WHSP blazars:}

   \subtitle{Fermi-LAT detection of 150 excess signal in the 0.3-500 GeV band.}

   \author{
          B. Arsioli  \inst{1,2,3} 
		  \and	 
		  Y-L. Chang \inst{1,2}       
          }
        
        \institute{ASI Science Data Center, ASDC, Agenzia Spaziale Italiana, via del Politecnico snc, 00133 Roma, Italy
        \and 
        Sapienza Universit\`a di Roma, Dipartimento di Fisica, Piazzale Aldo Moro 5, I-00185 Roma, Italy  
        \and
	    ICRANet-Rio, CBPF, Rua Dr. Xavier Sigaud 150, 22290-180 Rio de Janeiro, Brazil  \\ 
        \email{bruno.arsioli@asdc.asi.it}
        }
	          
\date{Accepted on September 1st, 2016}

\abstract
{}
{A direct search of $\gamma$-ray emission centered on multi-frequency selected candidates is a valuable complementary approach to the standard search adopted in current $\gamma$-ray Fermi-LAT catalogs. Our sources are part of the 2WHSP sample that was assembled with the aim of providing targets for Imaging Atmospheric Cherenkov Telescopes (IACT). A likelihood analysis based on their known position enabled us to detect 150 $\gamma$-ray excess signals that have not yet been reported in previous $\gamma$-ray catalogs (1FGL, 2FGL, 3FGL). By identifying new sources, we  solve a fraction of the extragalactic isotropic $\gamma$-ray background (IGRB) composition, improving the description of the $\gamma$-ray sky.
} %
{We perform data reduction with the Fermi Science Tools using positions from 400 high synchrotron peaked (HSP) blazars as seeds of tentative $\gamma$-ray sources; none of them have counterparts from previous 1FGL, 2FGL and 3FGL catalogs. Our candidates are part of the 2WHSP sample (currently the largest set of HSP blazars). We focus on HSPs characterised by bright synchrotron component with peak flux $\nu f_{(\nu)} \geq 10^{-12.1}$ ergs/cm$^{2}$/s, testing the hypothesis of having a $\gamma$-ray source in correspondence to the WHSP positions. Our likelihood analysis considers the 0.3-500 GeV energy band, integrating over 7.2 yrs of Fermi-LAT observation and making use of the Pass 8 data release.   
} 
{From the 400 candidates tested, a total of 150 2WHSPs showed excess $\gamma$-ray signature: 85 high-significance detections with test statistic (TS)$>$25, and 65 lower-significance detections with TS between 10 to 25. We assume a power law spectrum in the 0.3-500 GeV band and list in Table \ref{tableFermi1} the spectrum parameters describing all 150 new $\gamma$-ray sources. We study the $\gamma$-ray photon spectral index distribution, the likelihood of detection according to the synchrotron peak brightness (figure of merit parameter), and plot the measured $\gamma$-ray LogN-LogS of HSP blazars, also discussing the portion of the IGRB that has been resolved by the present work. We also report on four cases where we could resolve source confusion and find counterparts for unassociated 3FGL sources with the help of high-energy TS maps together with multi-frequency data. The 150 new $\gamma$-ray sources are named with the acronym 1BIGB for the first version of the Brazil ICRANet Gamma-ray Blazar catalog, in reference to the cooperation agreement supporting this work.  
} 
{}

 
 \maketitle
 
%

\section{Introduction}

Catalogs of $\gamma$-ray sources currently compiled by the Fermi-LAT team are based on  $\gamma$-ray data only, and their standard detection method is blind with respect to information coming from other wavelengths. This approach is clean and unbiased with respect to any class of potential $\gamma$-ray emitters. However, there are populations of astrophysical objects that are now known to emit  $\gamma$-rays, and the knowledge of their position in the sky can be used to facilitate the detection and identification of new $\gamma$-ray sources. Based on this principal, we select a sample of  candidates to be used as seeds for a direct search of $\gamma$-ray signatures using  likelihood analysis with the Fermi Science Tools.

Blazars are the most abundant $\gamma$-ray sources in the latest Fermi-LAT 3FGL catalog, being 1147 (660 BL Lacs and 487 Flat Spectrum Radio Quasars - FSRQ) of the total 3034 \citep{3FGL}. Even so, one third of the known blazars from 5BZcat\footnote{The 5BZcat \citep{5BZcat} is a large sample of 3 561 identified blazars. Multi-frequency  data for the 5BZcat is available at http://www.asdc.asi.it/bzcat with a direct link to the SED-builder tool.} are not confirmed as $\gamma$-ray emitters. Probably many of them are faint $\gamma$-ray sources that are hard to identify by automatic search methods only based on Fermi-LAT data. 
The blazar population has been extensively studied by means of a multi-frequency approach considering dedicated databases on radio, microwave, infra-red (IR), optical, ultra-violet (UV), and X-ray, since they are characterised by radiation emission extending along the whole electromagnetic spectrum, up to TeV energies.

A particular family of extreme sources with the synchrotron component peaking at frequencies $\nu_{peak}$ larger than 10$^{15}$  Hz is classified as a high synchrotron peak blazar \citep[HSP,][]{padgio95,abdo10} and is the dominant population associated with extragalactic very high-energy \citep[VHE: E$>$0.1 TeV,][]{TeVAstronomy} sources in the 2nd Catalog of Hard Fermi-LAT Sources \citep[2FHL,][]{2FHL}. Therefore, HSPs constitute a key population for the detection of point-like $\gamma$-ray sources within Fermi-LAT data. 


A large sample of HSP blazars was recently assembled using a multi-frequency selection procedure that exploits the unique features of their spectral energy distribution (SED). This sample is known as the 1WHSP catalog \citep{1WHSP} and was built using a primary source-selection based on IR colours \citep[following][]{WiseBlazars}, later demanding all potential candidates to have a radio, IR and X-ray counterpart. The sources had to satisfy broadband spectral slope criteria (from radio to X-rays) that were fine-tuned to match the SED of typical HSP blazars. In addition, their multi-frequency SEDs were inspected individually using the SED-builder tool (http://www.asdc.asi.it) fitting the synchrotron component with a third degree polynomial to determine the $\nu_{peak}$ parameter, only keeping cases with $\nu_{peak}>10^{15}$ Hz. The catalog name ``WHSP" stands for WISE High Synchrotron Peak blazars, since all sources have an IR counterpart from the WISE mission \citep{WISE}, which defines their positions. The 1WHSP catalog includes 992 objects at Galactic latitude $|b|>20^{\circ}$. A total of 299 1WHSPs have a confirmed $\gamma$-ray counterpart in 1FGL, 2FGL and 3FGL \citep{1WHSP}, but many HSPs with bright synchrotron peak are still not detected/confirmed in the $\gamma$-ray band.

Given the importance of finding new HSP blazars, an extension of the 1WHSP sample \citep[the 2WHSP,][]{2WHSP} has ben assembled. It considers sources located at latitudes as low as  $|b|=10^{\circ}$ with a total of $\approx$1 693 sources, 439 of which have counterparts within the error circles from the 3FGL catalog. The 2WHSP sample avoids the selection based on IR colours that was used as a primary step for the 1WHSP catalog. This brings an overall improvement in completeness\footnote{ \label{b10} Also to improve the completeness of the final sample, known HSP sources at $|b|<$10$^{\circ}$ were incorporate in the 2WHSP catalog.}, since some HSP blazars were out of the 1WHSP sample owing to the contamination of IR colours by the elliptical-galaxy thermal emission. Compared to the 1WHSP, the 2WHSP sample incorporates extra X-ray catalogs like Einstein IPC, IPC slew and Chandra \citep{Harris1993,Elvis1992,Evans2010} as well as updated versions from 3XMM-DR5 and XMM-slew catalogs \citep{Rosen2015,Saxton2008}. In addition, Swift-XRT alone performed a series of $\sim$160 new X-ray observation of WHSP sources (enabling us to better estimate synchrotron peak parameters) and an extensive study of X-ray extended sources helped to avoid contamination with spurious objects (more details are given by \cite{2WHSP}). The catalogs are available at: www.asdc.asi.it/1whsp or /2whsp; where multi-frequency SEDs can be quickly built using open access online tools. Since the 2WHSP catalog supersedes the 1WHSP (with improved selection and better estimate of synchrotron peak parameters), from now on we only refer to the 2WHSP sample.


\section{Brightness of the synchrotron peak and detectability by Fermi-LAT}
\label{bright-peak}

The HSP blazars are characterised by hard  $\gamma$-ray  spectrum with average photon index $ \langle \Gamma \rangle$=1.85$\pm$0.01 \citep{1WHSP,2lac,3LAC} favouring their detection in the high-energy band. Therefore, the 2WHSP catalog has collected an unprecedented number of remarkably rare and extreme sources that are expected to emit $\gamma$-rays. 

\begin{figure}[h]
   \centering
    \includegraphics[width=1.02\linewidth]{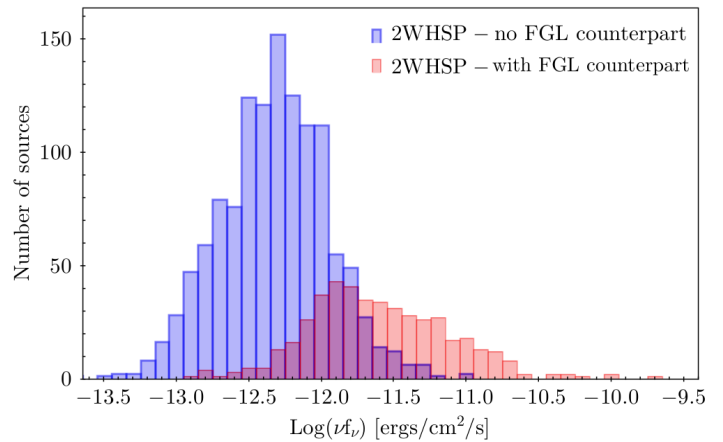}
     \caption{Distribution of Log(\nufnupeak) synchrotron peak flux with indigo bars that represent the $\gamma$-ray subsample of 439 2WHSP-FGL sources, and light red-bars representing the 1 255 $\gamma$-ray undetected 2WHSPs. The intersection between detected and undetected distributions suggests there may be numerous 2WHSP sources close to the detection threshold from Fermi-LAT. The 2WHSP catalog lists Log(\nufnupeak) values in 0.1 steps, and histogram-bins are centered on those values.}
      \label{histofermi}
\end{figure}

In Fig. \ref{histofermi} we plot the distribution of synchrotron  peak fluxes (\nufnupeak) for the 2WHSP detected\footnote{ We may use 2WHSP-FGL when referring to the subsample of 439 2WHSPs that have counterparts from the 1FGL, 2FGL, and 3FGL catalogs.} and undetected $\gamma$-ray sources. As seen, most of the bright 2WHSPs with Log($\nu_{\rm peak}$f$_{\nu_{\rm peak}})>-$11.2 ergs/cm$^2$/s have already been unveiled by Fermi-LAT. The range between $-$12.4$<$Log(\nufnupeak)$<-11.2$ ergs/cm$^{2}$/s where histograms for detected and undetected sources have significant overlap, tells us that there must be a population of undetected 2WHSP blazars that is within the reach of Fermi-LAT; especially when taking into consideration integration time greater than 4 yrs, as used to build the 3FGL catalog.       

As a first step for testing the efficiency of a dedicated $\gamma$-ray sources-search, we performed a series of likelihood analysis on bright HSPs that were not included in previous Fermi-LAT catalogs 1FGL, 2FGL and 3FGL (FGL). It soon became clear that,  when considering longer exposure time with improved event reconstruction of Femi-LAT Pass 8, a significant number of faint $\gamma$-ray sources could be detected. For the likelihood analysis, we define a subsample of 2WHSPs with a synchrotron component \nufnupeak$\geq 10^{-12.1}$ ergs/cm$^{2}$/s simply for limiting the number of seeds to 400, indeed showing its potential for wider studies with the whole 2WHSP sample.

Our present effort for unveiling new $\gamma$-ray sources not only provides targets for future follow-up and variability studies, but also helps us to enhance the current understanding on the nature of the VHE $\gamma$-ray background, which probably has a strong contribution from unresolved point-like sources \citep{OriginEGB}. Especially at the E$>$10 GeV band, unresolved HSP blazars may have increasing relevance \citep{EGBpaolo} as was confirmed by our results described in Sect. \ref{secIGRB}.

\section{Fermi-LAT data reduction}

The Fermi-LAT detector \citep[][]{FermiLAT} is characterised by  a point spread function (PSF) which contains 68\% of the 1 GeV events within 0.8$^{\circ}$. The PSF decreases with energy, following a trend $\propto E^{-0.8}$ up 10's GeV, and is roughly constant at  0.1$^{\circ}$ from there to the highest energies considered in this paper.

Based on the position of a potential $\gamma$-ray source, we downloaded Pass 8 processed events from the public Fermi-database\footnote{http://fermi.gsfc.nasa.gov/ssc/data/access/} that includes all photons recorded in a region of interest (ROI) of 25$^\circ$ radius from the candidate's position, for the whole 7.2 years of observations (MM/DD/YYYY: 08/04/2008 to 11/04/2015). In our analysis we used the Fermi Science Tools (v10r0p5), performing binned analysis to deal with the long integration time.

A series of quality cuts were applied to the raw data, starting with the selection of events having high probability of being photons (which is done by requiring $evclass=128$ and $evtype=3$ in the gtselect routine), working with maximum zenith angle-cut of 90$^\circ$ \footnote{The zenith angle-cut is used to avoid contamination with  Earth's limb $\gamma$-ray photons, which are induced by cosmic-ray  interactions with the atmosphere, and are known as strong source of background for the low-energy band covered by Fermi-LAT.}. Given the fact that HSPs are characterised by hard $\gamma$-ray spectrum (with an average photon index $ \langle \Gamma \rangle \approx 1.85$) we choose to work at E$>$300 MeV, avoiding the need to calculate energy-dispersion correction during the data analysis (which is necessary for E$<$300 MeV photons). With the gtmktime routine, we then generate a list of good time intervals (GTIs) to be considered in further analysis. In this step, some given flags ((DATA-QUAL$>$0) and (LAT-CONFIG==1)) assure that only events acquired by LAT instrument in normal science data-taking mode are considered. Using the gtbin routine, we generate counts maps (CMAP) and counts cubes (CCUBE), having $500 \times 500$ and $350 \times 350$ pixels with $0.1^\circ$/pixel, respectively. The CCUBE is a series of CMAPs, each one having photons within a given energy bin, and here we consider 37 logarithmically spaced energy bins along 0.3-500 GeV.

\begin{figure}[h]
   \centering
    \includegraphics[width=1.0\linewidth]{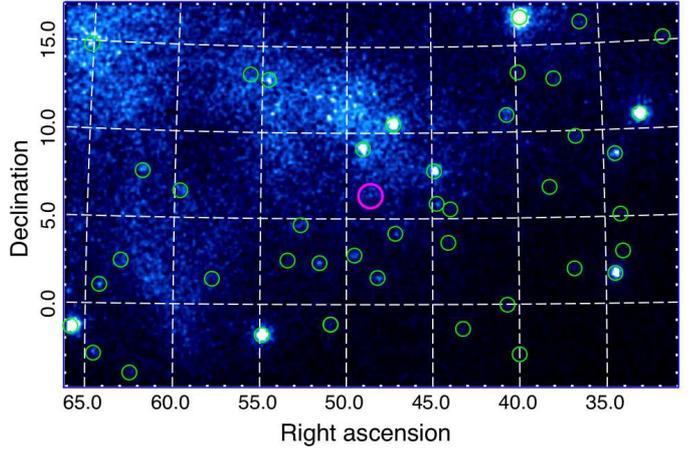}
     \caption{Fermi-LAT $\gamma$-ray counts map in the energy range 0.3-500 GeV, over 7.2 yrs, showing detected $\gamma$-ray sources, at the center of the green circles (only as indicative of their 3FGL positions). We highlight the source 2WHSP J031423.8+061955 (center of the magenta circle), which we detected in $\gamma$-rays with TS=69.9. As seen, not all relevant sources are easy to unveil with only the CMAP inspection.}
      \label{countsmap}
\end{figure}

As an example, the CMAP (Fig.\ref{countsmap}) have green circles corresponding to known 3FGL sources, and a magenta circle to mark the 2WHSP J031423.8+061955 position. As seen, together with our candidate, other faint $\gamma$-ray sources may be present but cannot easily be distinguished from the counts map, demanding a dedicated data reduction to test the point-like source hypothesis.

A livetime cube is then generated using the gtltcube routine, holding information about the sky coverage as a function of inclination with respect to the LAT z-axis. An important parameter to set in this step is the cos(theta) which is related to the angle-binning, summing incoming photons from a particular solid angle; here we use 0.025$^\circ$ (following recommendations from the Fermi-LAT team). Later, the source map is created using the gtsrcmaps routine, and take into account the models describing all previously known $\gamma$-ray sources and background emission that are within $25^\circ$ radius from the center candidate. The models that describe point-like and extended sources, as well as the diffuse Galactic and isotropic  background are included in a single .xml file. This was built using the script make3FGLxml.py\footnote{The make3FGLxml.py is a python routine written by T. Johnson, 2015, and provided by the Fermi-LAT team as an user contribution tool.} and considers the 3FGL catalog for describing the point-like and extended sources known so far, loading parameters from the source file gll-psc-v16.fit. For the diffuse Galactic background content, the high-resolution model from the source file gll-iem-v06.fit was considered, and for the  isotropic component we use the model from iso-source-v06.txt. We also considered the latest data and instrument response functions (IRFs) available at the time of work,  P8R2 SOURCE V6, and event selection Pass 8 processed Source: front+back.


Since our $\gamma$-ray candidates are not part of the latest Fermi-LAT catalog 3FGL, they have to be added to the source-input file (.xml) that contains the model parameters and the positions of all known $\gamma$-ray  sources. In this work, the spectrum of each $\gamma$-ray candidate is always modelled as a power law:

\begin{equation}
\frac{dN}{dE}  =  N_0  \Big( \frac{E}{E_0} \Big)^{-\Gamma}
\label{powerlaw}
\end{equation} 
where $E_0$ is a scale parameter (also known as pivot energy), $N_0$ is the pre-factor (corresponding to the flux density in units ph/cm$^{2}$/s/MeV at the pivot energy $E_0$), and $\Gamma$ is the photon spectral index for the energy range under consideration. Usually, the starting guess-values were chosen to be consistent with SEDs from HSP blazars, therefore: E$_0$=1000 MeV, $N_0 = 1.0 \times 10^{-12}$ ph/cm$^{2}$/s/MeV, $\Gamma$=1.8 . Both $\Gamma$ and $N_0$ are set as free parameters  and further adjusted by the gtlike fitting routine. The source position and the scaling  $E_0$ are set as fixed parameters. In the source-input xml file, all sources within 10$^\circ$ from the candidate  are set free to vary their spectral fitting parameters, therefore 3FGL models that are based on four years of observation are adjusted, since we now integrate over 7.2 years of data. This particular choice increases the computational burden of the analysis, but is crucial for adapting the model maps to the extra 3.2 years of exposure that is considered. 


A likelihood analysis is then performed by the gtlike routine, considering all the information from the livetime cube, source maps and source input files, together with the PSF and IRFs, in order to best fit the free parameter from the source input list. Finally it is calculated the Test Statistic (TS) parameter, which is defined as \citep{TSmapsConfidenceRadius}
\begin{equation}
{\rm TS}=-2ln \Big(  \frac{L_{(no ~ source)}}{L_{(source)}} \Big)
\end{equation} 
where $L_{(no ~ source)}$ is the likelihood of observing a certain photon-count for a model without the candidate source (the null hypothesis), and $L_{(source)}$ is the likelihood value for a model with the additional candidate source at the given location. The reported TS values correspond to a full band fitting, which constrains the whole spectral distribution along 0.3-500 GeV to vary smoothly with energy and assuming no spectral break. Considering we have a good description of Galactic and extragalactic diffuse components, this is a measure of how strong a source emerges from the background, also assessing the goodness of free parameters fit. A TS$\approx$25 is equivalent to a 4-5$\sigma$ detection \citep{FermiCat1} (depending on the strength of the background in the region), and only cases with TS$>$25 are considered by the Fermi-LAT team as a positive detection of point-like source. We first run the gtlike with the fitting optimiser mode DRMNFB, which generates an enhanced source-input list with all the free parameters recalculated (the first interaction of the fitting procedure). We again feed the gtlike routine with the enhanced source-input list, and use the fitting optimiser mode NEWMINUIT to generate the final model for all sources in the ROI.

\section{New $\gamma$-ray detections. Validation and population properties} 

Here we present the $\gamma$-ray detection of 85 2WHSP sources at TS$>$25 level; and we extend the $\gamma$-ray analysis by considering another 65 2WHSPs with lower significance $\gamma$-ray signal, with TS ranging between 10 and 25. In Table \ref{tableFermi1} we list relevant information for these 150 sources, including names, positions, redshift, $\gamma$-ray model parameters and their associated uncertainties. The new $\gamma$-ray sources are named with the acronym 1BIGB for the first version of the Brazil ICRANet Gamma-ray Blazar catalog, with source designation 1BIGB JHHMMSS.s$\pm$DDMMSS, and coordinates corresponding to the 2WHSP seed-positions. We also present a few examples of TS maps (Sect. \ref{tsmaps} and \ref{faintdetec}) both for high and low-significance $\gamma$-ray signatures, showing that they all emerge as point-like sources, and should not be taken as spurious signals. 

In Sect. \ref{prova} we test a direct source-search using the 3FGL catalog analysis setup, showing that we could successfully probe faint $\gamma$-ray emitters and add complementary $\gamma$-ray detections. In Sect. \ref{deteceffic} we calculate the $\gamma$-ray detection efficiency based on the brightness of the synchrotron peak ($\equiv$figure of merit). In Sect. \ref{slopeDist} we plot the photon spectral index ($\Gamma$) distribution for the newly-detected $\gamma$-ray sources and compare their spectral properties with the FGL counterparts from 2WHSP sources. We also show the $\Gamma$ vs. flux (1-100 GeV band) so that the improvement, with respect to the flux threshold when considering  detections down to TS=10, can be evaluated. In Sect. \ref{Eddington} we comment on flux-fluctuations associated with sources close to the flux-threshold (Eddington bias effect) showing that this effect is not severe in our context.

\subsection{High-significance $\gamma$-ray sources with TS$>$25}
\label{highTSsources}

In Sect. \ref{bright-peak} we show that the natural candidates for our analysis are the bright 2WHSP sources\footnote{By bright sources we mean: 2WHSPs with the largest flux density associated to the synchrotron peak $\nu$f$_{\nu-peak}$ component.} that have not yet been detected by Fermi-LAT (in previous 1FGL, 2FGL and 3FGL catalogs), as suggested from Fig. \ref{histofermi}. Therefore, by sorting the 2WHSP source based on their synchrotron peak brightness, we considered cases down to Log(\nufnupeak)=-12.1 ergs/cm$^2$/s, selecting 400 $\gamma$-candidates, from which 85 ($\sim$21\%) have shown high-significance $\gamma$-ray signature with TS$>$25. 

For each source of interest, we inspected a region of 50' radius  around it, checking for any previous $\gamma$-ray detections or for the presence of bright blazars that could also be potential high-energy counterparts. For this task we made use of the Sky Explorer Tool, available from the ASDC web site (tools.asdc.asi.it),  which displays all radio, optical, X-ray, and $\gamma$-ray detections for a given ROI. During the preparation of this work, the 2FHL catalog \citep{2FHL}, which contains only E$>$50 GeV detections was released, including six of the sources we were working with. Also \cite{MST1}, as well as \cite{MST-1WHSP}, reported on  possible counterparts of photon-clustering detected by Fermi-LAT at E$>$10 GeV, which included eight of our detections (two in common with the six 2FHL). We keep them in our Table \ref{tableFermi1}, indicated with ``a" and ``b" superscripts, respectively, since they constitute positive detections based on our primary approach, showing the intersection between valuable methods for unveiling new $\gamma$-ray sources.

The fact that few 2FHL-sources and Campana-sources are in common with our detections is certainly due to their analysis being based only on $\gamma$-rays, applied to E$>$50 GeV and E$>$10 GeV, respectively. Our energy threshold at E$>$300 MeV is much lower and well suited to the way we select our seeds (based on multi-frequency information from radio to X-rays, not only on $\gamma$-ray data). Therefore, we are able to probe hard $\gamma$-ray sources, even if they have low flux at E$>$10 GeV (we do not depend on $\gamma$-ray photon clustering to identify our seeds). In fact, both approaches are powerful and should be seen as complementary, since they all apply to the goal of enriching our description of the $\gamma$-ray sky. 


\subsection{The TS map and $\gamma$-ray spectrum}
\label{tsmaps}

\begin{figure}[h!]
\centerline{\includegraphics[width=1.0\linewidth]{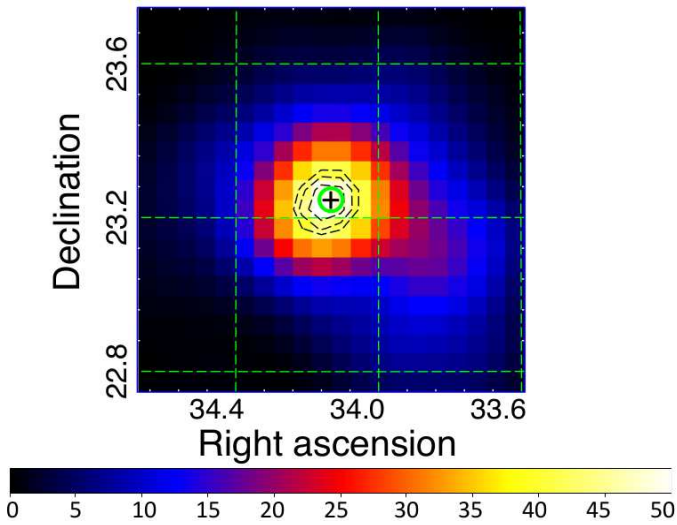}}
  \caption{TS map (3-500 GeV), having 20x20 pixels and built with resolution 0.05$^\circ$/pixel, integrating over 7 yrs of Fermi-LAT observations. The green circle centered on + highlights the position of 2WHSP J021631.9+231449; contour dashed lines correspond, respectively, to  the 68\%, 95\%, and 99\%  confinement regions (from inner to outer lines) for the $\gamma$-ray signature position.}
 \label{tsmap}
\end{figure}

A TS map consists of a pixel-grid where the existence of a point-like source is tested in each pixel. This is a demanding computational task when exposure time that is longer than few months are considered. Here we study the case of 2WHSP J021631.9+231449,  defining a 25x25 grid with 0.05$^\circ$/pixel, and evaluated each grid-bin using likelihood analysis from gttsmap routine. Given the fact that WHSP blazars are expected to be hard spectrum $\gamma$-ray sources, we built a TS map that only considers photons with energy larger than 3 GeV \footnote{For HSP sources with high-significance $\gamma$-ray signature, the cut at 3 GeV in many cases provided (and is therefore the reason why we choose it; see examples on Sect. \ref{sectConf}) a good balance between computation time and ability to solve the gamma-ray signature as a point-like source. Also, despite the fact we are dealing with HSP blazars (with mean $\gamma$-ray photon index $\sim$ 1.9) Fermi-LAT has relatively good sensitivity along the 1-100 GeV band, and improved PSF at $>$3 GeV, which helped to achieve better localization for the $\gamma$-ray sources when necessary.}. The cut in photon energy helps not only to save computation time, it also has another important purpose: Since the PSF improves with increasing energy, working with high-energy photons help us to determine the TS-peak position with better precision.  
When building the TS map from Fig. \ref{tsmap}, the input model (.xml file) does not contain our $\gamma$-candidate, so that the map alone can test the existence of  point-like sources (with no previous bias), which may manifest as a TS peak that emerges from the background. 

All sources within 7$^\circ$ from the grid center have their corresponding model parameters set as free to adjust for the current analysis. Also, we set as free parameters the Normalization (from the diffuse extragalactic background model) and ConstantValue (from the Galactic background model) to avoid having large TS values that are only due to an overestimated background flux. An overestimated background usually manifest as a smooth distribution of high TS values along the whole grid, therefore it is important to properly scale it in the studied region, and evaluate if the source emerges with high TS values from a low TS ($\approx 0.0$) background. When the background is not well described, it could affect (or mimic) a point-like source detection. Since we are mainly working at high Galactic latitude $|b|>10^{\circ}$, we avoid most of the Galactic diffuse emission (which is strong and highly structured at lower latitudes), preventing spurious detections. As seen in Fig. \ref{tsmap}, the 2WHSP source is well within the 68\% confinement region for $\gamma$-ray signature \citep[][]{TSmapsConfidenceRadius}, and TS values at the grid center contrasts with outer regions, ensuring that the observed TS peak is due to a point-like source rather than an overestimated background component.

\begin{figure}[h]
   \centering
\includegraphics[width=0.95\linewidth,angle=0]{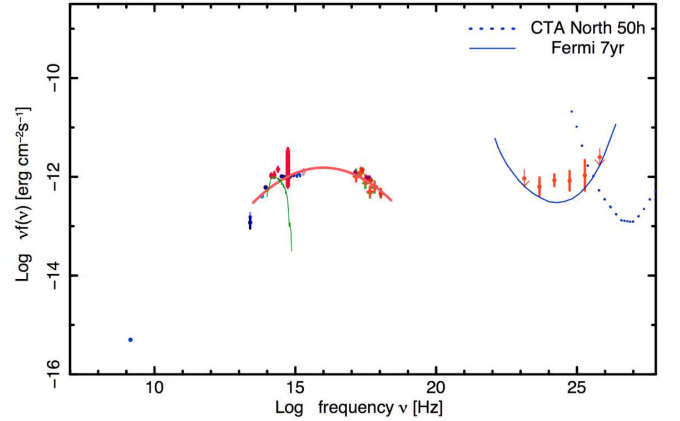}
     \caption{SED for 2WHSP J021631.9+231449 adding new $\gamma$-ray description in the full energy band 0.3-500 GeV. The red line is a fitting for the non-thermal component in the synchrotron peak, the green line is the giant elliptical host galaxy template for z=0.288, the blue line corresponds to the Fermi-LAT (7 yrs) broadband sensitivity, and blue dashed line to CTA-North \citep[50h exposure,][]{CTA50h}.}
      \label{sed0216}
\end{figure}

In Fig. \ref{sed0216} we show the multi-frequency SED
\footnote{Here we make use of the Sky-Explorer tool from www.asdc.asi.it, quickly retrieving multi-frequency information from public data bases, to cite: \cite{Radio1,Radio2,Radio3,Radio4,Radio5,Radio6First,Radio7,WISE,UKIDSDR2,IPC,ROSATBSC,ROSATFSC,XMM,Saxton2008,SWIFT,deliaswift,SDSSDR12}.} for 2WHSP J021631.9+231449. The $\gamma$-ray SED was calculated by dividing the full energy band 0.3-500 GeV into 6 bins, equally spaced in a logarithmic scale, to compensate for the lower number-counts when increasing the photon energy. The upper limits (u.l.) are computed only for energy bins with TS$<$9, and considering a 95$\%$ confidence level on the integrated flux along the whole energy bin. The broadband sensitivity at a certain energy E (thin blue line in Fig. \ref{sed0216}) is calculated as the maximum flux of a power law source at the LAT detection threshold, for any spectral index.


The source 2WHSP J021631.9+231449 ($\Gamma$=1.97$\pm$0.12) is clearly a promising candidate for observation in the VHE domain with the future Cherenkov Telescope Array \citep[CTA,][]{CTA}, or even in reach for present detectors during flaring events. This new $\gamma$-ray detection is only one example within the many cases listed in Table \ref{tableFermi1}, raising our expectations for future VHE studies. Currently, we do not have TS maps and $\gamma$-ray SEDs calculated for all sources\footnote{The computational demand for accomplishing such task is relatively large, and requires further planning together with our Computer-Cluster partners; to cite: Joshua-Cluster from ICRANet Italy, and Gauss-Cluster from CESUP Brazil.}. However, we plan to make them available in the near future as a natural extensions of the present work, given the importance of HSP blazars for upcoming observations with CTA.

\subsection{Lower significance $\gamma$-ray sources}  
\label{faintdetec}

Within the 400 2WHSPs studied, 65 had faint signatures of $\gamma$-ray emission with TS values ranging between 10 and 25, and are also listed in Table \ref{tableFermi1}. We call these lower-significance detections because these sources have TS$<$25 (the threshold-limit assumed by the Fermi-LAT team) but they still represent relevant findings considering the number of seed-positions used in our present approach.

As known, the significance $\sigma$ can be approximated as $\sigma \approx \sqrt{\rm TS}$ \citep{TSmapsConfidenceRadius} and, working with TS=10 threshold, implies our detections have significance of the order of $\sim$3$\sigma$. Since we performed a series of 400 binned likelihood analyses for positions only associated with WHSP sources, the number of spurious detections (N$_{spur}$) expected is N$_{spur}\approx$400$\times$10$^{-3}$=0.40; therefore we do not consider spurious detections as a concern in our work. In fact, we have individuated a total of 150 $\gamma$-ray excess within TS$>$10 level, which corresponds to $\sim 37\%$ positive signatures for all the 400 candidates tested. 

It should be clear that we are not performing a blind-analysis of the whole $\gamma$-ray sky, therefore the number of seed-positions we inspect is very small. The strong threshold cut (TS$>$25) adopted by the Fermi-LAT team manifest their rigour before validating any new populations of $\gamma$-ray emitters. In contrast, our candidates are a particularly small population of well-characterised (from radio to X-rays) blazars which are firmly established as a family of $\gamma$-ray emitters, therefore a 3 $\sigma$ detections threshold is suitable for our approach.

\begin{figure}[h!]
   \centering
    \includegraphics[width=1.0\linewidth]{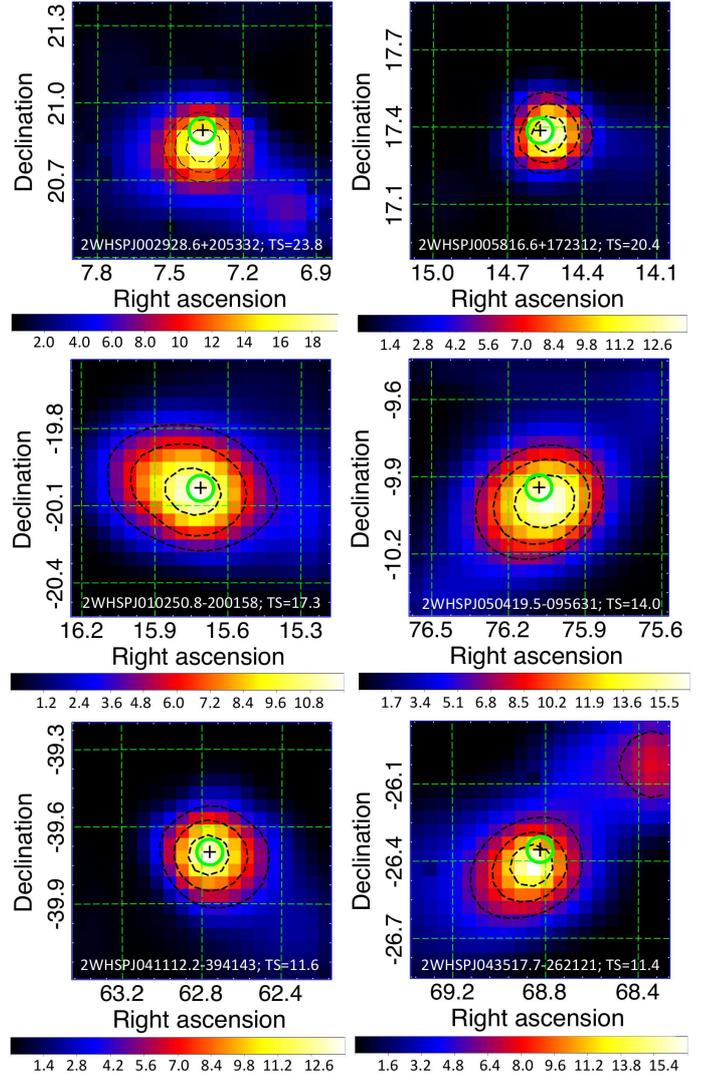}
     \caption{TS maps in the 0.6-500 GeV band for six sources representing the lower-significance detections with TS between 10 to 25. At the bottom of each map, we write the corresponding source name and the reported TS value for a binned likelihood analysis when integration is over 7.2 years of observations (along the full energy band 0.3-500 GeV). The 2WHSP positions are highlighted by thick green circles with their centered on +. The contour black dashed lines are TS surfaces representing 68\%, 95\%, and 99\% containment region for the $\gamma$-ray signature (from inner to outer lines). }
      \label{lowTSmaps}
\end{figure}

As a complementary test to evaluate if our lower-significance detections are consistent with point-like sources, we randomly choose six cases with TS in the range from 10 to 25 and calculate their corresponding TS maps (Fig. \ref{lowTSmaps}). { Since lower-significance sources are hard to detect based only on $>$3 GeV photons, to improve the photon counts we go to lower energies (0.6-500 GeV), which helps to individuate the gamma-ray signatures\footnote{Also, we should add that the overall computation time for lower-energy TS maps integrating over 7 yrs can easily became prohibitive ($>$weeks). Especially for bright sources; large photon-counts translate into large computational demand. Therefore, there is no absolute way to choose a working energy-range. We are always limited by the computation time, and many cases demand us to adapt (for example, see the lower-energy TS map from Sect. \ref{conf1}, where we had to work in a narrow energy range of 700-800 MeV to reach results in reasonable time).}. All six candidates studied clearly emerge from the background as point sources, and are consistent with the 2WHSP positions within the 68\% confident radius for the $\gamma$-ray signature. In fact, this reinforces our view that assuming a TS$>$10 threshold does not contaminate our results with spurious detections.

Given the variability of one order of magnitude is often observed for HSP blazars in the GeV-TeV bands\footnote{For dedicated studies on variability involving HSP blazars, see \cite{variabilityTeV-WHSP} reporting on 1ES 1959+650 $\equiv$ 2WHSP J195959.8+650853, or \cite{FlareMkr421-MultiWave,FlareMkr421-Hmodel} reporting on Mrk421 $\equiv$ 2WHSP J110427.3+381230.}, most of these lower-significance detections may be in reach of CTA during flaring episodes (see Fig. \ref{sed0216}). In fact, the validation of lower-significance detections associated with HSP blazars provide relevant hints about the population responsible for considerable portion of the high-energy isotropic $\gamma$-ray background (IGRB); and it is also important to account for their existence and imprints, since they add anisotropic contribution to the IGRB \citep[][]{PointSourcesBellowDetecLimit,GammaAnisotropy,GammaAnisotropyFermiTeam,blazarbackground}.

knowledge about the position and model-parameter for describing individual faint $\gamma$-ray sources may also help to improve tentative correlations of the IGRB with the large-scale structures/clusters \citep{dgrb-LSS,CountGammaCluster}, since their contribution could be subtracted from the currently unresolved background (i.e., to clean the IGRB from faint point-like sources before trying any sort of correlation). By relying on multi-frequency data to search for faint $\gamma$-ray source, we may improve our capability of resolving them. Moreover, since the present evaluation is primary driven by the position of HSP blazars, it is important to keep track of any case that shows a faint $\gamma$-ray signature even if not enough to fit in the current TS-limit for detection demanded by Fermi-LAT team.

\subsection{A direct source-search as a complementary approach to probe faint $\gamma$-ray emitters}
\label{prova}

To evaluate the potential of using direct source-search as a complementary method when building $\gamma$-ray catalogs, we select 30 objects with the highest significance $\gamma$-ray signatures from our list of new-detections (all sources having TS$>$45 in our analysis with Pass 8 data integrating over 7.2 yrs). We then test if these sources could be detectable with high/lower-significance based on the $\gamma$-ray analysis setup used to build the 3FGL catalog. We  download Pass 7 data\footnote{Pass 7 data: http://heasarc.gsfc.nasa.gov/FTP/fermi/data/ lat/weekly/p7v6/} corresponding to the first four yrs of observations (MM/DD/YYYY: 08/04/2008 to 08/04/2012) and proceed with the likelihood analysis that considers a background of extended and point-like sources built based on the gll-psc-v14.fit list, with information available at that time. For the diffuse Galactic background content we consider the source file gll-iem-v05-rev1.fit, and the iso-source-v05.txt model for the isotropic component. We also choose the IRF corresponding to the preparation of 3FGL catalog P7REP SOURCE V15, and event selection Pass 7 reprocessed source data (front+back). 

The results are listed in Table \ref{tablepass7}, showing four high-significance detections at TS$>$25 level, and 17 lower-significance cases with TS in between 10 to 25. Indeed, our test shows that a direct-source search can be used as a complementary method to refine the description of the $\gamma$-ray sky; not only revealing high-significance sources, but also allowing lower-significance sources to be successfully probed. The fact that we only present four extra sources that could fit the 3FGL detection threshold should not mislead us into thinking that these types of contributions are not worth to incorporate. As discussed, this test considers only a few sources, and an extended study over the whole blazar population could add a significant complementary contribution. Also, with increasing integration time from Fermi-LAT data, we reach a lower flux threshold (S), and the source number-count may improve $\propto \sim$S$^{-1.5}$; therefore the impact of a direct source-search probably has increasing relevance to the building of the next $\gamma$-ray catalogs.

Clearly, most of our new high-significance detections based on 7.2 yrs with Pass 8 data are mainly driven by a longer integration time (from four years, enhanced to 7.2 years) and improved event reconstruction (from Pass 7 to Pass 8). However, all 150 detections presented in our work (Table \ref{tableFermi1}) were only possible because we knew where to look, using selected seed positions selected when considering multi-frequency data. In this regard, multi-frequency selected seeds are indeed very promising for driving new $\gamma$-ray detections just as gamma-ray seeds are, and here we emphasise their complementarity. Also, a likelihood analysis based on seeds selected from populations of $\gamma$-ray emitter enabled us to successfully probe a population of lower-significance emitters (using Pass 7 - 4 yrs) that are later confirmed with TS$>$45 when working with 7.2 yrs of Pass 8 data.

\subsection{Detection efficiency according to FOM parameter}
\label{deteceffic}

The figure of merit (FOM) parameter \citep{1WHSP} is defined as the ratio between the synchrotron peak flux \nufnupeak\ of a given source and that of the faintest 1WHSP blazar already detected in the TeV band ( \nufnupeak= 10$^{-11.3}$ erg/cm$^2$/s); FOM = \nufnupeak\ / 10$^{-11.3}$. The FOM  then provides an objective way of assessing the likelihood for GeV-TeV detection of HSP blazars, based on the synchrotron peak brightness, and is not affected by absorption of VHE photons owing to the interaction with extragalactic background light \citep[EBL,][]{EBLabsorption}.

\begin{figure}[h!]
   \centering
    \includegraphics[width=1.0\linewidth]{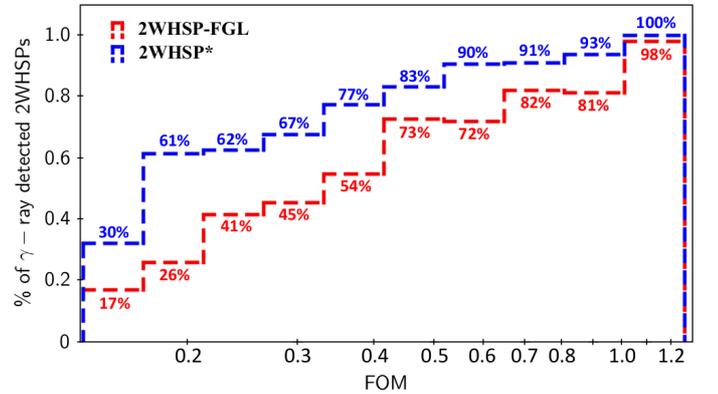}
     \caption{The $\gamma$-ray detection efficiency for each bin in FOM. Red represents the 2WHSP-FGL subsample (2WHSP sources with FGL counterparts: 439 objects), and blue the 2WHSP* subsample (2WHSP sources considering all 150 new + 439 FGL $\gamma$-ray counterparts). The first bin at FOM=1.2 condensate  all cases with FOM$>$1.2 (sources with the brightest synchrotron peak flux -11.2$<$Log(\nufnupeak)$<$-9.7) since almost all of them have been already $\gamma$-ray detected, having a 3FGL counterpart.}
      \label{FOMhisto}
\end{figure}

Figure \ref{FOMhisto} illustrates this concept, showing the fraction of 2WHSPs already detected in $\gamma$-ray according to different FOM bins, considering the subsample of 439 2WHSP-FGL sources (in red), and the subsample of 589 2WHSP* \footnote{We may use 2WHSP* when referring to the total 589 sources that include: 439 2WHSP-FGL + our 150 $\gamma$-ray detections at TS$>$10 level.}, which represents all $\gamma$-ray detections (in blue).  Clearly, the detection efficiency increases with increasing FOM, and there is a considerable increment in the fraction of sources detected for each FOM-bin when accounting for the 150 sources listed in Table \ref{tableFermi1}. Therefore, the 2WHSP sample shows its potential for unveiling high/lower-significance $\gamma$-ray sources, emphasising the power of considering multi-frequency information to select VHE $\gamma$-ray targets for Cherenkov Telescope Arrays, as discussed in \cite{1WHSP} and \cite{2WHSP}.
 In addition, given that $\gamma$-ray detected HSP blazars have been suggested as counterparts of IceCube astrophysical neutrinos \citep{Neutrino-HSP}, our present work may contribute to discussions in the realm of multi-messenger astrophysics, especially when studying  cross-correlations between extreme $\gamma$-ray blazars and astro-particles.

\subsection{The $\gamma$-ray spectral properties of 2WHSP blazars}
\label{slopeDist}

In Fig. \ref{histoindex} we present the photon spectral index ($\Gamma$) distribution for the 150 new $\gamma$-ray excess signals (indigo), and compare it with the $\Gamma$ distribution for the 439 2WHSP-FGL sources (red continuous line). The histogram is normalized with respect to the size of each subsample, so we can visualise their distribution-shape more accurately. A Kolmogorov-Smirnov (KS) test comparing both histograms gives a p$_{value}$=0.991, meaning the distributions are fully consistent with the same parent population and have similar $\gamma$-ray distribution properties. Also, the mean photon spectral index only associated with the 150 new $\gamma$-ray sources is $ \langle \Gamma \rangle_{new} =1.94 \pm 0.03$ in good agreement with that calculated for the 2WHSP-FGL sample $ \langle \Gamma \rangle_{2WHSP-FGL} =1.89 \pm 0.01$. Considering all $\gamma$-ray detections together (the 2WHSP* subsample) we have $ \langle \Gamma \rangle_{2WHSP^*} =1.90 \pm 0.01$.

\begin{figure}[h!]
   \centering
    \includegraphics[width=0.95\linewidth]{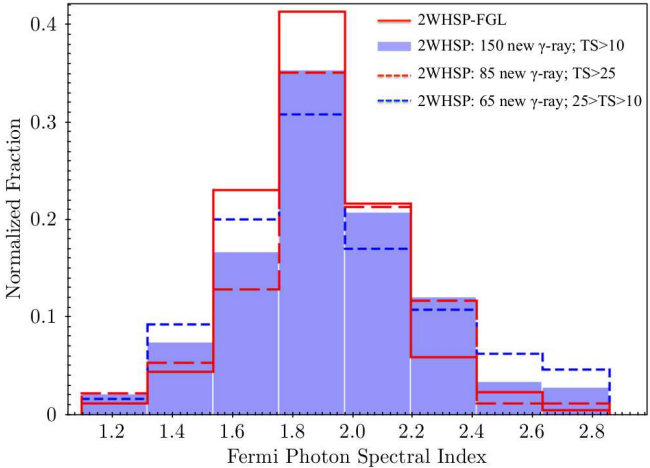}
     \caption{Distribution of photon spectral index $\Gamma$ for the 439 2WHSP-FGL sources (red-continuous line). For the new $\gamma$-ray signatures we have: solid indigo bars considers all the 150 detections at TS$>$10, red dashed line only for the 85 detections at TS$>$25 level, and blue dashed line only for the 65 lower-significance detections with TS in between 10 to 25. }
      \label{histoindex}
\end{figure}

When comparing the photon spectral index distribution of 2WHSP-FGL with the subsample only having our 85 new $\gamma$-ray detections at TS$>$25; also with the one only having our 65 lower-significance $\gamma$-ray detections, the p-values are respectively: p$_{value}^{({\rm TS}>25)}$=0.987 and p$_{value}^{(10<{\rm TS}<25)}$=0.763. Therefore, since all the cases we compared showed p$_{value}>$0.05, we should not reject the hypothesis that all distributions are similar, consistent with a single-parent population.

The $\Gamma$ vs. S$_{1-100 \ {\rm GeV}}$ plot (Fig. \ref{GammaFlux}) shows how we went into lower flux-limit (blue dashed line) compared to previous $\gamma$-ray catalogs (red dashed line). This improvement is a combination of many elements: Our dedicated search for $\gamma$-ray counterparts based on WHSP positions, the larger exposure time used (since we integrate over 7.2 yrs of observations), better events reconstruction (from Pass7 to Pass 8), and also the fact that we consider sources down to TS$>$10.

\begin{figure}[h]
   \centering
    \includegraphics[width=1.0\linewidth]{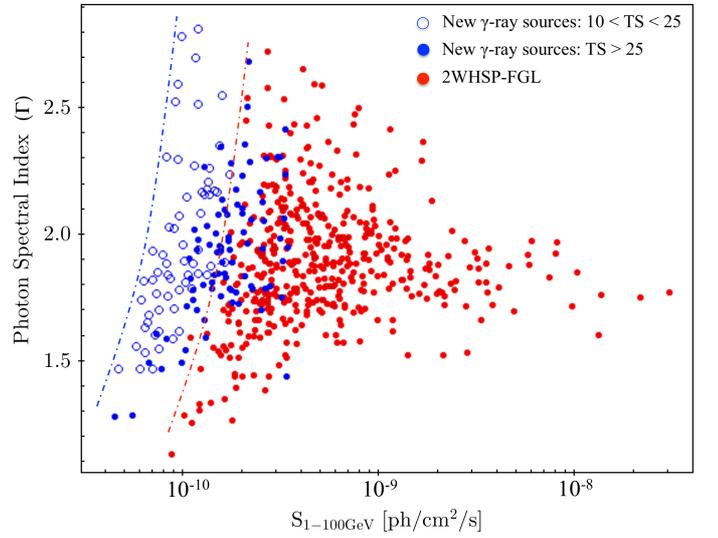}
     \caption{Photon spectral index $\Gamma$ plotted against total flux S$_{1-100 \ {\rm GeV}}$ for the 439 2WHSP-FGL sources (in red), the 85 new detections with TS$>$25 (filled-in blue), and for the 65 lower-significance detections with TS between 10 and 25 (blue outlines). The dashed lines represent the flux limit achieved by 3FGL-4 yrs (red) and by our direct search based on 7.2 yrs of data (blue) down to TS=10.}
      \label{GammaFlux}
\end{figure}

The overlapping between red and blue dots (Fig. \ref{GammaFlux}) in the range 1-4$\times$10$^{-10}$ ph/cm$^2$/s illustrates how we  improved completeness for our HSP $\gamma$-ray sample, when considering the new detections presented in Table \ref{tableFermi1}. We note that the $\gamma$-ray threshold sensitivity for HSP blazars has little dependence on the photon spectral index down to S$_{1-100 \ {\rm GeV}}^{limit} = 7 \times$10$^{-11}$ ph/cm$^2$/s, so that sub-samples with flux-limit $>$ S$_{1-100 \ {\rm GeV}}^{limit}$ have low bias arising from $\Gamma$. On the other hand, the threshold dependence on $\Gamma$ is much stronger when considering the integrated flux along the whole band 0.1-100 GeV, as reported in \cite{2FGL} and \cite{3FGL}. Therefore, the discussion in section \ref{secIGRB} considers the 1-100 GeV energy range\footnote{The integral flux for the energy range 1-100 GeV is commonly reported in all Fermi-LAT  catalogs (1FGL, 2FGL and 3FGL). For practical reasons we work in the same energy range, making it easy to combine flux information from our current list (Table \ref{tableFermi1}) with the 1-100 GeV flux  reported in Fermi-LAT catalogs. Moreover, since the 1-100 GeV band is covered with relatively good sensitivity by Fermi-LAT, the power law modelling of faint hard-spectrum $\gamma$-ray sources is more reliable in this range. } for the flux distribution histogram (Fig. \ref{hist-flux}) and also for the $\gamma$-ray LogN-LogS studies (Fig. \ref{gammaLogN} and \ref{diff-log}).

If we plot the histogram of $\gamma$-ray flux for the 2WHSP-FGL subsample, comparing it to our 150 $\gamma$-ray detections  with TS$>$10 (Fig. \ref{hist-flux}), we see that our sources dominate the faint-end. A KS test comparing both histograms gives a p-value of 0.062, which is relatively low, almost excluding the hypothesis that the histograms are similar with respect to the flux distribution. In fact, it shows that our new $\gamma$-ray detections (Table \ref{tableFermi1}) are part of a population of faint sources that was not probed before, and represents a contribution to the IGRB that was previously unresolved.

\begin{figure}[h]
   \centering
    \includegraphics[width=0.95\linewidth,angle=0]{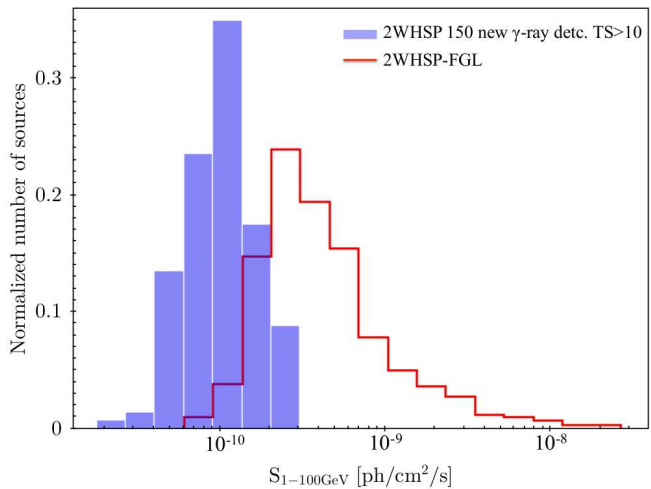}
     \caption{Histogram comparing the flux distribution S$_{1-100 \ {\rm GeV}}$ for the $\gamma$-ray subsamples of 439 2WHSP-FGL (red line), and the 150 newly detected WHSPs at TS$>$10 level (indigo box).}
      \label{hist-flux}
\end{figure}

As is clear from Fig. \ref{hist-flux}, there is a region in the S$_{1-100 \ {\rm GeV}}$ histogram where our new detections overlap the 2WHSP-FGL sources. For fluxes lower than $\sim$2.7$\times$10$^{-10}$ ph/cm$^{2}$/s the detection efficiency from the 3FGL catalog drops considerably (also shown by a sharp cut in differential number counts dN/dS, Fig. \ref{diff-log} in Sect. \ref{secIGRB}). As discussed in Sect. \ref{prova}, our new detections at the faint-end of S$_{1-100 \ {\rm GeV}}$ histogram are mainly driven by longer integration time, improved event reconstruction, and a lower TS threshold; not only, a direct search can bring complementary sources that improve detection efficiency close to the flux-limit. In addition, the Fermi-LAT exposure is not uniform \citep[][see their Fig. 1]{3FGL} and therefore sky-regions inspected with lower exposure (4-yrs  with Pass 7) now benefit from better sensitivity owing to longer exposure, revealing new sources at the same faintest flux levels probed by the 3FGL setup. Therefore, taking all of this into consideration, we naturally expect an overlap in the S$_{1-100 \ {\rm GeV}}$ faint-end.

\subsection{Comments on the Eddington bias effect}
\label{Eddington}

\cite{50GeVbackground} has called attention to the statistical fluctuations of photon flux, especially for faint $\gamma$-ray sources close to the Fermi-LAT detection limit, which could lead to overestimated flux-values. The statistical fluctuation of sources close to the flux threshold of any sample is known as an Eddington bias \citep{EddingtonBias} and has a direct impact on the number counts (LogN-LogS) or any other study relying upon the measured flux. For the 2FHL catalog it has been shown through simulations \citep{50GeVbackground} that the  measured fluxes along 50 GeV-2 TeV band could be overestimated up to 10$\times$ for the faintest sources. However, we should note that this factor also has a strong dependence on the $\gamma$-ray spectral properties from individual sources. 

Especially, the $\Gamma_{(50 GeV-2 TeV)}$ distribution for the 2FHL sample ranges from $\approx$ 1.0 to 5.5 (see Fig. \ref{dNdGamma}) with mean value $ \langle \Gamma \rangle_{2FHL} =3.20 \pm 0.08$. Naturally, statistical fluctuations on photon flux measurements are more extreme for the steepest $\gamma$-ray spectra (Fig. \ref{dNdGamma}, right side).

\begin{figure}[h]
   \centering
    \includegraphics[width=0.95\linewidth,angle=0]{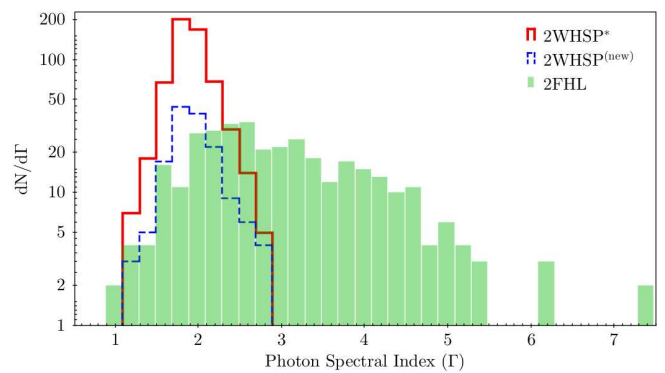}
     \caption{Histogram comparing the photon spectrum index ($\Gamma$) distribution for the $\gamma$-ray samples 2FHL (all sources), 2WHSP$^*$ (all $\gamma$-ray detected 2WHSPs down to TS=10), and 2WHSP$^{(new)}$ (only the 150 new $\gamma$-ray detections). Note that this is a qualitative comparison given that $\Gamma$ parameter for the 2FHL sample is measured in the 50 GeV-2 TeV energy band.}
      \label{dNdGamma}
\end{figure}

In this case, we may be subject to the same effect, however the $\Gamma_{(0.3-500 GeV)}$ distribution for the 2WHSP$^*$ sample is well confined to the 1.2 to 2.8 range, with mean value $ \langle \Gamma \rangle_{2WHSP^*} =1.90 \pm 0.01$. Since the mean $\gamma$-ray  spectrum for the 2WHSP$^*$ sources is close to flat, the effect of statistical flux fluctuations is much less severe in our sample, and does not compromise the measured-flux.

To estimate the effect of the Eddington bias in the faint-end of our sample we  compare the parameter S$_{1-100 \ {\rm GeV}}$ from Tables \ref{tableFermi1} and \ref{tablepass7} that were calculated for the same sources, but with different flux limits (see Sect. \ref{prova}). In the first case, the likelihood setup is based on 7.2 yrs Pass 8 data, and the second one is based on 4 yrs Pass 7 data; therefore different flux-thresholds. 

From Table \ref{tablepass7} let us assume that the 17 sources with TS$^{Pass \ 7}_{4 \ yrs}$ in between 10-25  are a good representative of our lower-significance detections, for which the measured fluxes S$^{meas}_{1-100 \ {\rm GeV}}$ could be overestimated. When analysing these same 17 sources with an advanced setup of 7.2 yrs with Pass 8 (Table \ref{tableFermi1}) all of them become $\gamma$-ray detected with relatively high-significance TS$^{Pass \ 8}_{7.2 \ yrs}>$45, and their measured fluxes can be considered as the true ones, S$^{true}_{1-100 \ {\rm GeV}}$, since the flux threshold is now relatively improved. 

We then calculate $\langle$S$^{meas}_{1-100 \ {\rm GeV}}$$\rangle$/$\langle$S$^{true}_{1-100 \ {\rm GeV}}$$\rangle$=0.79 as an estimate for the order of magnitude of flux fluctuations for 2WHSP sources close to the Fermi-LAT threshold. This is far from the 10$\times$ factor that could affect FHL sources, especially the ones with a steep $\gamma$-ray spectrum. Clearly, the effect is not representative for our sample and does not compromise further results. Moreover, in systematically overestimating the flux from faint sources, the Eddington bias would manifest as re-steepening in the number counts, which is not observed (see Fig. \ref{gammaLogN} and \ref{diff-log}).

In conclusion, the $\langle$S$^{meas}_{1-100 \ {\rm GeV}}$$\rangle$/$\langle$S$^{true}_{1-100 \ {\rm GeV}}$$\rangle$ value tells us that the $\gamma$-ray variability associated with HSP blazars probably dominates eventual oscillations of the $\langle$S$_{1-100 \ {\rm GeV}}\rangle$ parameter when the two likelihood analysis-setups are compared. Also, it shows that the overlapping between $\gamma$-ray subsamples in Fig. \ref{hist-flux} is mainly driven by better detection efficiency (from longer exposure time and improved event reconstruction from Pass 8) rather than statistical flux fluctuations.

\section{The isotropic $\gamma$-ray background: Contribution from HSP blazars to the diffuse component}
\label{secIGRB}

Since we unveil and model a relatively large number of $\gamma$-ray emitters down to TS=10, we try to evaluate quantitatively what is the impact of our approach for resolving the extragalactic $\gamma$-ray background (EGB), and isotropic $\gamma$-ray background (IGRB) components.

\begin{table*}[ht]
\centering
\caption{Integral contribution from our new $\gamma$-ray sources for different energy bins, compared to EGB and IGRB fluxes reported in \citep{dgrbFermi}. Columns headed $\%$EGB show the significance of new detections with respect to the total EGB component, and columns headed $\%$IGRB show the fraction of IGRB component we have solved in the present work. The superscripts TS identify the cases where we only consider our new detections at TS$>$25 level, and the case considering all new detections at TS$>$10 level. Here we take into account our detections at $|b|>$10$^{\circ}$, and all intensities reported are in units [ph/cm$^{2}$/s/sr].}  
\begin{tabular}{ccc|ccc|ccc}
E$_{bin}$ (GeV)  &  I$_{{\rm EGB}}$ &  I$_{{\rm IGRB}}$ &  I$_{new-detc}^{{\rm TS}>25}$  &  $\%$EGB & $\%$IGRB & I$_{new-detc}^{{\rm TS}>10}$ &  $\%$EGB & $\%$IGRB  \\
\hline

1.1-13  & 4.5$\times$10$^{-7}$  & 2.7$\times$10$^{-7}$   &  1.22$\times$10$^{-9}$  &  0.27\% &  0.45\%  & 1.73$\times$10$^{-9}$   & 0.38\% & 0.65\%  \\

13-36   & 1.4$\times$10$^{-8}$  & 8.2$\times$10$^{-9}$   &  8.48$\times$10$^{-11}$ &  0.61\% &  1.0\%   & 1.19$\times$10$^{-10}$  & 0.86\% & 1.4\%   \\

36-51   & 1.8$\times$10$^{-9}$  & 1.1$\times$10$^{-9}$   &  1.59$\times$10$^{-11}$ &  0.88\% &  1.4\%   & 2.24$\times$10$^{-11}$  & 1.2\%  & 2.0\%   \\

51-72   & 1.1$\times$10$^{-9}$  & 6.3$\times$10$^{-10}$  &  1.19$\times$10$^{-11}$ &  1.1\%  &  1.9\%   & 1.68$\times$10$^{-11}$  &  1.5\%  & 2.7\% \\

72-100  & 6.2$\times$10$^{-10}$ & 3.6$\times$10$^{-10}$  &  8.69$\times$10$^{-12}$ &  1.4\%  &  2.4\%   & 1.23$\times$10$^{-11}$  & 2.0\%  & 3.4\%   \\

100-140 & 3.1$\times$10$^{-10}$ & 1.5$\times$10$^{-10}$  &  6.89$\times$10$^{-12}$ &  2.2\%  &  4.6\%   & 9.82$\times$10$^{-12}$  & 3.2\%  & 6.5\%   \\

140-200 & 1.9$\times$10$^{-10}$ & 9.8$\times$10$^{-11}$  &  5.63$\times$10$^{-12}$ &  2.9\%  &  5.7\%   & 8.05$\times$10$^{-12}$  & 4.2\%  & 8.2\%  \\

\end{tabular}
\label{IGRB}
\end{table*}

Following the discussion from \cite{dgrbFermi}, we refer to EGB as the sum of all resolved and unresolved contributions from individual extragalactic sources (Blazars, misaligned AGNs and Starburst Galaxies) plus the diffuse emission coming from outer Milky Way regions (which could be related to dark-matter annihilation, intergalactic shocks, and $\gamma$-ray cascades induced by ultra high-energy cosmic rays). The exact EGB composition is a matter of intense debate\footnote{ \cite{dgrbFermi} also discuss the challenges for measuring the EGB component, which demands a proper modelling of the diffuse Galactic emission (DGE) especially as a result of cosmic rays interacting with the Milky Way gas and photon fields. The DGE has intensity comparable to the EGB, and to obtain the EGB, both the DGE and the known Galactic sources have to be subtracted from the total-sky $\gamma$-ray counts. The reported EGB flux (1.1-200 GeV) is I$_{{\rm EGB}} \approx 4.74 \times 10^{-7}$ ph/cm2/s/sr.}, and it is well known that contributions owing to unresolved sources may build-up a dominant fraction of the diffuse EGB. Therefore, resolving point-like sources translates directly into narrowing the window available for putative truly-diffusive components, especially when constraining the upper-limits of dark-matter annihilation cross-section as discussed by e.g. \cite{OriginEGB} and \cite{NatureIDGB}.

Another term commonly used is isotropic $\gamma$-ray background (IGRB), and it is obtained by subtracting the known extragalactic sources from the EGB. Therefore IGRB represents the sum of a true extragalactic diffuse component, plus the contribution of unresolved sources (which mimic and contaminate the diffuse component). According to \cite{DifuseGammaBLlacs}, \cite{IsoDifuseGammaBackGround} and \cite{EGBpaolo}, unresolved HSPs/BL lacs may be the dominant component of the IGRB at E$>$10 GeV and indeed, here we bring evidence of a population composed of faint $\gamma$-ray HSP blazars near the detectability threshold from Fermi-LAT, that was previously undetected.

Based on the model for individual sources, we calculate their corresponding fluxes for each energy bin (E$_{bin}$ are listed in Table \ref{IGRB}), and sum over our $\gamma$-ray detections at $|b|>$10$^{\circ}$ listed in Table \ref{tableFermi1}. We then normalize these values multiplying by A$_{sky}$/(4$\pi$ $\times$ A$_{|b|>10^{\circ}}$), where A$_{sky}$ = 41252.96 deg$^2$ is the total sky area, A$_{|b|>10^{\circ}}$ = 34110.3 deg$^{2}$ is the sky area out of the Galactic disk, and the factor 4$\pi$ normalize the integral flux per unit of steradian, [ph/cm$^2$/s/sr], written as I$_{new-detc}$. We compared I$_{new-detc}$ with the IGRB and EGB intensities (I$_{\rm {IGRB}}$ and I$_{{\rm EGB}}$) as reported by \cite{dgrbFermi}, following the same E$_{bin}$ steps as theirs. Since our source-modelling does not account for broken power law features that may arise from EBL absorption, especially for large redshift sources, we extend our calculations up to 200 GeV only.

Table \ref{IGRB} lists the corresponding IGRB and EGR fractions we resolved, showing that the subsample of previously undetected $\gamma$-ray HSP blazars has increasing relevance for the background composition at higher energies. We evaluate separately the impact owing to all new detections at TS$>$10 level, and also owing only to the cases reported with TS$>$25. This helped us understand what to expect (in terms of ability to solve the IGRB) from dedicated source-searches based on catalogs of potential $\gamma$-ray candidates, and also to evaluate the importance of taking into account lower-significance detections from faint $\gamma$-ray blazars. As can be seen, their contribution is not negligible, showing an increment of the order of 40$\%$ larger $\%$IGRB solved for each energy bin if we compare the subsamples of our new $\gamma$-ray sources detected at TS$>$25 and TS$>$10.


We also report on the LogN-LogS for HSP blazars using all $\gamma$-ray information currently available for the 2WHSP sample. Especially, we incorporate a complementary description for the HSP population at lower $\gamma$-ray fluxes by considering our 150 new $\gamma$-ray detections down to TS$=$10 level. We define the 2WHSP$^*$ $\gamma$-ray sample (which encloses all 2WHSP-FGL sources together with our new detections) and compare it to the 2WHSP-FGL. 
\begin{figure}[h!]
   \centering
    \includegraphics[width=1.0\linewidth,angle=0]{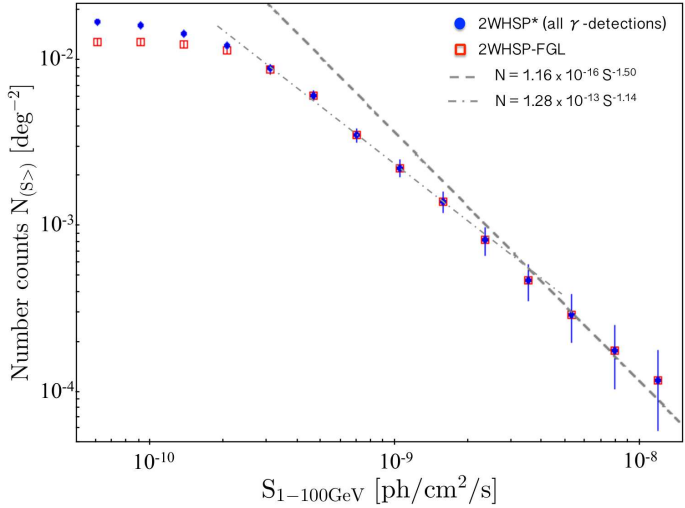}
     \caption{Measured $\gamma$-ray LogN-LogS of 2WHSP sources, plotting the cumulative number counts with integrated fluxes larger than S$_{1-100 \ {\rm GeV}}$, at $|b|>10^{\circ}$. The dashed lines represent a broken power law fit to the 2WHSP* sample, with an early break at S$_{break-1}$=3.5$\times$10$^{-9}$ ph/cm$^{2}$/s and fitting parameters given in Eq. \ref{logNfit}. This plot is not corrected for non-uniform exposure from Fermi-LAT.}
      \label{gammaLogN}
\end{figure}

In Fig.\ref{gammaLogN} we plot the cumulative number counts N [deg$^{-2}$] with flux S$_{1-100 \ {\rm GeV}}$ larger than the corresponding value on the x-axis. By fitting the $\gamma$-ray LogN-LogS with a broken power law (Fig. \ref{gammaLogN}), we consider an Euclidian behaviour for the bright-end, and an early break\footnote{To confirm the presence of this early break in the number counts, we  extracted the LogN-LogS data from \cite{3FGL} paper (their Fig. 29) plotting the cumulative energy flux $S_{energy}$ distribution for the clean sample of HSP blazars (also uncorrected for non-uniform sensitivity and detection efficiency). Although they do not mention the fitting parameters, we found good agreement with a broken power law that has similar slopes as ours: N$_{(S>S_{break-1})}$ = 3.2 $\times$ 10$^{-19}$ S$_{energy}^{-1.5}$, N$_{(S<S_{break-1})}$ = 1.8 $\times$ 10$^{-15}$ S$_{energy}^{-1.14}$; In this case S$_{break-1} \approx$ 3.5 $\times$ 10$^{-11}$ ergs/cm$^2$/s, but there is also a strong cut at S$_{break-2} \approx$ 7.0 $\times$ 10$^{-12}$ ergs/cm$^2$/s. Therefore also manifesting two breaks that probably have the same origin as in our case.} at S$_{break-1}$=3.5$\times$10$^{-9}$ ph/cm$^{2}$/s: 

  \begin{equation}
     {\rm N_{(S)} }=\left\{
                \begin{array}{ll}
             1.16 \times 10^{-16} {\rm S}^{-1.50}  \ \ \ \  \ \ {\rm S>S}_{break-1}  \\
             1.28 \times 10^{-13} {\rm S}^{-1.14}  \ \ \ \ \ \ {\rm S<S}_{break-1}  \\
                \end{array}
              \right. 
              \label{logNfit}
  \end{equation}

As a test of consistency, we calculate the number of sources n predicted by the fitting (Eq. \ref{logNfit}), which have flux in the interval $S_{min} < S < S_{max}$ (with S$_{max}$=1.0$\times$10$^{-8}$ and S$_{min}$=3.0$\times$10$^{-10}$ ph/cm$^2$/s) where the LogN-LogS is well described by the broken power law. The number of sources predicted is: $ n_{(fit)} = A_{|b|>10^{\circ}} \int_{S_{min}}^{S_{max}} dN/dS \times dS  $, where the parameter $A_{|b|>10^{\circ}}$=34110.3 deg$^{2}$ is the sky area at high Galactic latitudes $|b|>10^{\circ}$, resulting in n$_{(fit)}\approx$ 309.5, which is in very good agreement with the number of $\gamma$-ray sources expected from the 2WHSP$^{*}$ sample in this same interval, n$_{(2WHSP*)}$ = 311.

An important point to mention is that our new detections only add improvements to the LogN-LogS after the second break. The region in between the first and the second break is not affected by the incompleteness of the 2WHSP $\gamma$-ray sample. However, even if the LogN-LogS for S$<$S$_{break-1}$ deviates from the Euclidian prediction, it is early to argue we are probing evolution of HSP blazars; further studies need to introduce corrections owing to non-uniform exposure from Fermi-LAT.

When plotting the differential number counts -dN/dS vs. S$_{1-100 \ {\rm GeV}}$ (Fig. \ref{diff-log}), the low flux threshold (S$_{break-2}$) is evidenced. This manifests as a sharp cut in dN/dS at S$\sim$2.7$\times$10$^{-10}$ ph/cm$^{2}$/s, showing that incompleteness becomes severe for both samples (2WHSP-FGL and 2WHSP$^*$) at that particular flux level. Figure \ref{diff-log} also clearly presents how we increment the $\gamma$-ray sample completeness, specially at the faint-end where the 2WHSP* (blue) detaches from the 2WHSP-FGL (red). 

\begin{figure}[h]
   \centering
    \hspace*{-0.4cm}\includegraphics[width=1.0\linewidth,angle=0]{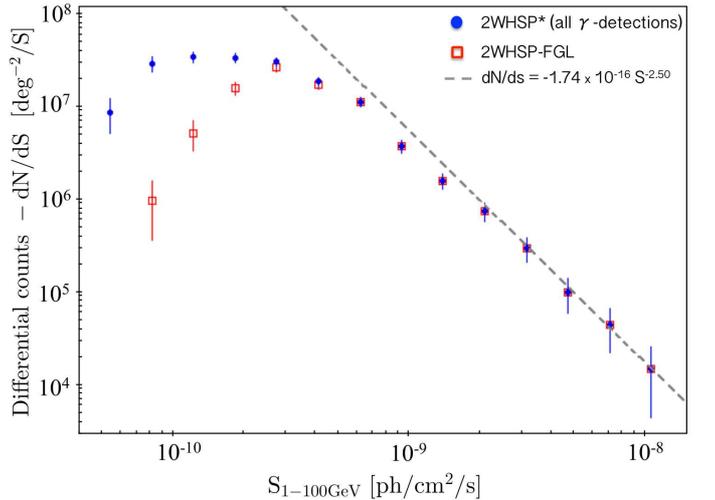}
     \caption{Differential number counts (-dN/dS) with respect to S$_{1-100 \ {\rm GeV}}$ for HSP blazars at $|b|>10^{\circ}$. The dashed and dash-dotted lines represent the derivative for the power law fit from Eq. \ref{logNfit}.}
      \label{diff-log}
\end{figure}

If we assume there is no cut owing to the flux threshold at S$_{break-2}$, so that the power law (Eq. \ref{logNfit}) is a good description for the number counts when extrapolating to the faint-end (from S$_{max}$=1.0$\times$10$^{-8}$ down to S$_{min}$=6.0$\times$10$^{-11}$ ph/cm$^2$/s)\footnote{We choose the faint-end to be $S_{min}=6.0 \times 10^{-11}$ ph/cm$^2$/s since this is consistent with our flux threshold; as can be seen in Fig. \ref{hist-flux}, there is a sharp cut in the number of $\gamma$-ray sources for fluxes lower than that.}, it is possible to estimate the integral contribution of HSP blazars to the EGB in the 1-100 GeV band, I$_{1-100 \ {\rm GeV}}$:

\begin{equation}
{\rm I^{HSPs}_{(1-100 \ {\rm GeV})} = \frac{A_{sky}}{4 \pi} \int^{S_{max}}_{S_{min}}  S \frac{dN}{dS} dS  \ \ \ \ [ph/cm^{2}/s/sr] };
\label{intensity}
\end{equation}
where A$_{sky}$ = 41252.96 deg$^2$, and the factor 4$\pi$ is for normalizing the total flux per unit of sky-steradian. The total 1-100 GeV flux generate by HSP blazars is of the order I$^{HSPs}_{(1-100 \ {\rm GeV})}\approx$ 4.80 $\times$ 10$^{-8}$ ph/cm$^{2}$/s/sr, which represents $\approx$ 8.5$\%$ when compared to the total EGB content for the same energy band I$^{{\rm EGB}}_{(1-100 \ {\rm GeV})}$ = 5.63 $\times$ 10$^{-7}$ ph/cm$^2$/s/sr \cite[][their Table 3]{dgrbFermi}.

We should note (from Fig. \ref{gammaLogN}) that the fitting presented in Eq. \ref{logNfit} is suitable for the flux range S$>$S$_{break-2}$ ($\approx$2.7$\times$10$^{-10}$ ph/cm$^{2}$/s), which is well described by the 2WHSP-FGL subsample even without incorporating the 150 new $\gamma$-ray detections. In fact, our new detections mainly address the problem of incompleteness at S$_{1-100 \ {\rm GeV}}<$ 2.7$\times$10$^{-10}$ ph/cm$^2$/s, as evidenced from Fig. \ref{diff-log}. However, we are now confident of extrapolating Eq. \ref{logNfit} down to S$_{min}$=6.0$\times$10$^{-11}$ ph/cm$^2$/s only because our new detections push to a lower flux threshold. We emphasise that the measured $\gamma$-ray LogN-LogS was calculated without corrections for non-uniform Fermi-LAT exposure. Therefore, the total flux estimated when extrapolating the LogN-LogS to lower fluxes should be regarded as a lower bound to the true contribution of HSP blazars in the 1-100 GeV band, bearing in mind that HSPs have increasing relevance for the high-energy channels (Table \ref{IGRB}).

Other intervening factors to mention, that may add corrections to the LogN-LogS fitting are: 

\begin{itemize}

\item The point spread function (PSF) and effective area from Fermi-LAT depends on energy, therefore the true sensitivity-limits rely on the intrinsic source spectrum properties. 

\item The data taken mode is turned off during Fermi-LAT passages along the South Atlantic Anomaly inducing $\approx$15\% sensitivity differences between North and South hemisphere.

\item Another bias is related to the incompleteness introduced by the poor all-sky coverage in X-rays, and probably an extra component owing to limitations imposed by current radio surveys (SUMMS and NVSS) that were used when building the 2WHSP sample. Evolution of the HSP Population could play an important role as well and demands further investigation.

\end{itemize}

Therefore, a refined representation of the $\gamma$-ray LogN-LogS for HSP blazars demands further corrections (see \cite{50GeVbackground} for a practical example) that needs to incorporate parameters like the Fermi-LAT detection efficiency, sensitivity non-uniformities along the sky, and selection-efficiency of current blazar catalogs.

\section{Addressing unassociated sources and confusion} 
\label{sectConf}

In Sect. \ref{bright-peak} we described the selection of $\sim$400 2WHSP sources to search for their $\gamma$-ray  signatures using the Fermi Science Tools. Before any likelihood analysis takes place, we inspect the region within 60' radius from all  candidates, considering multi-frequency databases from radio to $\gamma$-rays (using the Sky-Explorer Tool at tools.asdc.asi.it). In this process, we found four cases where the 2WHSP $\gamma$-ray candidates were close to one of the 3FGL sources, but outside, or at the border of, 3FGL error-circles. In the following we study these fields in more details by working with energy dependent TS maps, trying to improve the $\gamma$-ray signature description and confirm the association. 

The cases studied are 3FGL J0536.4-3347, 3FGL J0935.1-1736, 3FGL J0421.6+1950, and 3FGL J1838.5-6006; well representative examples of how a multi-frequency approach can lead to refined scientific products, especially for $\gamma$-ray confused sources.

In particular we draw attention to 3FGL J0935.1-1736 and 3FGL J0421.6+1950 which are currently unassociated. As known, a large number of 3FGL objects (1058) have no official association to date, despite the fact that many of them have blazars and AGNs as main association-candidates (especially the 541 unassociated $\gamma$-ray sources out of the Galactic plane $|b|>$10$^\circ$ \citep{Fujinaga07052015,UnassocFGL-AGN}). Although a clear picture that accounts for the large fraction of unassociated 3FGL source is yet to be build, there is evidence of sources that are not clearly related to pulsars nor to AGNs \citep{Unassoc2FGL}. In this context, any new association may help to clarify the true nature of current unassociated $\gamma$-ray source, and therefore we report on those two cases for which we propose new associations.

\subsection{3FGL J0536.4-3347: a case of source confusion}
\label{conf1}

Source 3FGL J0536.4-3347 is one of the unassociated $\gamma$-ray detections in the 3FGL catalog. At this sky position the 2FHL catalog \citep{2FHL} reports the source 2FHLJ0536.4-3342, which has been associated with 5BZBJ0536-3343 (=2WHSPJ053628.9-334301) with SED shown in Fig. \ref{sed0536m33}. The $\gamma$-ray description of this source is very rich, being detected in the 1FGL, 2FGL and 3FGL catalogs (blue/red/green points); pink dots and u.l. correspond to the 2FHL counterpart at E$>$50 GeV. For this case, there is a steep+hard component (fig \ref{sed0536m33}) in the $\gamma$-ray SED, which may be hard to explain as intrinsic emission from a single source. 

\begin{figure}[h]
   \centering
    \includegraphics[width=0.9\linewidth]{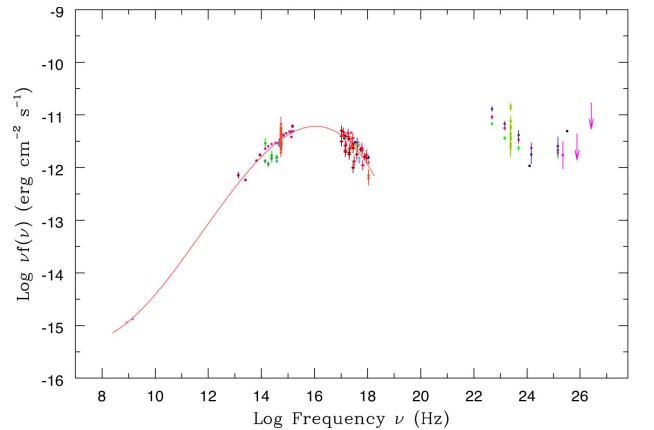}
     \caption{SED for 2WHSP J053628.9-334301 ($\equiv$5BZBJ0536-3343) with red thin line showing a fitting for the synchrotron component (radio to X-rays), and its corresponding $\gamma$-ray spectrum from 3FGL J0536.4-3347 ($\equiv$2FHL J0536.4-3342).}
      \label{sed0536m33}
\end{figure}

Exploring the sky area around 3FGL J0536.4-3347 with the ASDC error circle tool (Fig. \ref{1WHSP0536m33}) we note that blazar 2WHSP J053628.9-334301 is just outside the $\gamma$-ray error ellipse; Also, there is a bright FSRQ (5BZQ J0536-3401) within 15' from the 3FGL source, and it could contribute to the overall $\gamma$-ray flux that is observed.

\begin{figure}[h]
   \centering
    \includegraphics[width=1.0\linewidth]{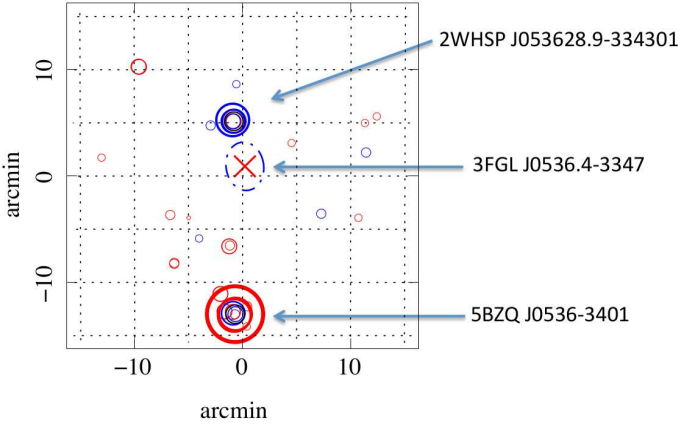}
     \caption{Sky-Explorer view around 3FGL J0536.4-3347 with position indicated by $\times$, and $\gamma$-ray error-circle shown as dotted line. The 2WHSP J053628.9-334301 (top), and 5BZQ J0536-3401 (bottom) are indicated. X-ray and radio  detections in the field are represented by blue and red circles, respectively.}
      \label{1WHSP0536m33}
\end{figure}

A likelihood analysis, assuming two $\gamma$-ray sources instead of one (with position corresponding to 5BZQ J0536-3401 and 2WHSP J053628.9-334301), results in a model adjustment with a very large statistical significance for both, as reported in Table \ref{solved1}. Each of the resolved sources is associated with a distinct $\gamma$-ray spectral component (one steep and one hard) in agreement with  expectations: hard for the 2WHSP source 
and steep for the BZQ object, as shown in Fig. \ref{confSED}. We then consider 2WHSP J053628.9-334301 as part of the 2WHSP-FGL sample (with updated $\gamma$-ray parameters), but 5BZQ J0536-3401 does not count as part of the 150 new detections associated with 2WHSP blazars (since this source is not an HSP).

\begin{table}[h]
\centering
\caption{Source model parameters from Fermi Science Tools, assuming a power law to describe the $\gamma$-ray spectrum within 0.3-500 GeV, with $N_0$ given in [ph/cm$^{2}$/s/MeV].}  
\label{tableFermi2}
\begin{tabular}{c|ccc}
Source  & $N_0$ (10$^{-13}$) & $\Gamma$  & TS   
\\
\hline

5BZQ J0536-3401           &  17.82$\pm$0.97 & 2.75$\pm$0.05 &  813.9  \\

2WHSP J053628.9-334301    &  7.08$\pm$0.84  & 1.76$\pm$0.05 &  482.8  \\
\end{tabular}
\label{solved1}
\end{table}

To validate our modelling, we also calculated TS maps for the region, taking into consideration different energy bands: 3-500 GeV (high-energy map), and 700-800 MeV (lower-energy map\footnote{In the lower-energy map we use lower resolution, to account for the larger PSF with respect to high-energy photons. In this case, the particular lower-energy range was chosen to try to balance between ``going to the lowest energies probed by Fermi-LAT" and ``still acceptable computation time" of the order of two weeks. A lower energy range could be used, but since it is a bright $\gamma$-ray source, photon-counts (and therefore computation time) escalate very rapidly.}).

\begin{figure}[h]
   \centering
    \hspace*{-0.2cm} \includegraphics[width=0.48\textwidth]{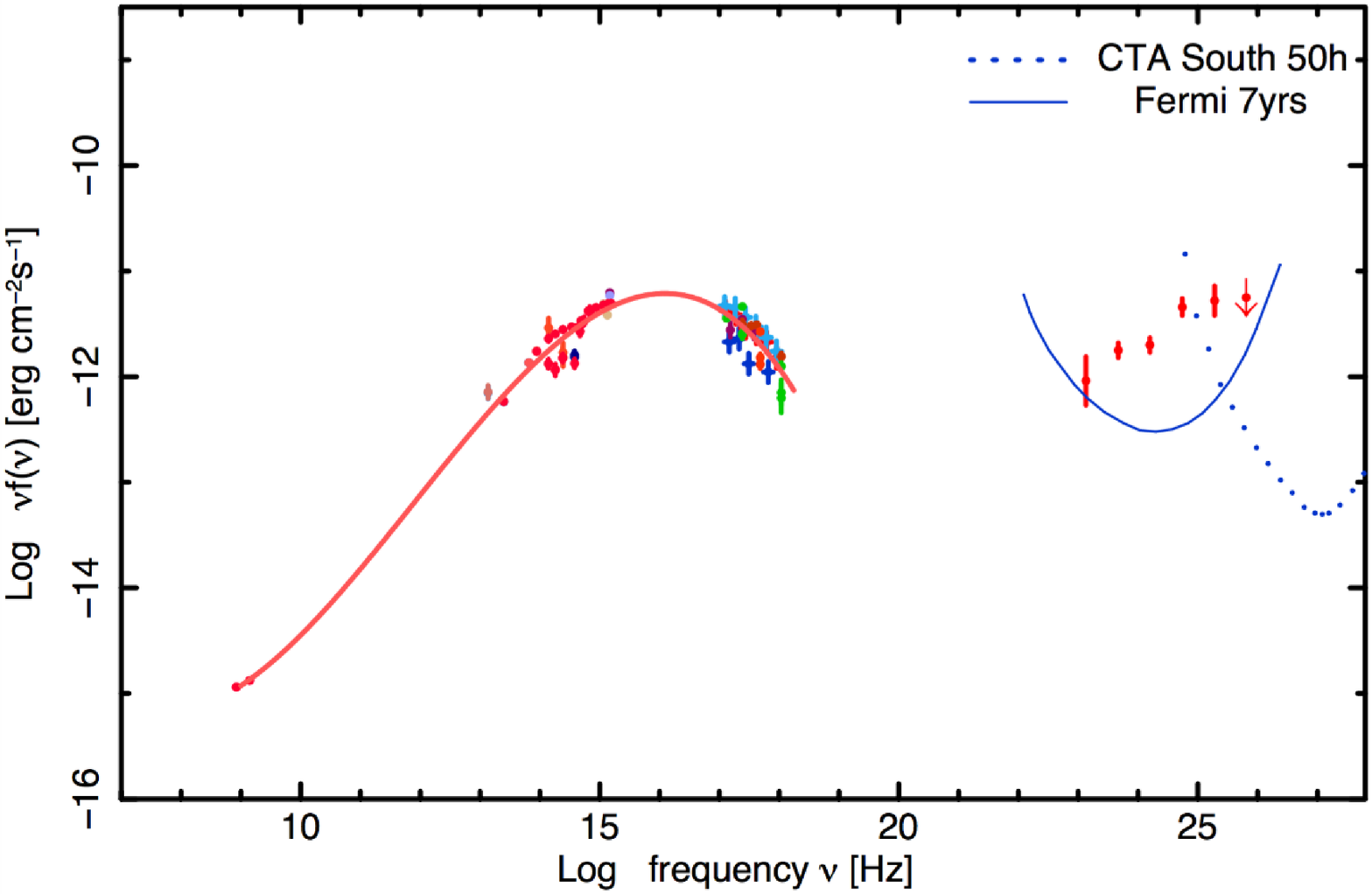}
       \centering
     \hspace*{-0.17cm}\includegraphics[width=0.48\textwidth]{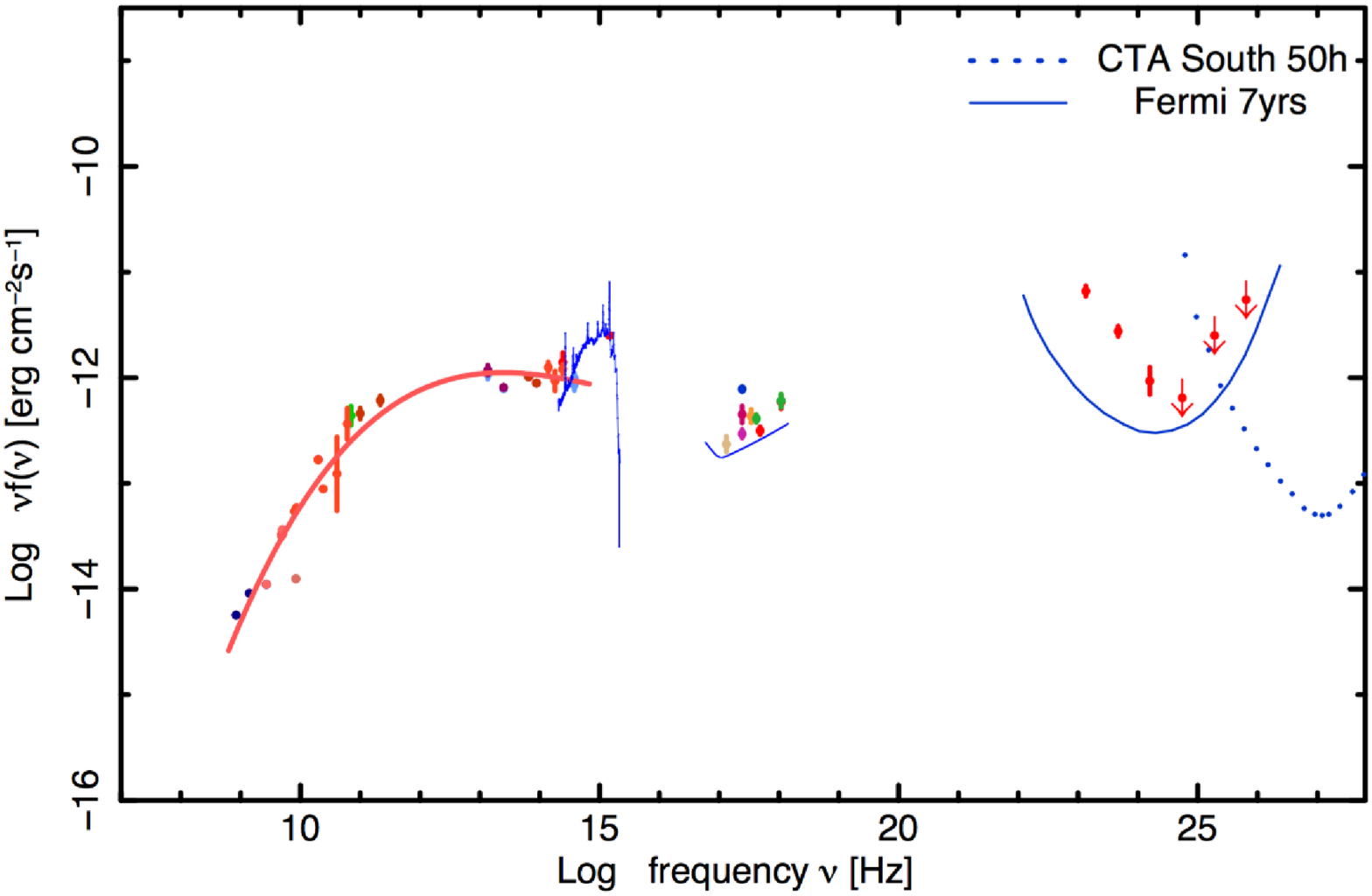}
     \caption{Top figure: multifrequency SED for 2WHSP J053628.9$-$334301; source with hard $\gamma$-ray spectrum $\Gamma_{WHSP}$=1.76. Bottom figure: multifrequency SED for 5BZQ J0536$-$3401; source with  steep $\gamma$-ray spectrum $\Gamma_{BZQ}$=2.75 (also showing the FSRQ template \citep{FSRQtemplate} as a thin blue-line along the optical to X-ray band). In the GeV-TeV band, both plots show the sensitivity curve for Fermi-LAT 7 yrs broadband detection, and for CTA-South considering 50h of exposure.}
      \label{confSED}
\end{figure}

We build the high-energy map so that the hard $\gamma$-ray spectrum source dominates, driving the TS-map peak over the 2WHSP J053628.9-334301, left side of Fig. \ref{tsmap0536m33}. The lower-energy map was build  so that the steeper source dominates, driving the TS peak over 5BZQ J0536$-$3401, right side of Fig. \ref{tsmap0536m33}. This approach can be applied for disentangling confused $\gamma$-ray components within 10'-15', just as shown in Fig. \ref{confSED} where we plot the resolved SED for both sources. One of the main reasons for source confusion is related to the PSF strong dependence on photon energy. The final position associated to the confused $\gamma$-ray sources is misplaced from their real counterparts, since the arrival direction of photons originating from distinct source are competing. In the case of close-by sources with steep/hard components, the improved PSF at high-energies may favour the association with the hard $\gamma$-ray spectrum sources (as seen in Fig. \ref{1WHSP0536m33}).

\begin{figure*}[ht]
   \centering
    \hspace*{-0.0cm}\includegraphics[width=0.94\textwidth]{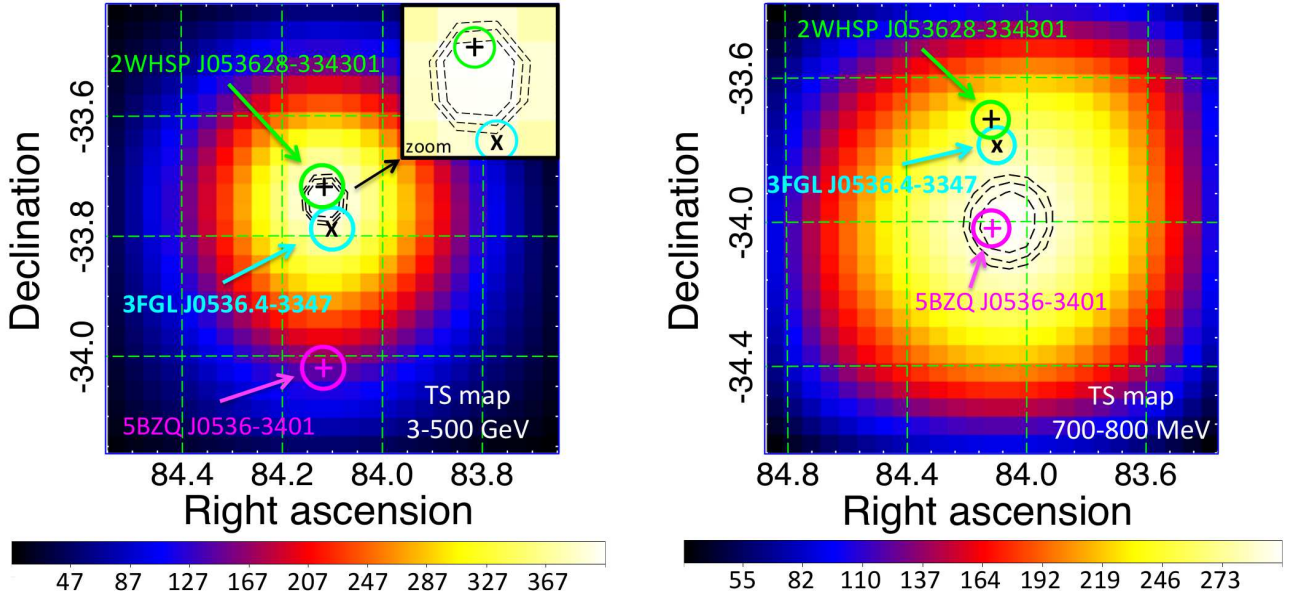}
         \caption{Energy-dependent TS maps indicating three objects in the studied field; 3FGL J0536.4-3347's position is highlighted with a cyan circle, centered on $\times$, 2WHSP J053628.9-334301 and 5BZQ J0536-3401 are highlighted with green and magenta circles, centered on + symbol. Contour dashed lines in black (from inner to outer lines) represent the 68\%, 95\%, and 99\% containment region for the $\gamma$-ray signature  Left: high-energy TS map (20x20, 0.05$^\circ$/pixel) taking into consideration only 3-500 GeV photons; zooming into central TS-peak region, the source 2WHSP J053628.9-334301 (hard $\gamma$-ray spectrum, $\Gamma_{2WHSP}$=1.76) is within the 68\% containment region for the high-energy $\gamma$-ray signature. 
         Right: lower-energy TS map (20x20, 0.08$^\circ$/pixel) considering only 700-800 MeV photons. In this case, the TS-peak position is dominated by the 5BZQ J0536-3401 (steep $\gamma$-ray spectrum source, $\Gamma_{BZQ}$=2.75) well within the 68\% containment region for the lower-energy $\gamma$-ray signature. }
      \label{tsmap0536m33}
\end{figure*}

The source 2WHSP J053628.9$-$334301 is a promising candidate for observation with Imaging Atmosphere Cherenkov Telescope (IACTs) and probably a future targets for the CTA-South array \citep{TeVAstronomy}. Therefore, any dedicated $\gamma$-ray analysis has strong motivations, especially when modelling the high-energy component of TeV candidates properly. In this case we have combined multi-frequency knowledge of potential GeV-TeV emitters with information from the TS-maps, showing  that higher quality scientific products can be extracted from the currently available data bases.

Other cases that have their $\gamma$-ray SED characterised by steep + hard component may help to identify potential cases of confusion. Source confusion between objects with similar photon spectral index seems harder to identify, but a multi-frequency study in the vicinity of each $\gamma$-ray detection is useful to evaluate the presence of potential $\gamma$-ray emitters. In a hypothetical case, where a steep+hard $\gamma$-ray spectrum could emerge from a single source (multiple blobs scenario), the present treatment would help to evaluate/rule-out the possibility of source confusion for any candidate under study.

\subsection{Solving a case of source confusion for the unassociated source 3FGL J0935.1-1736}
\label{un1}

The source 3FGL J0935.1-1736 is one of the unassociated 3FGL objects. Here we provide strong evidence for source confusion in $\gamma$-rays involving two objects: a blazar candidate (brighter $\gamma$-ray source in the field) and 2WHSP J093430.1-172120 (fainter $\gamma$-ray source in the field). In Fig. \ref{unassos1} we study the $\gamma$-ray signature in the 3-500 GeV band, to revel the TS distribution that is based on improved PSF photons. For the left-grid marked with TS map, we removed the 3FGL source from the input-model, to show the TS distribution without any bias from the 3FGL catalog. As can be seen, the TS peak matches the 3FGL position (within 99\% confinement radius, shown as dashed green line), but the $\gamma$-ray signature clearly extends towards the 2WHSP source, embracing it with high-significance $TS_{surfaces}>30$. 

To test if the extended signature is due to an extra $\gamma$-ray source in the field, we built a residual map, as shown in Fig.\ref{unassos1}, right. It corresponds to a TS map that considers the 3FGL J0935.1-1736 source is in the background (therefore, it is part of the input-model and positioned at the center of the magenta circle). In this case, the TS distribution is a result of excess photons with respect to the modelled background, which includes point sources and the diffuse component. Clearly, the residual map shows our 2WHSP blazar matching the TS peak, well within the 68\% containment region for the $\gamma$-ray signature.

\begin{figure*}[ht]
   \centering
   \hspace*{-0.0cm} 
    \includegraphics[width=0.8\linewidth]{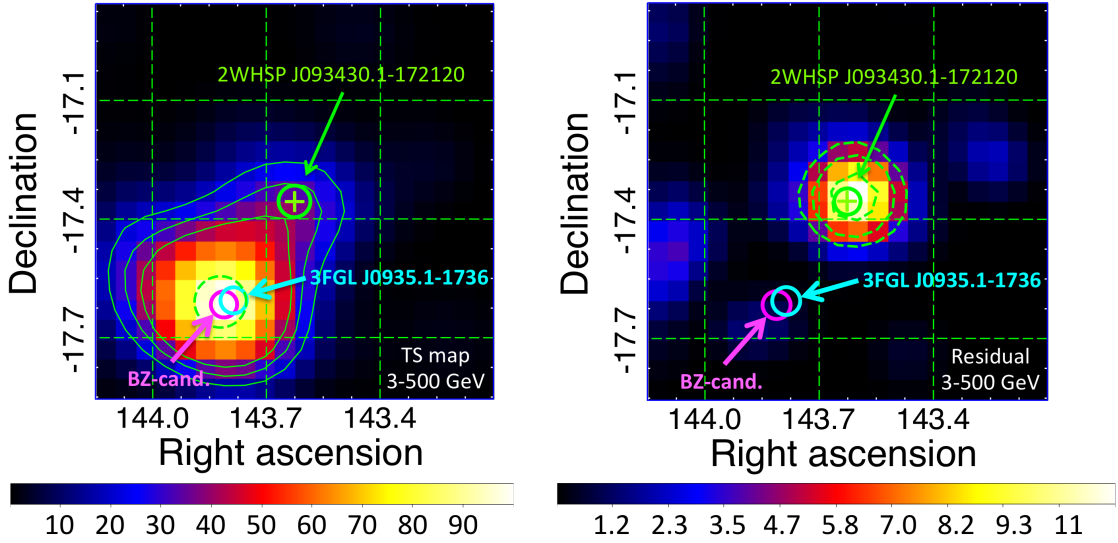}
     \caption{High-energy TS-maps 3-500 GeV with 20x20 grid 0.05$^{\circ}$/pixel, integrating over 7 yrs of data. In both maps the 2WHSP J093430.1-172120 position is highlighted by a thick green circle centered on +; the 3FGL J0935.1-1736 is in the center of the thick cyan circle, and a blazar candidate (possible counterpart for the 3FGL source) is indicated with a thick magenta circle. Left side: For this map, the 3FGL source is removed from the model input, showing the pure shape of TS distribution in the region. Green dashed line represents the 99\% containment region for the $\gamma$-ray signature, which is compatible both with 3FGL and blazar candidate positions; The thin contour lines refer to TS surfaces of 40-30-20, only to show how the $\gamma$-ray signature is extended, matching the 2WHSP J093430.1-172120. Right side: Residual TS map (built using the 3FGL in the background) showing excess signal consistent with a point-like source; The green dashed lines correspond to the 68\%, 95\%, and 99\% containment region for the $\gamma$-ray signature (from inner to outer lines).}
      \label{unassos1}
\end{figure*}

This is clearly a case of source confusion where the 2WHSP is a counterpart for the residual $\gamma$-ray signature, and the brighter source is a counterpart of an, as yet, unidentified object. However, when inspecting this region searching information from other wavelengths, we find the radio-source NVSS J093514-173658 (Fig.\ref{BZcandXUV}, in red). Recent measurements with {\it Swift} satellite XRT/UVOT show both UV and an X-ray signatures matching the radio source within their error ellipses (Fig.\ref{BZcandXUV}). Together with IR, optical, UV and X-ray counterparts we study the multi-frequency SED for NVSS J093514-173658 which turns to be a blazar candidate\footnote{The term blazar candidate refers to a source with multi-frequency SED characteristic of blazars, but missing optical identification (optical spectrum is not available).} with synchrotron-peak parameters $\nu_{peak}\approx$10$^{15.0}$ Hz and $\nu f_{\nu}$=10$^{-12.0}$ ergs/cm$^2$/s, marked as BZ-cand in Fig. \ref{unassos1}, and probable counterpart for the high-significance TS peak. 

\begin{table}[h]
\centering
\caption{Source model parameters from Fermi Science Tools, assuming a power law to describe 3FGL J0935.1-1736 $\gamma$-ray spectrum within 0.3-500 GeV, with $N_0$ given in ph/cm$^{2}$/s/MeV.}  
\label{tableFermi2}
\begin{tabular}{c|ccc}
Source  & $N_0$ (10$^{-13}$) & $\Gamma$  & TS   
\\
\hline

NVSSJ093514-173658     & 3.46$\pm$0.87 & 1.92$\pm$0.12 & 102.9\\

2WHSP J093430.1-172120 & 1.26$\pm$0.73 & 1.87$\pm$0.24 & 21.7 \\
\end{tabular}
\label{solved2}
\end{table}

A likelihood analysis that considers two sources (with positions corresponding to the 2WHSP and  NVSS, removing the 3FGL from the field) results in a relatively good adjustment, as shown in Table \ref{solved2}. Although the blazar candidate could be of HSP type, it is not part of the 2WHSP catalog (because the X-ray data was not available at the time of the sample selection) and therefore we do not count it within the 2WHSP-FGL $\gamma$-ray subsample; the source 2WHSP J093430.1-172120 is considered a lower-significance $\gamma$-ray detection, so we count it within the 150 sources listed in Table \ref{tableFermi1} .

\begin{figure}[h]
   \centering
    \includegraphics[width=1.0\linewidth]{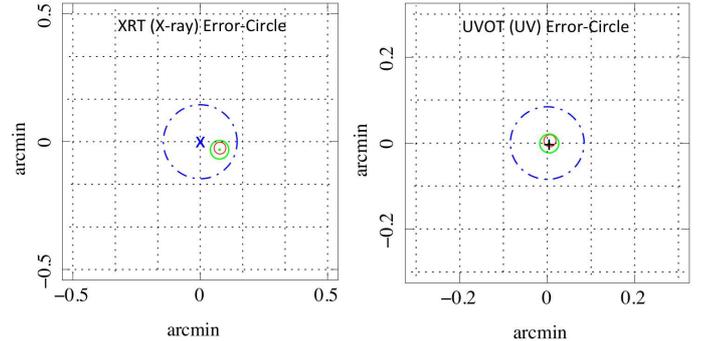}
     \caption{Sky-Explorer view around NVSS J093514-173658 (blazar candidate). Left side is shows the XRT field with the x-ray detection marked as $\times$; The right side is shows the UVOT detection indicated as +. The X-ray and UV error-circle are shown with a dotted blue line. The radio-source is marked in red, and its optical counterpart USNOB1.0 (J2000 ra,dec: 143.8116$^{\circ}$,-17.6163$^{\circ}$) is shown in green.}
      \label{BZcandXUV}
\end{figure}

\subsection{Improved position for the unassociated source 3FGL J0421.6+1950}
\label{un2}

In Fig. \ref{unassos} we show a chart that includes the 3FGL J0421.6+1950 source (indicated with a red $\times$) and its corresponding error-ellipse, which is determined over the entire Fermi-LAT energy range. However, it is well known that the detection of lower-energy photons (E$<$1 GeV) from point-like sources have large PSF, so that the $\gamma$-ray signature can spread along a region of the order of 1$^\circ$, as shown in the right side of Fig. \ref{tsmap0536m33}. 
\begin{figure}[h]
   \centering
    \includegraphics[width=0.65\linewidth]{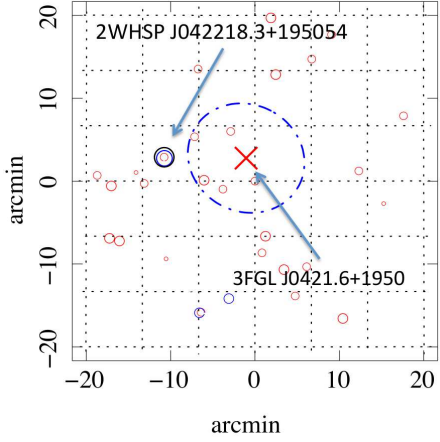}
     \caption{Sky-Explorer view around 3FGL J0421.6+1950 positions indicated as a red cross. The blue dash-dotted line represents the error circle for the $\gamma$-ray detection reported in the 3FGL catalog. As shown, within 15' from the unassociated $\gamma$-ray source there is a 2WHSP blazar. X-ray and radio  detections in the field are represented by blue and red circles, respectively.}
     \label{unassos}
\end{figure}
Since the unassociated 3FGL J0421.6+1950 is $\approx$10' away from a 2WHSP source (see Fig. \ref{unassos}), we tried to better evaluate the $\gamma$-ray signature localization by studying the high-energy TS map (E$>$3 GeV), which benefits from smaller PSF with respect to lower-energy photons.

For the TS map in Fig. \ref{Unass0422}, we removed 3FGL J0421.6+1950 from the model-input so that the TS distribution has no bias from previous $\gamma$-ray catalogs. The TS map peaks at the position of 2WHSP J042218.3+195054 (thick green circle centered on +) well within the 68\% confinement region for the $\gamma$-ray signature, which is $\approx$9.7' away from the position reported in the 3FGL catalog (thick circle centered on $\times$). This source is taken as part of the 2WHSP-FGL subsample, using 3FGL parameters to describe it (since there is no $\gamma$-ray confusion in this particular case).

\begin{figure}[h]
   \centering
    \includegraphics[width=0.75\linewidth]{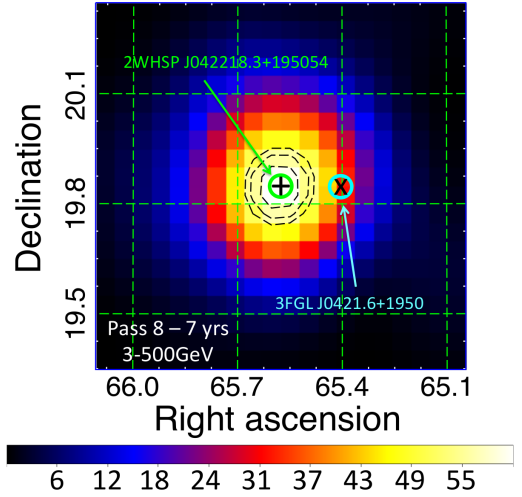}
     \caption{High-energy 3-500 GeV TS-map, integrating 7 yrs of data. The contour black dashed lines are TS surfaces representing 68\%, 95\% and 99\% containment region for the $\gamma$-ray signature (from inner to outer lines). The 2WHSP J042218.3+195054 position is highlighted by a thick green circle centered on +. The 3FGL J0421.6+1950 position is highlighted by the cyan circle centered on $\times$, $\approx$9.7' away from the high-energy TS peak.} 
      \label{Unass0422}
\end{figure}

Although the 3FGL positions are based on information associated with the full energy band 0.1-300 GeV, we attempted to improve the $\gamma$-ray signature localization by selecting only high-energy photons that are know to have better PSF. In fact, we also improve the localization owing to longer exposure time (since we now integrate along seven years of Fermi-LAT observations instead of four years in the 3FGL), but it is important to mention that the high-energy maps, integrated over four years of Pass 7 data (with the same analysis setup used to build the 3FGL catalog, as described in Sect. \ref{prova}) already enabled this kind of study, as seen in Fig. \ref{Unass0422pass7}. Probably, when using the full energy range from Fermi-LAT (0.1-500 GeV), lower-energy photons with the largest position uncertainties could be degrading the final source localization. Indeed, based on currently available data, there is room for improvements  which may bring complementary and relevant information for describing the $\gamma$-ray sky.

\begin{figure}[h]
   \centering
    \includegraphics[width=0.7\linewidth]{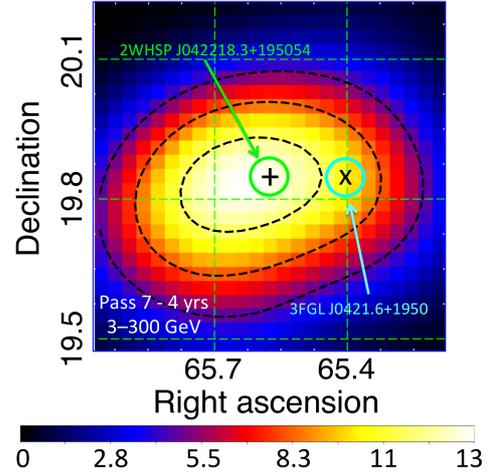}
     \caption{High-energy 3-300 GeV TS map, integrating 4 yrs of data (in this case, the highest energy used is 300 GeV following recommendations for the use of Pass 7 data). The contour black dashed lines are TS surfaces representing 68\%, 95\%, and 99\% containment region for the $\gamma$-ray signature (from inner to outer lines). The 2WHSP J042218.3+195054 position is highlighted by a thick green circle centered on +, well within the 68\% containment, while the 3FGL J0421.6+1950 source (position highlighted by the cyan circle centered on $\times$) is localized within the 95\% containment region.} 
      \label{Unass0422pass7}
\end{figure}

Working with high-energy TS maps proved to be very useful when searching for candidate-counterparts of current unassociated $\gamma$-ray sources, and could be applied systematically as a complementary refinement for the building of upcoming catalogs with the potential to improve source localization for the whole $\gamma$-ray sample. In the next subsection, we discuss a case where the 3FGL association has  already been done, but we still improve the $\gamma$-ray localization using high-energy TS maps.

\subsection{Improved position for the 3FGL J1838.5-6006 source}
\label{improvedposition}
\label{un3}

Here we study 3FGL J1838.5-6006, which is associated with the radio-source SUMSS J183806-600033 ($\equiv$2WHSP J183806.7-600032). 
\begin{figure*}[ht]
   \centering
    \includegraphics[width=0.75\linewidth]{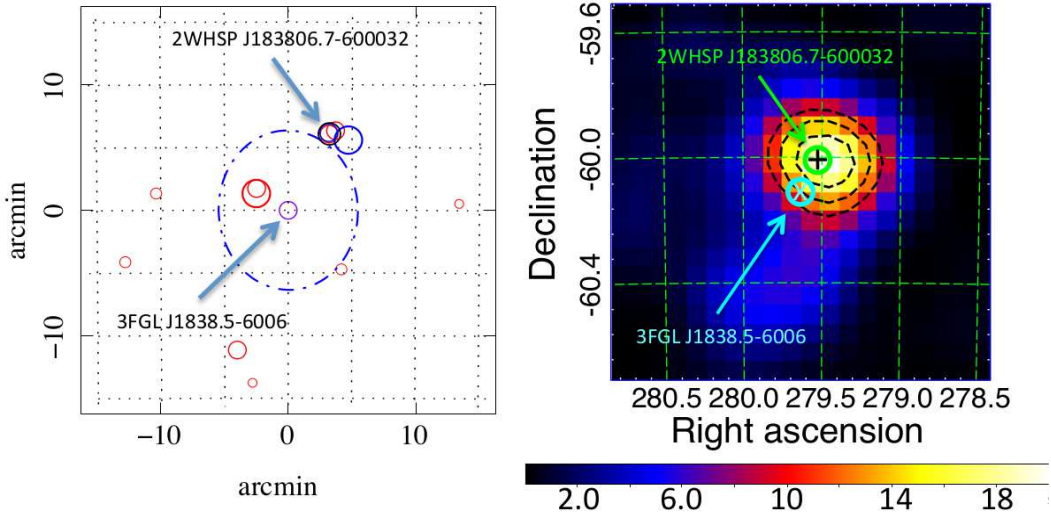}
     \caption{Left side showing the error-circle (dash-dotted) associated with the 3FGL source. Right side showing the high-energy 3-500 GeV TS map, integrating over 7 yrs of Fermi-LAT observations. The contour black dashed lines are TS surfaces representing 68\%, 95\% and 99\% containment region for the $\gamma$-ray signature (from inner to outer lines). The 2WHSP J183806.7-600032 (thick green circle centered on +). The 3FGL J1838.5-6006 position is shows in cyan (centered on $\times$), $\approx$6.9' away from the high-energy TS peak.}
      \label{better-posit}
\end{figure*}
In this case, the 2WHSP blazar is just at the border of the 3FGL error-circle (Fig. \ref{better-posit}, left) and therefore we try  improving the $\gamma$-ray source localization by working only with high-energy photons only. As shown in Fig. \ref{better-posit}, the high-energy TS peak is few arcminutes away from the 3FGL position, and matches our 2WHSP source (which is within the 68\% confinement radius for the $\gamma$-ray signature).

Although the association is correct, this is another example where we could improve the $\gamma$-ray signature localization ($\approx$6.9' drift) just by working with E$>$3 GeV photons. We also study this region at a lower-energy band (850-950 MeV) and there is no evidence of another close-by source that could be the cause of the offset position. Therefore, it could be that, for some cases the determination of the source position based on the broadband counts 0.1-500 GeV is non optimal, probably because of the large PSF associated with lower-energy photons. Building high-energy TS maps may help to improve source positioning, especially for cases with hard $\gamma$-ray spectrum, as shown for 3FGL J0421.6+1950 and 3FGL J1838.5-6006.

\section{Conclusions $\&$ perspectives}

The 2WHSP catalog was built to select promising VHE candidates for the present and future generation of Cherenkov Telescope Arrays, therefore we have tested the efficiency of a direct search for $\gamma$-ray signatures associated with 2WHSP blazars, achieving significant results.

We have detected 150 $\gamma$-ray excess signals out of 400 seed positions based on 2WHSP sources that had no counterpart in previous 1FGL, 2FGL, and 3FGL catalogs. A total of 85 sources were found with high-significance with TS$>$25, and we also report on 65 lower-significance detections with TS between 10 to 25. The 150 new $\gamma$-ray sources presented in Table \ref{tableFermi1} are named with acronym 1BIGB (first version of the Brazil ICRANet Gamma-ray Blazar catalog) which corresponds with the 2WHSP seed-positions used for our likelihood analysis. Clearly, the subsample of 2WHSP blazars that have not yet been detected by Fermi-LAT is a  key representative population of faint $\gamma$-emitters, and we show how the new detections down to TS$>$10 level can probe the faint-end of the flux-distribution (see Fig. \ref{GammaFlux} and \ref{hist-flux}). As discussed in Sect. \ref{faintdetec}, a $\gamma$-ray source-search based on the seed positions from HSP blazars can be used to unveil faint HE sources down to TS=10 without compromising the $\gamma$-ray sample with spurious detections.

Our current work enabled us to associate a relevant fraction of the IGRB to a population of faint $\gamma$-ray emitters that had been previously unresolved. Moreover, we show the increasing relevance of faint-HSPs for the IGRB composition with respect to energy (see Table \ref{IGRB}), specially for E$>$10 GeV, reaching 6-8$\%$ in the 100-200 GeV band. Motivated by this first assessment, we plan to perform a complete $\gamma$-ray analysis of the 2WHSP sample, down to the lowest fluxes, and probably extend the search to other blazar families with potential to improve the $\gamma$-ray description of lower-significance $\gamma$-ray blazars, also helping to constrain the origins of the extragalactic diffuse $\gamma$-ray background.

We have worked out the possibility of solving source confusion when considering multi-frequency data for identifying potential $\gamma$-ray emitters in a certain ROI, and building energy dependent TS maps to help disentangle hard-steep components from confused sources. 

We also addressed cases of unassociated 3FGL sources by studying high-energy TS maps to evaluate possible counterparts. This could be a key for solving cases of unassociated $\gamma$-ray sources (just as discussed in Sect. \ref{un1}, \ref{un2} and \ref{un3}) showing that we can improve the $\gamma$-ray signature localization based on currently available databases. Certainly, it is  interesting to evaluate if this kind of approach could be applied systematically as a complementary refinement for the building of upcoming $\gamma$-ray catalogs.

\begin{acknowledgements}
     
BA is supported by the Brazilian Scientific Program Ci\^{e}ncias sem Fronteiras from Cnpq, YLC is supported by the Government of the Republic of China (Taiwan). We thank ICRANet and Prof. Carlo Bianco for the cooperation that enabled us to perform part of the data reduction at Joshua Cluster, Rome-Italy. We thank the Centro Nacional de Supercomputação (CESUP) Porto Alegre-Brazil, and Carlos Brandt, for the cooperation that enabled us to perform part of the data reduction using CESUP machines;  This work was supported by the ASDC, Agenzia Spaziale Italiana Science Data Center; and University La Sapienza of Rome, Department of Physics. We thank the guidance and comments from Prof. Paolo Giommi, and the special attention from Dr. Dario Gasparrini, Dr. Sara Cutini,  Dr. Stefano Ciprini and Prof. Toby Brunett. This publication makes use of public data products and software from Fermi-LAT collaboration. We also make use of archival data and bibliographic information obtained from the NASA/IPAC Extragalactic Database (NED), data and software facilities from the ASDC managed by the Italian Space Agency (ASI).    

\end{acknowledgements}

\bibliographystyle{aa}
\bibliography{150NewGammaRaySources}

\include{longtable}

\include{Pass7table}

\end{document}

%% file: longtable.tex
\begin{longtable}{llrc|cccc}
\caption{The 150 new $\gamma$-ray signatures detected down to TS$>$10 level are named with the acronym 1BIGB (for the first version of ``Brazil ICRANet Gamma-ray Blazar" catalog) with coordinates corresponding to those of 2WHSP seeds used for the likelihood analysis. The first three columns show respectively the 1BIGB source names, right ascension ``R.A." and declination ``Dec." in degrees (J2000). The fourth column shows the reported redshifts from literature \citep{bllacz2,pita2013,pks1424p240HighZ,PG1553,shaw2,masetti2013,bll,5BZcat}, flag ``?" is used for values reported as uncertain; lower limit are marked with ``$>$" \citep[all lower-limits shown here were derived in][]{1WHSP,2WHSP}, and sources with currently absent redshift were given ``0." value. The $\gamma$-ray model parameters from Fermi Science Tools assume a power law to describe the spectrum within the studied energy range 0.3-500 GeV. The parameter ``$N_0$" (see eq. \ref{powerlaw}) is given in units of [ph/cm$^{2}$/s/MeV], and ``$\Gamma$" is the spectral photon index, which are direct outputs from the likelihood analysis over 7.2 years of Fermi-LAT data in the 0.3-500 GeV band; those results consider the pivot energy fixed as E$_0$=1 GeV. The column ``Flux" gives the photon counts in units of [ph/cm$^2$/s] calculated by integrating eq. \ref{powerlaw} along the energy range 1-100 GeV. Sources marked with (a) are also reported in the 2FHL catalog \citep{2FHL}, and sources marked with (b) were reported as possible counterparts of photon clustering detected by Fermi-LAT at E$>$10 GeV \citep{MST1}. Source 2WHSPJ194356.2+211821 is the single case located at $|$b$|<$10$^{\circ}$ (see footnote \ref{b10}).}\\
\hline\hline
1BIGB Source name  & R.A.(deg.)  &  Dec.(deg.)  &  z  &  $\Gamma$ & $N_0$ (10$^{-13}$)  & TS & Flux$_{1-100GeV}^{( \times 10^{-10})}$ \\
\hline
\endfirsthead
\caption{continued.}\\
\hline\hline
1BIGB Source name  & R.A.(deg) &  Dec.(deg) &  z  &  $\Gamma$ & $N_0$ (10$^{-13}$)  & TS & Flux$_{1-100GeV}^{( \times 10^{-10})}$ \\
\hline
\endhead
\hline
\endfoot
 1BIGB J000949.6$-$431650        &  2.45708   &  $-$43.28056    & $>$0.56 & 2.10 $\pm$ 0.15  & 2.22 $\pm$ 0.48  & 67.0  & 2.0 $^{+0.81}_{-0.62}$  \\             
 1BIGB J001328.8+094929          &  3.37000   &  9.82500        &  0.     & 2.35 $\pm$ 0.23  & 2.07 $\pm$ 0.68  & 16.0  & 1.53$^{+0.93}_{-0.66}$  \\             
 1BIGB J001527.8+353638          &  3.86625   &  35.61083       & $>$0.57 & 1.98 $\pm$ 0.23  & 1.17 $\pm$ 0.57  & 25.1  & 1.18$^{+1.09}_{-0.69}$  \\             
 1BIGB J002928.6+205332          &  7.36917   &  20.89250       &  0.     & 1.46 $\pm$ 0.22  & 0.31 $\pm$ 0.23  & 23.8  & 0.6 $^{+0.94}_{-0.48}$  \\             
 1BIGB J004146.9$-$470136        &  10.44583  &  $-$47.02667    &  0.     & 1.73 $\pm$ 0.27  & 0.46 $\pm$ 0.28  & 14.0  & 0.6 $^{+0.81}_{-0.43}$  \\             
 1BIGB J005816.6+172312          &  14.56958  &  17.38694       &  0.     & 1.80 $\pm$ 0.25  & 0.85 $\pm$ 0.52  & 20.4  & 1.02$^{+1.25}_{-0.72}$  \\             
 1BIGB J010250.8$-$200158        &  15.71208  &  $-$20.03278    & $>$0.38 & 1.55 $\pm$ 0.22  & 0.34 $\pm$ 0.22  & 17.3  & 0.57$^{+0.78}_{-0.42}$  \\             
 1BIGB J011501.6$-$340027        &  18.75708  &  $-$34.00750    &  0.48   & 1.49 $\pm$ 0.17  & 0.53 $\pm$ 0.29  & 54.0  & 0.97$^{+1.04}_{-0.62}$  \\             
 1BIGB J012657.1+330730          &  21.73833  &  33.12500       &  0.     & 2.27 $\pm$ 0.26  & 1.45 $\pm$ 0.54  & 14.2  & 1.13$^{+0.81}_{-0.54}$  \\             
 1BIGB J014040.8$-$075849        &  25.17000  &  $-$7.98028     & $>$0.49 & 1.78 $\pm$ 0.14  & 1.40 $\pm$ 0.43  & 49.0  & 1.73$^{+0.96}_{-0.70}$  \\             
 1BIGB J020106.1+003400          &  30.27542  &  0.56667        &  0.298  & 1.82 $\pm$ 0.28  & 0.58 $\pm$ 0.38  & 12.0  & 0.7 $^{+0.95}_{-0.51}$  \\             
 1BIGB J020412.9$-$333339$^{(b)}$&  31.05375  &  $-$33.56111    &  0.617  & 1.78 $\pm$ 0.20  & 0.76 $\pm$ 0.36  & 22.8  & 0.95$^{+0.87}_{-0.54}$  \\             
 1BIGB J021205.6$-$255757        &  33.02375  &  $-$25.96611    &  0.     & 1.93 $\pm$ 0.16  & 1.63 $\pm$ 0.47  & 44.4  & 1.71$^{+0.91}_{-0.66}$  \\             
 1BIGB J021216.8$-$022155        &  33.07000  &  $-$2.36528     &  0.     & 1.94 $\pm$ 0.19  & 1.48 $\pm$ 0.53  & 35.2  & 1.54$^{+1.05}_{-0.72}$  \\             
 1BIGB J021631.9+231449          &  34.13333  &  23.24722       &  0.288  & 1.89 $\pm$ 0.11  & 2.90 $\pm$ 0.64  & 102.0 & 3.19$^{+1.20}_{-0.96}$  \\             
 1BIGB J022048.4$-$084250        &  35.20167  &  $-$8.71389     & $>$0.43 & 1.74 $\pm$ 0.19  & 0.85 $\pm$ 0.40  & 32.4  & 1.11$^{+0.98}_{-0.63}$  \\             
 1BIGB J023340.9+065611          &  38.42042  &  6.93639        &  0.     & 2.05 $\pm$ 0.19  & 2.72 $\pm$ 1.10  & 52.0  & 2.56$^{+1.82}_{-1.27}$  \\             
 1BIGB J023430.5+804336          &  38.62750  &  80.72694       &  0.     & 1.65 $\pm$ 0.20  & 0.44 $\pm$ 0.28  & 18.2  & 0.65$^{+0.78}_{-0.46}$  \\             
 1BIGB J030103.7+344100          &  45.26542  &  34.68361       &  0.24   & 2.30 $\pm$ 0.15  & 3.43 $\pm$ 0.69  & 43.1  & 2.63$^{+0.95}_{-0.75}$  \\             
 1BIGB J030330.1+055429          &  45.87542  &  5.90833        &  0.196  & 1.46 $\pm$ 0.24  & 0.36 $\pm$ 0.31  & 22.0  & 0.69$^{+1.25}_{-0.61}$  \\             
 1BIGB J030433.9$-$005403$^{(a)}$&  46.14125  &  $-$0.90111     &  0.511  & 1.70 $\pm$ 0.17  & 0.85 $\pm$ 0.41  & 30.4  & 1.17$^{+1.03}_{-0.67}$  \\             
 1BIGB J030544.1+403509          &  46.43375  &  40.58611       &  0.     & 1.84 $\pm$ 0.28  & 0.71 $\pm$ 0.53  & 12.4  & 0.83$^{+1.23}_{-0.67}$  \\             
 1BIGB J031103.1$-$440227        &  47.76333  &  $-$44.04111    &  0.     & 2.02 $\pm$ 0.32  & 0.93 $\pm$ 0.44  & 13.5  & 0.9 $^{+0.99}_{-0.53}$  \\             
 1BIGB J031423.8+061955          &  48.59958  &  6.33222        &  0.62?  & 1.79 $\pm$ 0.12  & 2.26 $\pm$ 0.65  & 69.9  & 2.76$^{+1.38}_{-1.04}$  \\             
 1BIGB J032009.1$-$704533        &  50.03833  &  $-$70.75917    &  0.     & 1.75 $\pm$ 0.21  & 0.65 $\pm$ 0.33  & 21.7  & 0.83$^{+0.83}_{-0.50}$  \\             
 1BIGB J032037.9+112451          &  50.15833  &  11.41444       &  0.     & 2.17 $\pm$ 0.28  & 1.75 $\pm$ 0.98  & 10.0  & 1.48$^{+1.56}_{-0.96}$  \\             
 1BIGB J032056.2+042447          &  50.23458  &  4.41333        &  0.     & 2.68 $\pm$ 0.21  & 3.62 $\pm$ 0.71  & 32.3  & 2.15$^{+0.80}_{-0.62}$  \\             
 1BIGB J032647.2$-$340446        &  51.69708  &  $-$34.07972    &  0.     & 1.99 $\pm$ 0.14  & 1.70 $\pm$ 0.42  & 52.7  & 1.7 $^{+0.76}_{-0.58}$  \\             
 1BIGB J032852.6$-$571605        &  52.21917  &  $-$57.26806    &  0.     & 1.60 $\pm$ 0.20  & 0.47 $\pm$ 0.26  & 29.6  & 0.73$^{+0.82}_{-0.48}$  \\             
 1BIGB J033623.7$-$034738$^{(b)}$&  54.09875  &  $-$3.79389     &  0.162  & 1.61 $\pm$ 0.23  & 0.49 $\pm$ 0.33  & 22.0  & 0.75$^{+1.03}_{-0.56}$  \\             
 1BIGB J033831.9$-$570447        &  54.63333  &  $-$57.08000    &  0.     & 1.63 $\pm$ 0.23  & 0.42 $\pm$ 0.27  & 20.7  & 0.63$^{+0.84}_{-0.46}$  \\             
 1BIGB J035856.1$-$305447        &  59.73375  &  $-$30.91306    &  0.65?  & 1.88 $\pm$ 0.14  & 1.60 $\pm$ 0.44  & 57.6  & 1.77$^{+0.89}_{-0.66}$  \\             
 1BIGB J041112.2$-$394143        &  62.80125  &  $-$39.69528    & $>$0.7  & 1.81 $\pm$ 0.22  & 0.52 $\pm$ 0.28  & 11.6  & 0.62$^{+0.64}_{-0.39}$  \\             
 1BIGB J041238.3$-$392629        &  63.16000  &  $-$39.44139    &  0.     & 1.93 $\pm$ 0.24  & 0.65 $\pm$ 0.34  & 11.4  & 0.69$^{+0.69}_{-0.43}$  \\             
 1BIGB J042900.1$-$323641        &  67.25042  &  $-$32.61139    & $>$0.51 & 1.89 $\pm$ 0.21  & 0.92 $\pm$ 0.38  & 18.2  & 1.01$^{+0.82}_{-0.53}$  \\             
 1BIGB J043517.7$-$262121        &  68.82375  &  $-$26.35611    &  0.     & 2.52 $\pm$ 0.31  & 1.39 $\pm$ 0.48  & 11.4  & 0.91$^{+0.63}_{-0.41}$  \\             
 1BIGB J044127.4+150454          &  70.36417  &  15.08194       &  0.109  & 2.06 $\pm$ 0.17  & 3.49 $\pm$ 1.30  & 45.3  & 3.26$^{+2.06}_{-1.50}$  \\             
 1BIGB J044240.6+614039$^{(a)}$  &  70.66917  &  61.67750       &  0.     & 1.95 $\pm$ 0.12  & 3.30 $\pm$ 0.92  & 94.5  & 3.42$^{+1.59}_{-1.24}$  \\             
 1BIGB J044328.3$-$415156        &  70.86833  &  $-$41.86556    & $>$0.39 & 1.94 $\pm$ 0.20  & 1.10 $\pm$ 0.30  & 33.2  & 1.16$^{+0.67}_{-0.45}$  \\             
 1BIGB J050335.3$-$111506        &  75.89750  &  $-$11.25167    & $>$0.57 & 1.83 $\pm$ 0.13  & 1.76 $\pm$ 0.55  & 59.3  & 2.06$^{+1.13}_{-0.84}$  \\             
 1BIGB J050419.5$-$095631        &  76.08125  &  $-$9.94222     & $>$0.46 & 2.15 $\pm$ 0.22  & 1.49 $\pm$ 0.62  & 14.9  & 1.28$^{+0.96}_{-0.65}$  \\             
 1BIGB J050601.6$-$382054        &  76.50667  &  $-$38.34861    &  0.182  & 2.07 $\pm$ 0.14  & 1.81 $\pm$ 0.40  & 41.5  & 1.67$^{+0.67}_{-0.52}$  \\             
 1BIGB J050727.1$-$334635        &  76.86333  &  $-$33.77639    &  0.     & 1.72 $\pm$ 0.10  & 1.44 $\pm$ 0.36  & 78.7  & 1.91$^{+0.83}_{-0.65}$  \\             
 1BIGB J053626.8$-$254748        &  84.11167  &  $-$25.79667    &  0.     & 1.97 $\pm$ 0.17  & 1.67 $\pm$ 0.52  & 38.4  & 1.69$^{+0.98}_{-0.70}$  \\             
 1BIGB J053645.2$-$255841        &  84.18875  &  $-$25.97806    &  0.     & 1.81 $\pm$ 0.18  & 1.15 $\pm$ 0.49  & 33.0  & 1.39$^{+1.10}_{-0.73}$  \\             
 1BIGB J055716.7$-$061706        &  89.32000  &  $-$6.28500     &  0.     & 1.79 $\pm$ 0.15  & 1.80 $\pm$ 0.73  & 44.1  & 2.21$^{+1.56}_{-1.10}$  \\             
 1BIGB J060714.2$-$251859        &  91.80958  &  $-$25.31639    &  0.275  & 1.93 $\pm$ 0.16  & 1.39 $\pm$ 0.43  & 35.0  & 1.47$^{+0.84}_{-0.61}$  \\             
 1BIGB J062149.6$-$341148        &  95.45667  &  $-$34.19694    &  0.529  & 2.50 $\pm$ 0.17  & 3.17 $\pm$ 0.64  & 33.9  & 2.1 $^{+0.75}_{-0.60}$  \\             
 1BIGB J062626.2$-$171045        &  96.60917  &  $-$17.17944    & $>$0.7  & 2.05 $\pm$ 0.19  & 2.35 $\pm$ 0.83  & 35.7  & 2.2 $^{+1.42}_{-0.99}$  \\             
 1BIGB J063014.9$-$201236        &  97.56250  &  $-$20.21000    &  0.     & 1.84 $\pm$ 0.27  & 1.13 $\pm$ 0.95  & 14.9  & 1.32$^{+2.11}_{-1.16}$  \\             
 1BIGB J065932.8$-$674350        &  104.88708 &  $-$67.73056    &  0.     & 1.53 $\pm$ 0.21  & 0.37 $\pm$ 0.26  & 23.8  & 0.63$^{+0.89}_{-0.49}$  \\             
 1BIGB J071745.0$-$552021        &  109.4375  &  $-$55.33944    &  0.     & 2.16 $\pm$ 0.15  & 2.87 $\pm$ 0.64  & 53.3  & 2.44$^{+0.97}_{-0.76}$  \\             
 1BIGB J073152.6+280432          &  112.96958 &  28.07583       &  0.248  & 2.07 $\pm$ 0.20  & 1.63 $\pm$ 0.54  & 27.7  & 1.5 $^{+0.93}_{-0.65}$  \\             
 1BIGB J073329.5+351542          &  113.37292 &  35.26167       &  0.177  & 2.59 $\pm$ 0.30  & 1.49 $\pm$ 0.48  & 11.7  & 0.93$^{+0.59}_{-0.40}$  \\             
 1BIGB J075936.1+132116          &  119.90042 &  13.35472       &  0.     & 1.78 $\pm$ 0.13  & 1.81 $\pm$ 0.54  & 82.1  & 2.24$^{+1.17}_{-0.88}$  \\             
 1BIGB J080015.4+561107          &  120.06458 &  56.18528       &  0.     & 1.96 $\pm$ 0.11  & 2.06 $\pm$ 0.42  & 74.8  & 2.11$^{+0.77}_{-0.61}$  \\             
 1BIGB J080135.8+463824          &  120.39958 &  46.64000       &  0.369  & 2.30 $\pm$ 0.36  & 1.06 $\pm$ 0.52  & 10.2  & 0.81$^{+0.86}_{-0.49}$  \\             
 1BIGB J082904.7+175415          &  127.27000 &  17.90417       &  0.089  & 2.41 $\pm$ 0.12  & 4.69 $\pm$ 0.61  & 94.6  & 3.3 $^{+0.79}_{-0.66}$  \\             
 1BIGB J083724.5+145819          &  129.35250 &  14.97222       &  0.278  & 2.00 $\pm$ 0.22  & 1.48 $\pm$ 0.61  & 30.7  & 1.46$^{+1.15}_{-0.75}$  \\             
 1BIGB J085749.8+013530          &  134.45750 &  1.59167        &  0.281  & 2.51 $\pm$ 0.24  & 1.80 $\pm$ 0.55  & 13.9  & 1.18$^{+0.65}_{-0.48}$  \\             
 1BIGB J090802.2$-$095936        &  137.00917 &  $-$9.99361     &  0.053  & 1.82 $\pm$ 0.25  & 0.71 $\pm$ 0.47  & 13.7  & 0.85$^{+1.08}_{-0.62}$  \\             
 1BIGB J090953.2+310602          &  137.47167 &  31.10083       &  0.272  & 1.91 $\pm$ 0.23  & 0.73 $\pm$ 0.35  & 13.2  & 0.78$^{+0.73}_{-0.45}$  \\             
 1BIGB J091322.3+813305          &  138.34292 &  81.55139       &  0.639? & 1.49 $\pm$ 0.16  & 0.36 $\pm$ 0.17  & 39.6  & 0.66$^{+0.62}_{-0.39}$  \\             
 1BIGB J091651.8+523827          &  139.21625 &  52.64111       &  0.19   & 1.79 $\pm$ 0.15  & 1.04 $\pm$ 0.34  & 54.9  & 1.26$^{+0.77}_{-0.54}$  \\             
 1BIGB J093239.2+104234$^{(b)}$  &  143.16375 &  10.70972       &  0.361  & 2.00 $\pm$ 0.16  & 2.06 $\pm$ 0.58  & 46.7  & 2.02$^{+1.05}_{-0.77}$  \\             
 1BIGB J093430.1$-$172120        &  143.62542 &  $-$17.35583    &  0.     & 1.87 $\pm$ 0.24  & 1.26 $\pm$ 0.73  & 21.7  & 1.42$^{+1.56}_{-0.94}$  \\
 1BIGB J095224.1+750212$^{(a)}$$^{(b)}$ &  148.10042 & 75.03694 &  0.181  & 1.28 $\pm$ 0.16  & 0.21 $\pm$ 0.12  & 55.0  & 0.54$^{+0.64}_{-0.37}$  \\             
 1BIGB J095507.9+355100          &  148.78292 &  35.85000       &  0.834  & 1.88 $\pm$ 0.25  & 0.73 $\pm$ 0.40  & 20.0  & 0.82$^{+0.90}_{-0.53}$  \\             
 1BIGB J095628.2$-$095719        &  149.11750 &  $-$9.95528     &  0.     & 1.68 $\pm$ 0.25  & 0.47 $\pm$ 0.33  & 14.6  & 0.66$^{+0.95}_{-0.51}$  \\             
 1BIGB J095849.8+703959          &  149.70750 &  70.66639       &  0.     & 2.03 $\pm$ 0.20  & 1.36 $\pm$ 0.51  & 36.6  & 1.3 $^{+0.91}_{-0.62}$  \\             
 1BIGB J102100.3+162554          &  155.25125 &  16.43167       &  0.556  & 2.34 $\pm$ 0.25  & 2.08 $\pm$ 0.61  & 25.6  & 1.54$^{+0.92}_{-0.63}$  \\             
 1BIGB J104303.7+005420          &  160.76583 &  0.90556        &  0.     & 1.73 $\pm$ 0.17  & 1.33 $\pm$ 0.55  & 48.3  & 1.76$^{+1.36}_{-0.91}$  \\             
 1BIGB J104857.6+500945          &  162.24000 &  50.16250       &  0.402  & 2.26 $\pm$ 0.17  & 1.63 $\pm$ 0.40  & 28.0  & 1.28$^{+0.57}_{-0.43}$  \\             
 1BIGB J105534.3$-$012616        &  163.89292 &  $-$1.43778     &  0.     & 1.78 $\pm$ 0.12  & 2.09 $\pm$ 0.59  & 77.8  & 2.59$^{+1.27}_{-0.96}$  \\             
 1BIGB J111717.5+000633          &  169.32292 &  0.10917        &  0.451  & 2.17 $\pm$ 0.20  & 2.18 $\pm$ 0.63  & 29.4  & 1.84$^{+1.03}_{-0.73}$  \\             
 1BIGB J112317.9$-$323217        &  170.82500 &  $-$32.53833    &  0.     & 1.88 $\pm$ 0.21  & 1.01 $\pm$ 0.46  & 17.9  & 1.12$^{+0.98}_{-0.62}$  \\             
 1BIGB J112611.8$-$203723        &  171.54958 &  $-$20.62333    &  0.     & 2.16 $\pm$ 0.24  & 1.58 $\pm$ 0.70  & 13.7  & 1.36$^{+1.11}_{-0.73}$  \\             
 1BIGB J113046.0$-$313807$^{(a)}$&  172.69208 &  $-$31.63528    &  0.151  & 1.28 $\pm$ 0.21  & 0.17 $\pm$ 0.13  & 28.0  & 0.44$^{+0.79}_{-0.38}$  \\             
 1BIGB J113105.2$-$094405        &  172.77167 &  $-$9.73500     &  0.     & 1.84 $\pm$ 0.17  & 1.23 $\pm$ 0.48  & 34.3  & 1.43$^{+1.02}_{-0.70}$  \\             
 1BIGB J113444.6$-$172900        &  173.68625 &  $-$17.48361    &  0.571  & 1.58 $\pm$ 0.20  & 0.55 $\pm$ 0.33  & 24.3  & 0.88$^{+1.04}_{-0.61}$  \\             
 1BIGB J113755.6$-$171041$^{(a)}$$^{(b)}$ &  174.48167 &  $-$17.17833    &  0.6    & 1.69 $\pm$ 0.10  & 1.80 $\pm$ 0.45  & 90.7  & 2.48$^{+1.10}_{-0.85}$  \\             
 1BIGB J121158.6+224233          &  182.99417 &  22.70917       &  0.45   & 1.78 $\pm$ 0.18  & 0.90 $\pm$ 0.38  & 27.6  & 1.12$^{+0.90}_{-0.59}$  \\             
 1BIGB J121510.9+073203          &  183.79542 &  7.53444        &  0.137  & 1.78 $\pm$ 0.16  & 1.12 $\pm$ 0.41  & 38.6  & 1.39$^{+0.93}_{-0.65}$  \\             
 1BIGB J121603.1$-$024304$^{(b)}$&  184.01333 &  $-$2.71778     &  0.169  & 2.30 $\pm$ 0.12  & 3.95 $\pm$ 0.64  & 65.4  & 3.01$^{+0.85}_{-0.71}$  \\             
 1BIGB J124141.4+344029          &  190.42250 &  34.67500       & $>$0.7  & 1.96 $\pm$ 0.18  & 1.31 $\pm$ 0.45  & 32.6  & 1.34$^{+0.86}_{-0.60}$  \\             
 1BIGB J125015.4+315559          &  192.56458 &  31.93306       &  0.     & 1.76 $\pm$ 0.32  & 0.57 $\pm$ 0.44  & 16.7  & 0.72$^{+1.26}_{-0.60}$  \\             
 1BIGB J125341.2$-$393159        &  193.42167 &  $-$39.53306    &  0.179  & 1.83 $\pm$ 0.18  & 1.40 $\pm$ 0.59  & 37.0  & 1.64$^{+1.26}_{-0.85}$  \\             
 1BIGB J125847.9$-$044744        &  194.70000 &  $-$4.79583     &  0.586? & 1.84 $\pm$ 0.20  & 1.40 $\pm$ 0.50  & 32.9  & 1.62$^{+1.17}_{-0.76}$  \\             
 1BIGB J130145.6+405623          &  195.44000 &  40.94000       &  0.652  & 2.08 $\pm$ 0.14  & 2.16 $\pm$ 0.46  & 66.6  & 1.98$^{+0.78}_{-0.60}$  \\             
 1BIGB J130713.3$-$034430        &  196.80542 &  $-$3.74194     &  0.     & 2.55 $\pm$ 0.20  & 2.44 $\pm$ 0.12  & 20.5  & 1.57$^{+0.31}_{-0.24}$  \\             
 1BIGB J132541.8$-$022809        &  201.42417 &  $-$2.46944     &  0.8?   & 2.16 $\pm$ 0.20  & 1.45 $\pm$ 0.70  & 18.2  & 1.23$^{+0.95}_{-0.68}$  \\             
 1BIGB J132617.7+122957          &  201.57375 &  12.49944       &  0.204  & 2.14 $\pm$ 0.29  & 1.20 $\pm$ 0.55  & 12.9  & 1.04$^{+0.96}_{-0.59}$  \\             
 1BIGB J132833.4+114520          &  202.13958 &  11.75556       &  0.811  & 2.10 $\pm$ 0.20  & 1.78 $\pm$ 0.57  & 28.0  & 1.61$^{+0.96}_{-0.68}$  \\             
 1BIGB J133612.1+231958          &  204.05042 &  23.33278       &  0.267  & 1.92 $\pm$ 0.14  & 1.43 $\pm$ 0.38  & 43.2  & 1.52$^{+0.73}_{-0.55}$  \\             
 1BIGB J135328.0+560056$^{(b)}$  &  208.36667 &  56.01556       &  0.404  & 2.28 $\pm$ 0.18  & 2.23 $\pm$ 0.51  & 48.7  & 1.73$^{+0.74}_{-0.56}$  \\             
 1BIGB J140629.9$-$393508        &  211.62500 &  $-$39.58583    &  0.37   & 1.46 $\pm$ 0.15  & 0.40 $\pm$ 0.21  & 26.2  & 0.76$^{+0.74}_{-0.47}$  \\             
 1BIGB J143342.7$-$730437        &  218.42792 &  $-$73.07722    &  0.     & 1.26 $\pm$ 0.23  & 0.10 $\pm$ 0.09  & 13.6  & 0.27$^{+0.57}_{-0.25}$  \\             
 1BIGB J143825.4+120418          &  219.60625 &  12.07167       &  0.     & 2.02 $\pm$ 0.33  & 0.87 $\pm$ 0.57  & 11.2  & 0.83$^{+1.15}_{-0.62}$  \\             
 1BIGB J144236.4$-$462300        &  220.65167 &  $-$46.38361    &  0.103  & 1.98 $\pm$ 0.20  & 3.08 $\pm$ 0.08  & 61.2  & 3.1 $^{+0.82}_{-0.57}$  \\             
 1BIGB J145508.2+192014          &  223.78417 &  19.33750       &  0.115  & 2.78 $\pm$ 0.36  & 1.75 $\pm$ 0.51  & 13.9  & 0.98$^{+0.61}_{-0.40}$  \\             
 1BIGB J145543.6$-$760051        &  223.93167 &  $-$76.01444    &  0.     & 1.54 $\pm$ 0.14  & 0.60 $\pm$ 0.26  & 34.6  & 1.02$^{+0.81}_{-0.54}$  \\             
 1BIGB J145603.5+504825          &  224.01500 &  50.80722       & $>$0.49 & 2.17 $\pm$ 0.31  & 1.56 $\pm$ 0.78  & 23.6  & 1.32$^{+1.35}_{-0.80}$  \\             
 1BIGB J150637.0$-$054004        &  226.65458 &  $-$5.66778     &  0.518  & 1.89 $\pm$ 0.19  & 1.42 $\pm$ 0.61  & 28.2  & 1.57$^{+1.22}_{-0.83}$  \\             
 1BIGB J151041.0+333503          &  227.67125 &  33.58444       &  0.114  & 1.47 $\pm$ 0.28  & 0.24 $\pm$ 0.21  & 15.6  & 0.46$^{+0.97}_{-0.42}$  \\             
 1BIGB J151136.8$-$165326        &  227.90375 &  $-$16.89056    & $>$0.56 & 2.69 $\pm$ 0.35  & 1.95 $\pm$ 0.65  & 10.1  & 1.15$^{+0.78}_{-0.51}$  \\             
 1BIGB J151618.7$-$152344        &  229.07792 &  $-$15.39556    & $>$0.54 & 2.23 $\pm$ 0.22  & 2.12 $\pm$ 0.74  & 18.5  & 1.7 $^{+1.08}_{-0.76}$  \\             
 1BIGB J151826.5+075222          &  229.61083 &  7.87278        &  0.41   & 1.70 $\pm$ 0.23  & 0.54 $\pm$ 0.34  & 15.0  & 0.74$^{+0.94}_{-0.53}$  \\             
 1BIGB J151845.7+061355          &  229.69042 &  6.23222        &  0.102  & 1.84 $\pm$ 0.19  & 1.30 $\pm$ 0.58  & 31.9  & 1.5 $^{+1.26}_{-0.82}$  \\             
 1BIGB J152646.6$-$153025        &  231.69417 &  $-$15.50722    & $>$0.43 & 2.12 $\pm$ 0.17  & 2.86 $\pm$ 0.78  & 40.0  & 2.51$^{+1.24}_{-0.92}$  \\             
 1BIGB J152913.5+381216          &  232.30625 &  38.20472       & $>$0.59 & 1.82 $\pm$ 0.19  & 0.91 $\pm$ 0.37  & 27.5  & 1.08$^{+0.84}_{-0.55}$  \\             
 1BIGB J154202.9$-$291509        &  235.51250 &  $-$29.2525     &  0.     & 1.75 $\pm$ 0.10  & 2.44 $\pm$ 0.54  & 100.0 & 3.14$^{+1.16}_{-0.94}$  \\             
 1BIGB J154625.0$-$285723        &  236.60417 &  $-$28.95639    & $>$0.6  & 1.70 $\pm$ 0.19  & 0.65 $\pm$ 0.36  & 14.2  & 0.89$^{+0.91}_{-0.58}$  \\             
 1BIGB J155053.2$-$082245        &  237.72167 &  $-$8.37944     &  0.     & 1.93 $\pm$ 0.23  & 1.47 $\pm$ 0.76  & 25.7  & 1.55$^{+1.50}_{-0.94}$  \\             
 1BIGB J155432.5$-$121324        &  238.63542 &  $-$12.22361    &  0.     & 1.78 $\pm$ 0.14  & 1.40 $\pm$ 0.50  & 30.2  & 1.75$^{+1.07}_{-0.78}$  \\             
 1BIGB J160218.0+305108          &  240.57500 &  30.8525        & $>$0.47 & 1.95 $\pm$ 0.21  & 0.96 $\pm$ 0.40  & 22.7  & 0.99$^{+0.79}_{-0.52}$  \\             
 1BIGB J160519.0+542058          &  241.32917 &  54.34972       &  0.212  & 2.02 $\pm$ 0.20  & 1.17 $\pm$ 0.30  & 35.4  & 1.13$^{+0.61}_{-0.42}$  \\             
 1BIGB J160618.4+134532          &  241.57667 &  13.75889       &  0.29   & 2.35 $\pm$ 0.24  & 2.81 $\pm$ 0.71  & 32.4  & 2.07$^{+1.08}_{-0.76}$  \\             
 1BIGB J161327.1$-$190835        &  243.36292 &  $-$19.14333    &  0.     & 2.23 $\pm$ 0.16  & 4.11 $\pm$ 1.00  & 34.3  & 3.31$^{+1.42}_{-1.10}$  \\             
 1BIGB J162115.1$-$003140        &  245.31333 &  $-$0.52778     & $>$0.52 & 1.84 $\pm$ 0.21  & 0.91 $\pm$ 0.49  & 13.9  & 1.06$^{+1.06}_{-0.66}$  \\             
 1BIGB J162330.4+085724          &  245.87708 &  8.95667        &  0.533  & 1.94 $\pm$ 0.26  & 1.12 $\pm$ 0.61  & 21.6  & 1.18$^{+1.27}_{-0.76}$  \\             
 1BIGB J162646.0+630047          &  246.69167 &  63.01333       &  0.     & 1.95 $\pm$ 0.15  & 1.47 $\pm$ 0.42  & 61.5  & 1.52$^{+0.79}_{-0.58}$  \\             
 1BIGB J164220.2+221143          &  250.58458 &  22.19528       &  0.592  & 2.13 $\pm$ 0.21  & 1.79 $\pm$ 0.66  & 29.8  & 1.56$^{+1.06}_{-0.73}$  \\             
 1BIGB J164419.9+454644          &  251.08333 &  45.77889       &  0.225  & 1.90 $\pm$ 0.19  & 0.97 $\pm$ 0.36  & 31.9  & 1.06$^{+0.75}_{-0.51}$  \\             
 1BIGB J165517.8$-$224045        &  253.82458 &  $-$22.67917    &  0.     & 1.89 $\pm$ 0.20  & 1.47 $\pm$ 0.70  & 15.4  & 1.62$^{+1.40}_{-0.93}$  \\             
 1BIGB J171108.5+024403          &  257.78583 &  2.73444        &  0.     & 1.87 $\pm$ 0.19  & 1.22 $\pm$ 0.55  & 18.5  & 1.36$^{+1.11}_{-0.74}$  \\             
 1BIGB J174419.7+185218          &  266.08250 &  18.87167       &  0.     & 1.61 $\pm$ 0.19  & 0.61 $\pm$ 0.33  & 20.4  & 0.94$^{+1.00}_{-0.60}$  \\             
 1BIGB J174702.5+493800          &  266.76042 &  49.63361       &  0.46?  & 2.12 $\pm$ 0.17  & 2.13 $\pm$ 0.56  & 45.5  & 1.89$^{+0.90}_{-0.67}$  \\             
 1BIGB J184822.4+653656          &  282.09375 &  65.61583       &  0.364  & 1.58 $\pm$ 0.16  & 0.51 $\pm$ 0.24  & 37.0  & 0.81$^{+0.73}_{-0.47}$  \\             
 1BIGB J185023.9+263153          &  282.60000 &  26.53139       &  0.     & 1.77 $\pm$ 0.16  & 1.47 $\pm$ 0.61  & 45.6  & 1.84$^{+1.37}_{-0.95}$  \\             
 1BIGB J185813.3+432451          &  284.55583 &  43.41417       &  0.     & 2.28 $\pm$ 0.16  & 2.81 $\pm$ 0.58  & 43.1  & 2.18$^{+0.82}_{-0.64}$  \\             
 1BIGB J193412.7$-$241919        &  293.55292 &  $-$24.32222    &  0.     & 1.59 $\pm$ 0.14  & 0.82 $\pm$ 0.33  & 41.2  & 1.29$^{+0.96}_{-0.66}$  \\             
 1BIGB J194356.2+211821$^{(a)}$  &  295.98417 &  21.30611       &  0.     & 1.43 $\pm$ 0.20  & 1.69 $\pm$ 0.40  & 128.0 & 3.34$^{+2.49}_{-1.42}$  \\             
 1BIGB J200204.0$-$573644        &  300.51708 &  $-$57.61250    &  0.     & 1.94 $\pm$ 0.11  & 2.42 $\pm$ 0.53  & 69.6  & 2.52$^{+0.95}_{-0.76}$  \\             
 1BIGB J201200.9$-$771219        &  303.00375 &  $-$77.20528    &  0.     & 2.07 $\pm$ 0.29  & 1.02 $\pm$ 0.69  & 10.1  & 0.94$^{+1.20}_{-0.70}$  \\             
 1BIGB J205242.4+081040          &  313.17708 &  8.17778        &  0.     & 1.46 $\pm$ 0.23  & 0.3  $\pm$ 0.20  & 20.8  & 0.57$^{+0.84}_{-0.43}$  \\             
 1BIGB J214533.3$-$043438        &  326.38875 &  $-$4.57750     &  0.069  & 2.81 $\pm$ 0.26  & 2.16 $\pm$ 0.56  & 18.9  & 1.19$^{+0.56}_{-0.42}$  \\             
 1BIGB J215214.0$-$120540        &  328.05875 &  $-$12.09472    &  0.121  & 2.30 $\pm$ 0.16  & 4.10 $\pm$ 0.91  & 62.8  & 3.12$^{+1.23}_{-0.96}$  \\             
 1BIGB J220107.3$-$590639        &  330.28042 &  $-$59.11111    &  0.     & 1.85 $\pm$ 0.21  & 0.64 $\pm$ 0.31  & 13.0  & 0.74$^{+0.67}_{-0.43}$  \\             
 1BIGB J220155.8$-$170700        &  330.48250 &  $-$17.11667    &  0.169  & 2.20 $\pm$ 0.32  & 1.62 $\pm$ 0.85  & 19.3  & 1.33$^{+1.42}_{-0.83}$  \\             
 1BIGB J221029.5+362159          &  332.62333 &  36.36639       &  0.     & 2.08 $\pm$ 0.29  & 1.33 $\pm$ 0.81  & 14.7  & 1.22$^{+1.43}_{-0.84}$  \\             
 1BIGB J221108.2$-$000302$^{(b)}$&  332.78458 &  $-$0.05056     &  0.326  & 1.66 $\pm$ 0.16  & 0.88 $\pm$ 0.38  & 34.3  & 1.27$^{+1.02}_{-0.68}$  \\             
 1BIGB J223301.0+133601          &  338.25458 &  13.60028       &  0.214  & 1.60 $\pm$ 0.23  & 0.45 $\pm$ 0.32  & 19.6  & 0.71$^{+1.03}_{-0.56}$  \\             
 1BIGB J223626.2+370713          &  339.10958 &  37.12028       &  0.     & 1.84 $\pm$ 0.19  & 1.01 $\pm$ 0.50  & 22.1  & 1.17$^{+1.04}_{-0.68}$  \\             
 1BIGB J224910.6$-$130002        &  342.29458 &  $-$13.00056    & $>$0.5  & 2.23 $\pm$ 0.19  & 2.47 $\pm$ 0.68  & 33.8  & 1.99$^{+1.02}_{-0.75}$  \\             
 1BIGB J225147.5$-$320611        &  342.94792 &  $-$32.10333    &  0.246  & 1.75 $\pm$ 0.17  & 1.21 $\pm$ 0.47  & 54.9  & 1.55$^{+1.14}_{-0.77}$  \\             
 1BIGB J225613.3$-$330338        &  344.05542 &  $-$33.06056    &  0.243  & 2.29 $\pm$ 0.33  & 1.22 $\pm$ 0.54  & 13.1  & 0.94$^{+0.86}_{-0.52}$  \\             
 1BIGB J230634.9$-$110347        &  346.64583 &  $-$11.06333    &  0.     & 2.26 $\pm$ 0.26  & 1.71 $\pm$ 0.60  & 20.1  & 1.35$^{+0.95}_{-0.62}$  \\             
 1BIGB J232039.7$-$630918        &  350.16583 &  $-$63.15500    &  0.2    & 1.71 $\pm$ 0.17  & 0.77 $\pm$ 0.33  & 37.8  & 1.03$^{+0.81}_{-0.54}$  \\             
 1BIGB J233112.8$-$030129        &  352.80375 &  $-$3.02500     &  0.     & 2.17 $\pm$ 0.23  & 1.69 $\pm$ 0.65  & 17.3  & 1.43$^{+1.02}_{-0.69}$  \\             
 1BIGB J235320.9$-$145856        &  358.33750 &  $-$14.98250    &  0.     & 1.54 $\pm$ 0.22  & 0.44 $\pm$ 0.28  & 20.8  & 0.74$^{+0.98}_{-0.54}$  \\             
\label{tableFermi1}
\end{longtable}

%% file: Pass7table.tex
\begin{longtable}{l|cccc}
\caption{Here we list 30 2WHSPs detected with the largest significance within our $\gamma$-ray likelihood analysis, all having TS$>$45 based Pass 8 Fermi-LAT data when integrating over 7.2 yrs of observations. All those cases were re-evaluated based on Pass 7 data, integrating over the first 4 yrs of mission; in order to reproduce the setup used to prepare the 3FGL catalog (see sec. \ref{prova} for the setup description). First column show the 2WHSP source name, and following columns list the fitting parameter assuming a power-law model to describe the $\gamma$-ray spectrum in the 0.3-300 GeV band (we integrated up to 300 GeV following Fermi-LAT team recommendations to work with Pass 7); second column lists the photon spectral index $\Gamma$, third column lists the pre-factor parameter  $N_0$ in 10$^{-13}$ph/cm$^{2}$/s/MeV, fourth column list the TS values, and fifth column lists the integral flux [ph/cm$^2$/s] in the 1-100 GeV energy band. For two sources (marked with $*$) the power-law fitting does not converge well given their significance is too low.} \\
\hline\hline
2WHSP Source name & $\Gamma$ & $N_0$ (10$^{-13}$)  &  TS$^{(Pass \ 7)}_{4 \ yrs}$ & Flux$_{1-100GeV}^{( \times 10^{-10})}$ \\
\hline
\endfirsthead
\caption{continued.}\\
\hline\hline
2WHSP Source name & $\Gamma$ & $N_0$ (10$^{-13}$)  &  TS$^{(Pass \ 7)}_{4 \ yrs}$ & Flux$_{1-100GeV}^{( \times 10^{-10})}$ \\
\hline
\endhead
\hline
\endfoot
  2WHSPJ144236.4$-$462300   &  1.96$\pm$0.24  & 2.50$\pm$1.34   &  15.0  &  2.57$^{+2.50}_{-1.61}$   \\  
  2WHSPJ154202.9$-$291509   &  1.52$\pm$0.31  & 0.52$\pm$0.05   &   8.9  &  0.90$^{+0.77}_{-0.35}$   \\  
  2WHSPJ000949.6$-$431650   &  2.36$\pm$0.30  & 2.27$\pm$0.77   &  13.8  &  1.66$^{+1.1}_{-0.76}$    \\ 
  2WHSPJ011501.6$-$340027   &  1.57$\pm$0.28  & 0.55$\pm$0.43   &  11.1  &  0.89$^{+1.5}_{-0.75}$    \\ 
  2WHSPJ014040.8$-$075849   &  1.87$\pm$0.31  & 1.03$\pm$0.72   &  9.1   &  1.16$^{+1.7}_{-0.90}$    \\ 
  2WHSPJ021631.9+231449     &  2.00$\pm$0.21  & 2.81$\pm$1.08   &  26.3  &  2.78$^{+2.0}_{-1.35}$    \\ 
  2WHSPJ023340.9+065611     &  2.56$\pm$0.81  & 1.87$\pm$1.98   &  3.9   &  1.19$^{+3.7}_{-1.24}$    \\ 
  2WHSPJ031423.8+061955     &  1.90$\pm$0.24  & 1.99$\pm$1.03   &  16.3  &  2.17$^{+2.1}_{-1.33}$    \\ 
  2WHSPJ032647.2$-$340446   &  1.89$\pm$0.19  & 1.65$\pm$0.62   &  20.3  &  1.82$^{+1.2}_{-0.87}$    \\ 
  2WHSPJ035856.1$-$305447   &  1.66$\pm$0.36  & 0.64$\pm$0.57   &  8.8   &  0.92$^{+2.0}_{-0.85}$    \\ 
  2WHSPJ044240.6+614039     &  2.27$\pm$0.31  & 3.26$\pm$1.77   &  12.2  &  2.55$^{+2.6}_{-1.61}$    \\ 
  2WHSPJ050335.3$-$111506   &  1.96$\pm$0.32  & 1.60$\pm$1.02   &  9.6   &  1.64$^{+2.2}_{-1.19}$    \\ 
  2WHSPJ050727.1$-$334635   &  1.66$\pm$0.30  & 0.64$\pm$0.47   &  12.0  &  0.92$^{+1.5}_{-0.74}$    \\ 
  2WHSPJ071745.0$-$552021   &  2.24$\pm$0.40  & 1.48$\pm$1.01   &  5.8   &  1.18$^{+1.7}_{-0.90}$    \\ 
  2WHSPJ075936.1+132116     &  1.61$\pm$0.31  & 0.66$\pm$0.58   &  12.2  &  1.01$^{+2.0}_{-0.93}$    \\ 
  2WHSPJ080015.4+561107     &  1.85$\pm$0.19  & 1.36$\pm$0.6    &  19.8  &  1.56$^{+1.2}_{-0.84}$    \\ 
  2WHSPJ082904.7+175415     &  2.34$\pm$0.21  & 3.79$\pm$0.98   &  25.6  &  2.82$^{+1.3}_{-1.01}$    \\ 
  2WHSPJ091651.8+523827     &  1.76$\pm$0.33  & 0.66$\pm$0.53   &  10.0  &  0.84$^{+1.5}_{-0.72}$    \\ 
  2WHSPJ104303.7+005420     &  1.90$\pm$0.30  & 1.69$\pm$0.99   &  17.5  &  1.84$^{+2.3}_{-1.26}$    \\ 
  2WHSPJ105534.3$-$012616   &  2.06$\pm$0.23  & 2.48$\pm$0.96   &  19.5  &  2.32$^{+1.7}_{-1.14}$    \\ 
  2WHSPJ113755.6$-$171041   &  1.78$\pm$0.22  & 1.53$\pm$0.75   &  19.6  &  1.90$^{+1.8}_{-1.13}$    \\ 
  2WHSPJ121603.1$-$024304   &  2.16$\pm$0.22  & 3.38$\pm$1.24   &  25.4  &  2.89$^{+1.9}_{-1.35}$    \\ 
  2WHSPJ130145.6+405623     &  1.75$\pm$0.22  & 1.24$\pm$0.60   &  22.6  &  1.60$^{+1.5}_{-0.94}$    \\ 
  2WHSPJ135328.0+560056     &  2.11$\pm$0.24  & 1.73$\pm$0.76   &  15.7  &  1.54$^{+1.2}_{-0.83}$    \\ 
  2WHSPJ162646.0+630047     &  2.46$\pm$0.36  & 1.67$\pm$0.66   &  8.5   &  1.14$^{+0.96}_{-0.58}$   \\  
  2WHSPJ194356.2+211821     &  1.55$\pm$0.17  & 1.71$\pm$0.91   &  18.1  &  2.86$^{+2.8}_{-1.79}$    \\ 
  2WHSPJ200204.0$-$573644   &  2.13$\pm$0.29  & 1.55$\pm$0.77   &  5.7   &  1.36$^{+1.3}_{-0.81}$    \\ 
  2WHSPJ215214.0$-$120540   &  2.46$\pm$0.26  & 4.86$\pm$1.55   &  29.5  &  3.32$^{+1.9}_{-1.40}$    \\ 
  2WHSPJ095224.1+750212*    &  0.76$\pm$0.39  & 0.02$\pm$0.02   &   8.3  &  0.16$^{+1.0}_{-0.20}$    \\
  2WHSPJ225147.5$-$320611*  &  0.93$\pm$0.43  & 0.05$\pm$0.09   &  6.0   &  0.27$^{+2.2}_{-0.36}$    \\ 
  \label{tablepass7}
\end{longtable}

%% file: 150NewGammaRaySources.bbl
\begin{thebibliography}{70}
\expandafter\ifx\csname natexlab\endcsname\relax\def\natexlab#1{#1}\fi

\bibitem[{{Abdo} {et~al.}(2010){Abdo}, {Ackermann}, {Agudo}, {Ajello}, {Aller},
  {Aller}, {Angelakis}, {Arkharov}, {Axelsson}, {Bach}, \& et~al.}]{abdo10}
{Abdo}, A.~A., {Ackermann}, M., {Agudo}, I., {et~al.} 2010, \apj, 716, 30

\bibitem[{Abdo {et~al.}(2010)Abdo, Ackermann, Ajello, Allafort, Antolini,
  Atwood, Axelsson, Baldini, Ballet, Barbiellini, Bastieri, Baughman, Bechtol,
  Bellazzini, Belli, Berenji, Bisello, Blandford, Bloom, Bonamente, Bonnell,
  Borgland, Bouvier, Bregeon, Brez, Brigida, Bruel, Burnett, Busetto, Buson,
  Caliandro, Cameron, Campana, Canadas, Caraveo, Carrigan, Casandjian,
  Cavazzuti, Ceccanti, Cecchi, Çelik, Charles, Chekhtman, Cheung, Chiang,
  Cillis, Ciprini, Claus, Cohen-Tanugi, Conrad, Corbet, Davis, DeKlotz, den
  Hartog, Dermer, de~Angelis, de~Luca, de~Palma, Digel, Dormody, do~Couto~e
  Silva, Drell, Dubois, Dumora, Fabiani, Farnier, Favuzzi, Fegan, Ferrara,
  Focke, Fortin, Frailis, Fukazawa, Funk, Fusco, Gargano, Gasparrini, Gehrels,
  Germani, Giavitto, Giebels, Giglietto, Giommi, Giordano, Giroletti, Glanzman,
  Godfrey, Grenier, Grondin, Grove, Guillemot, Guiriec, Gustafsson, Hadasch,
  Hanabata, Harding, Hayashida, Hays, Healey, Hill, Horan, Hughes, Iafrate,
  Jóhannesson, Johnson, Johnson, Johnson, Johnson, Kamae, Katagiri, Kataoka,
  Kawai, Kerr, Knödlseder, Kocevski, Kuss, Lande, Landriu, Latronico, Lee,
  Lemoine-Goumard, Lionetto, Garde, Longo, Loparco, Lott, Lovellette, Lubrano,
  Madejski, Makeev, Marangelli, Marelli, Massaro, Mazziotta, McConville,
  McEnery, Michelson, Minuti, Mitthumsiri, Mizuno, Moiseev, Mongelli, Monte,
  Monzani, Moretti, Morselli, Moskalenko, Murgia, Nakajima, Nakamori,
  Naumann-Godo, Nolan, Norris, Nuss, Ohno, Ohsugi, Omodei, Orlando, Ormes,
  Ozaki, Paccagnella, Paneque, Panetta, Parent, Pelassa, Pepe, Pesce-Rollins,
  Pinchera, Piron, Porter, Poupard, Rainò, Rando, Ray, Razzano, Razzaque, Rea,
  Reimer, Reimer, Reposeur, Ripken, Ritz, Rochester, Rodriguez, Romani, Roth,
  Sadrozinski, Salvetti, Sanchez, Sander, Parkinson, Scargle, Schalk, Scolieri,
  Sgrò, Shaw, Siskind, Smith, Smith, Spandre, Spinelli, Starck, Stephens,
  Striani, Strickman, Strong, Suson, Tajima, Takahashi, Takahashi, Tanaka,
  Thayer, Thayer, Thompson, Tibaldo, Tibolla, Tinebra, Torres, Tosti,
  Tramacere, Uchiyama, Usher, Etten, Vasileiou, Vilchez, Vitale, Waite,
  Wallace, Wang, Watters, Winer, Wood, Yang, Ylinen, \& Ziegler}]{FermiCat1}
Abdo, A.~A., Ackermann, M., Ajello, M., {et~al.} 2010, ApJS, 188, 405

\bibitem[{{Acero} {et~al.}(2015){Acero}, {Ackermann}, {Ajello}, {Albert},
  {Atwood}, {Axelsson}, {Baldini}, {Ballet}, {Barbiellini}, {Bastieri},
  {Belfiore}, {Bellazzini}, {Bissaldi}, {Blandford}, {Bloom}, {Bogart},
  {Bonino}, {Bottacini}, {Bregeon}, {Britto}, {Bruel}, {Buehler}, {Burnett},
  {Buson}, {Caliandro}, {Cameron}, {Caputo}, {Caragiulo}, {Caraveo},
  {Casandjian}, {Cavazzuti}, {Charles}, {Chaves}, {Chekhtman}, {Cheung},
  {Chiang}, {Chiaro}, {Ciprini}, {Claus}, {Cohen-Tanugi}, {Cominsky}, {Conrad},
  {Cutini}, {D'Ammando}, {de Angelis}, {DeKlotz}, {de Palma}, {Desiante},
  {Digel}, {Di Venere}, {Drell}, {Dubois}, {Dumora}, {Favuzzi}, {Fegan},
  {Ferrara}, {Finke}, {Franckowiak}, {Fukazawa}, {Funk}, {Fusco}, {Gargano},
  {Gasparrini}, {Giebels}, {Giglietto}, {Giommi}, {Giordano}, {Giroletti},
  {Glanzman}, {Godfrey}, {Grenier}, {Grondin}, {Grove}, {Guillemot}, {Guiriec},
  {Hadasch}, {Harding}, {Hays}, {Hewitt}, {Hill}, {Horan}, {Iafrate}, {Jogler},
  {J{\'o}hannesson}, {Johnson}, {Johnson}, {Johnson}, {Johnson}, {Kamae},
  {Kataoka}, {Katsuta}, {Kuss}, {La Mura}, {Landriu}, {Larsson}, {Latronico},
  {Lemoine-Goumard}, {Li}, {Li}, {Longo}, {Loparco}, {Lott}, {Lovellette},
  {Lubrano}, {Madejski}, {Massaro}, {Mayer}, {Mazziotta}, {McEnery},
  {Michelson}, {Mirabal}, {Mizuno}, {Moiseev}, {Mongelli}, {Monzani},
  {Morselli}, {Moskalenko}, {Murgia}, {Nuss}, {Ohno}, {Ohsugi}, {Omodei},
  {Orienti}, {Orlando}, {Ormes}, {Paneque}, {Panetta}, {Perkins},
  {Pesce-Rollins}, {Piron}, {Pivato}, {Porter}, {Racusin}, {Rando}, {Razzano},
  {Razzaque}, {Reimer}, {Reimer}, {Reposeur}, {Rochester}, {Romani},
  {Salvetti}, {S{\'a}nchez-Conde}, {Saz Parkinson}, {Schulz}, {Siskind},
  {Smith}, {Spada}, {Spandre}, {Spinelli}, {Stephens}, {Strong}, {Suson},
  {Takahashi}, {Takahashi}, {Tanaka}, {Thayer}, {Thayer}, {Thompson},
  {Tibaldo}, {Tibolla}, {Torres}, {Torresi}, {Tosti}, {Troja}, {Van Klaveren},
  {Vianello}, {Winer}, {Wood}, {Wood}, {Zimmer}, \& {Fermi-LAT
  Collaboration}}]{3FGL}
{Acero}, F., {Ackermann}, M., {Ajello}, M., {et~al.} 2015, \apjs, 218, 23

\bibitem[{{Acero} {et~al.}(2013){Acero}, {Donato}, {Ojha}, {Stevens},
  {Edwards}, {Ferrara}, {Blanchard}, {Lovell}, \& {Thompson}}]{Unassoc2FGL}
{Acero}, F., {Donato}, D., {Ojha}, R., {et~al.} 2013, \apj, 779, 133

\bibitem[{{Ackermann} {et~al.}(2015{\natexlab{a}}){Ackermann}, {Ajello},
  {Albert}, {Atwood}, {Baldini}, {Ballet}, {Barbiellini}, {Bastieri},
  {Bechtol}, \& {Bellazzini}}]{dgrbFermi}
{Ackermann}, M., {Ajello}, M., {Albert}, A., {et~al.} 2015{\natexlab{a}}, \apj,
  799, 86

\bibitem[{{Ackermann} {et~al.}(2016{\natexlab{a}}){Ackermann}, {Ajello},
  {Albert}, {Atwood}, {Baldini}, {Ballet}, {Barbiellini}, {Bastieri},
  {Bechtol}, {Bellazzini}, {Bissaldi}, {Blandford}, {Bloom}, {Bonino},
  {Bregeon}, {Britto}, {Bruel}, {Buehler}, {Caliandro}, {Cameron}, {Caragiulo},
  {Caraveo}, {Cavazzuti}, {Cecchi}, {Charles}, {Chekhtman}, {Chiang}, {Chiaro},
  {Ciprini}, {Cohen-Tanugi}, {Cominsky}, {Costanza}, {Cutini}, {D'Ammando}, {de
  Angelis}, {de Palma}, {Desiante}, {Digel}, {Di Mauro}, {Di Venere},
  {Dom{\'{\i}}nguez}, {Drell}, {Favuzzi}, {Fegan}, {Ferrara}, {Franckowiak},
  {Fukazawa}, {Funk}, {Fusco}, {Gargano}, {Gasparrini}, {Giglietto}, {Giommi},
  {Giordano}, {Giroletti}, {Godfrey}, {Green}, {Grenier}, {Guiriec}, {Hays},
  {Horan}, {Iafrate}, {Jogler}, {J{\'o}hannesson}, {Kuss}, {La Mura},
  {Larsson}, {Latronico}, {Li}, {Li}, {Longo}, {Loparco}, {Lott}, {Lovellette},
  {Lubrano}, {Madejski}, {Magill}, {Maldera}, {Manfreda}, {Mayer}, {Mazziotta},
  {Michelson}, {Mitthumsiri}, {Mizuno}, {Moiseev}, {Monzani}, {Morselli},
  {Moskalenko}, {Murgia}, {Negro}, {Nuss}, {Ohsugi}, {Okada}, {Omodei},
  {Orlando}, {Ormes}, {Paneque}, {Perkins}, {Pesce-Rollins}, {Petrosian},
  {Piron}, {Pivato}, {Porter}, {Rain{\`o}}, {Rando}, {Razzano}, {Razzaque},
  {Reimer}, {Reimer}, {Reposeur}, {Romani}, {S{\'a}nchez-Conde}, {Schmid},
  {Schulz}, {Sgr{\`o}}, {Simone}, {Siskind}, {Spada}, {Spandre}, {Spinelli},
  {Suson}, {Takahashi}, {Thayer}, {Tibaldo}, {Torres}, {Troja}, {Vianello},
  {Yassine}, \& {Zimmer}}]{50GeVbackground}
{Ackermann}, M., {Ajello}, M., {Albert}, A., {et~al.} 2016{\natexlab{a}},
  Physical Review Letters, 116, 151105

\bibitem[{{Ackermann} {et~al.}(2012){Ackermann}, {Ajello}, {Albert}, {Baldini},
  {Ballet}, {Barbiellini}, \& {Bastieri}}]{GammaAnisotropyFermiTeam}
{Ackermann}, M., {Ajello}, M., {Albert}, A., {et~al.} 2012, \prd, 85, 083007

\bibitem[{{Ackermann} {et~al.}(2011){Ackermann}, {Ajello}, {Allafort},
  {Antolini}, {Atwood}, {Axelsson}, {Baldini}, {Ballet}, {Barbiellini},
  {Bastieri}, {Bechtol}, {Bellazzini}, {Berenji}, {Blandford}, {Bloom},
  {Bonamente}, {Borgland}, {Bottacini}, {Bouvier}, {Bregeon}, {Brigida},
  {Bruel}, {Buehler}, {Burnett}, {Buson}, {Caliandro}, {Cameron}, {Caraveo},
  {Casandjian}, {Cavazzuti}, {Cecchi}, {Charles}, {Cheung}, {Chiang},
  {Ciprini}, {Claus}, {Cohen-Tanugi}, {Conrad}, {Costamante}, {Cutini}, {de
  Angelis}, {de Palma}, {Dermer}, {Digel}, {Silva}, {Drell}, {Dubois},
  {Escande}, {Favuzzi}, {Fegan}, {Ferrara}, {Finke}, {Focke}, {Fortin},
  {Frailis}, {Fukazawa}, {Funk}, {Fusco}, {Gargano}, {Gasparrini}, {Gehrels},
  {Germani}, {Giebels}, {Giglietto}, {Giommi}, {Giordano}, {Giroletti},
  {Glanzman}, {Godfrey}, {Grenier}, {Grove}, {Guiriec}, {Gustafsson},
  {Hadasch}, {Hayashida}, {Hays}, {Healey}, {Horan}, {Hou}, {Hughes},
  {Iafrate}, {J{\'o}hannesson}, {Johnson}, {Johnson}, {Kamae}, {Katagiri},
  {Kataoka}, {Kn{\"o}dlseder}, {Kuss}, {Lande}, {Larsson}, {Latronico},
  {Longo}, {Loparco}, {Lott}, {Lovellette}, {Lubrano}, {Madejski}, {Mazziotta},
  {McConville}, {McEnery}, {Michelson}, {Mitthumsiri}, {Mizuno}, {Moiseev},
  {Monte}, {Monzani}, {Moretti}, {Morselli}, {Moskalenko}, {Murgia},
  {Nakamori}, {Naumann-Godo}, {Nolan}, {Norris}, {Nuss}, {Ohno}, {Ohsugi},
  {Okumura}, {Omodei}, {Orienti}, {Orlando}, {Ormes}, {Ozaki}, {Paneque},
  {Parent}, {Pesce-Rollins}, {Pierbattista}, {Piranomonte}, {Piron}, {Pivato},
  {Porter}, {Rain{\`o}}, {Rando}, {Razzano}, {Razzaque}, {Reimer}, {Reimer},
  {Ritz}, {Rochester}, {Romani}, {Roth}, {Sanchez}, {Sbarra}, {Scargle},
  {Schalk}, {Sgr{\`o}}, {Shaw}, {Siskind}, {Spandre}, {Spinelli}, {Strong},
  {Suson}, {Tajima}, {Takahashi}, {Takahashi}, {Tanaka}, {Thayer}, {Thayer},
  {Thompson}, {Tibaldo}, {Tinivella}, {Torres}, {Tosti}, {Troja}, {Uchiyama},
  {Vandenbroucke}, {Vasileiou}, {Vianello}, {Vitale}, {Waite}, {Wallace},
  {Wang}, {Winer}, {Wood}, {Wood}, \& {Zimmer}}]{2lac}
{Ackermann}, M., {Ajello}, M., {Allafort}, A., {et~al.} 2011, \apj, 743, 171

\bibitem[{{Ackermann} {et~al.}(2015{\natexlab{b}}){Ackermann}, {Ajello},
  {Atwood}, \& {Baldini}}]{3LAC}
{Ackermann}, M., {Ajello}, M., {Atwood}, W.~B., \& {Baldini}, e.~a.
  2015{\natexlab{b}}, \apj, 810, 14

\bibitem[{{Ackermann} {et~al.}(2016{\natexlab{b}}){Ackermann}, {Ajello},
  {Atwood}, {Baldini}, {Ballet}, {Barbiellini}, {Bastieri}, {Becerra Gonzalez},
  {Bellazzini}, {Bissaldi}, {Blandford}, {Bloom}, {Bonino}, {Bottacini},
  {Brandt}, {Bregeon}, {Bruel}, {Buehler}, {Buson}, {Caliandro}, {Cameron},
  {Caputo}, {Caragiulo}, {Caraveo}, {Cavazzuti}, {Cecchi}, {Charles},
  {Chekhtman}, {Cheung}, {Chiang}, {Chiaro}, {Ciprini}, {Cohen},
  {Cohen-Tanugi}, {Cominsky}, {Conrad}, {Cuoco}, {Cutini}, {D'Ammando}, {de
  Angelis}, {de Palma}, {Desiante}, {Di Mauro}, {Di Venere},
  {Dom{\'{\i}}nguez}, {Drell}, {Favuzzi}, {Fegan}, {Ferrara}, {Focke},
  {Fortin}, {Franckowiak}, {Fukazawa}, {Funk}, {Furniss}, {Fusco}, {Gargano},
  {Gasparrini}, {Giglietto}, {Giommi}, {Giordano}, {Giroletti}, {Glanzman},
  {Godfrey}, {Grenier}, {Grondin}, {Guillemot}, {Guiriec}, {Harding}, {Hays},
  {Hewitt}, {Hill}, {Horan}, {Iafrate}, {Hartmann}, {Jogler},
  {J{\'o}hannesson}, {Johnson}, {Kamae}, {Kataoka}, {Kn{\"o}dlseder}, {Kuss},
  {La Mura}, {Larsson}, {Latronico}, {Lemoine-Goumard}, {Li}, {Li}, {Longo},
  {Loparco}, {Lott}, {Lovellette}, {Lubrano}, {Madejski}, {Maldera},
  {Manfreda}, {Mayer}, {Mazziotta}, {Michelson}, {Mirabal}, {Mitthumsiri},
  {Mizuno}, {Moiseev}, {Monzani}, {Morselli}, {Moskalenko}, {Murgia}, {Nuss},
  {Ohsugi}, {Omodei}, {Orienti}, {Orlando}, {Ormes}, {Paneque}, {Perkins},
  {Pesce-Rollins}, {Petrosian}, {Piron}, {Pivato}, {Porter}, {Rain{\`o}},
  {Rando}, {Razzano}, {Razzaque}, {Reimer}, {Reimer}, {Reposeur}, {Romani},
  {S{\'a}nchez-Conde}, {Saz Parkinson}, {Schmid}, {Schulz}, {Sgr{\`o}},
  {Siskind}, {Spada}, {Spandre}, {Spinelli}, {Suson}, {Tajima}, {Takahashi},
  {Takahashi}, {Takahashi}, {Thayer}, {Thompson}, {Tibaldo}, {Torres}, {Tosti},
  {Troja}, {Vianello}, {Wood}, {Wood}, {Yassine}, {Zaharijas}, \&
  {Zimmer}}]{2FHL}
{Ackermann}, M., {Ajello}, M., {Atwood}, W.~B., {et~al.} 2016{\natexlab{b}},
  \apjs, 222, 5

\bibitem[{{Actis} {et~al.}(2011){Actis}, {Agnetta}, {Aharonian},
  {Akhperjanian}, {Aleksi{\'c}}, {Aliu}, {Allan}, {Allekotte}, {Antico},
  {Antonelli}, \& et~al.}]{CTA}
{Actis}, M., {Agnetta}, G., {Aharonian}, F., {et~al.} 2011, Experimental
  Astronomy, 32, 193

\bibitem[{{Ajello} {et~al.}(2015){Ajello}, {Gasparrini}, {S{\'a}nchez-Conde},
  {Zaharijas}, {Gustafsson}, {Cohen-Tanugi}, {Dermer}, {Inoue}, {Hartmann},
  {Ackermann}, {Bechtol}, {Franckowiak}, {Reimer}, {Romani}, \&
  {Strong}}]{OriginEGB}
{Ajello}, M., {Gasparrini}, D., {S{\'a}nchez-Conde}, M., {et~al.} 2015, \apjl,
  800, L27

\bibitem[{{Alam} {et~al.}(2015){Alam}, {Albareti}, {Allende Prieto}, {Anders},
  {Anderson}, {Anderton}, {Andrews}, {Armengaud}, {Aubourg}, {Bailey}, \&
  et~al.}]{SDSSDR12}
{Alam}, S., {Albareti}, F.~D., {Allende Prieto}, C., {et~al.} 2015, \apjs, 219,
  12

\bibitem[{{Ando} {et~al.}(2014){Ando}, {Benoit-L{\'e}vy}, \&
  {Komatsu}}]{dgrb-LSS}
{Ando}, S., {Benoit-L{\'e}vy}, A., \& {Komatsu}, E. 2014, \prd, 90, 023514

\bibitem[{{Arsioli} {et~al.}(2015){Arsioli}, {Fraga}, {Giommi}, {Padovani}, \&
  {Marrezi}}]{1WHSP}
{Arsioli}, B., {Fraga}, B., {Giommi}, P., {Padovani}, P., \& {Marrezi}, P.~M.
  2015, A\&A

\bibitem[{{Atwood} {et~al.}(2009){Atwood}, {Abdo}, {Ackermann}, {Althouse},
  {Anderson}, {Axelsson}, {Baldini}, {Ballet}, {Band}, {Barbiellini}, \&
  et~al.}]{FermiLAT}
{Atwood}, W.~B., {Abdo}, A.~A., {Ackermann}, M., {et~al.} 2009, \apj, 697, 1071

\bibitem[{Bernlöhr {et~al.}(2013)Bernlöhr, Barnacka, Becherini, Bigas,
  Carmona, Colin, Decerprit, Pierro, Dubois, Farnier, Funk, Hermann, Hinton,
  Humensky, Khélifi, Kihm, Komin, Lenain, Maier, Mazin, Medina, Moralejo,
  Nolan, Ohm, de~Oña~Wilhelmi, Parsons, Arribas, Pedaletti, Pita, Prokoph,
  Rulten, Schwanke, Shayduk, Stamatescu, Vallania, Vorobiov, Wischnewski,
  Yoshikoshi, \& Zech}]{CTA50h}
Bernlöhr, K., Barnacka, A., Becherini, Y., {et~al.} 2013, Astroparticle
  Physics, 43, 171

\bibitem[{{B{\l}a{\.z}ejowski} {et~al.}(2005){B{\l}a{\.z}ejowski}, {Blaylock},
  {Bond}, {Bradbury}, {Buckley}, {Carter-Lewis}, {Celik}, {Cogan}, {Cui},
  {Daniel}, {Duke}, {Falcone}, {Fegan}, {Fegan}, {Finley}, {Fortson},
  {Gammell}, {Gibbs}, {Gillanders}, {Grube}, {Gutierrez}, {Hall}, {Hanna},
  {Holder}, {Horan}, {Humensky}, {Kenny}, {Kertzman}, {Kieda}, {Kildea},
  {Knapp}, {Kosack}, {Krawczynski}, {Krennrich}, {Lang}, {LeBohec}, {Linton},
  {Lloyd-Evans}, {Maier}, {Mendoza}, {Milovanovic}, {Moriarty}, {Nagai}, {Ong},
  {Power-Mooney}, {Quinn}, {Quinn}, {Ragan}, {Reynolds}, {Rebillot}, {Rose},
  {Schroedter}, {Sembroski}, {Swordy}, {Syson}, {Valcarel}, {Vassiliev},
  {Wakely}, {Walker}, {Weekes}, {White}, {Zweerink}, {VERITAS Collaboration},
  {Mochejska}, {Smith}, {Aller}, {Aller}, {Ter{\"a}sranta}, {Boltwood},
  {Sadun}, {Stanek}, {Adams}, {Foster}, {Hartman}, {Lai}, {B{\"o}ttcher},
  {Reimer}, \& {Jung}}]{FlareMkr421-MultiWave}
{B{\l}a{\.z}ejowski}, M., {Blaylock}, G., {Bond}, I.~H., {et~al.} 2005, \apj,
  630, 130

\bibitem[{{Campana} {et~al.}(2016){Campana}, {Massaro}, \&
  {Bernieri}}]{MST-1WHSP}
{Campana}, R., {Massaro}, E., \& {Bernieri}, E. 2016, \apss, 361, 183

\bibitem[{{Campana} {et~al.}(2015){Campana}, {Massaro}, {Bernieri}, \&
  {D'Amato}}]{MST1}
{Campana}, R., {Massaro}, E., {Bernieri}, E., \& {D'Amato}, Q. 2015, \apss,
  360, 19

\bibitem[{{Chang} {et~al.}(2016){Chang}, {Arsioli}, {Giommi}, \&
  {Padovani}}]{2WHSP}
{Chang}, Y.-L., {Arsioli}, B., {Giommi}, P., \& {Padovani}, P. 2016, ArXiv
  e-prints

\bibitem[{{Cohen} {et~al.}(2007){Cohen}, {Lane}, {Cotton}, {Kassim}, {Lazio},
  {Perley}, {Condon}, \& {Erickson}}]{Radio3}
{Cohen}, A.~S., {Lane}, W.~M., {Cotton}, W.~D., {et~al.} 2007, \aj, 134, 1245

\bibitem[{{Condon} {et~al.}(1998){Condon}, {Cotton}, {Greisen}, {Yin},
  {Perley}, {Taylor}, \& {Broderick}}]{Radio4}
{Condon}, J.~J., {Cotton}, W.~D., {Greisen}, E.~W., {et~al.} 1998, \aj, 115,
  1693

\bibitem[{{Cuoco} {et~al.}(2012){Cuoco}, {Komatsu}, \&
  {Siegal-Gaskins}}]{GammaAnisotropy}
{Cuoco}, A., {Komatsu}, E., \& {Siegal-Gaskins}, J.~M. 2012, \prd, 86, 063004

\bibitem[{{Danforth} {et~al.}(2010){Danforth}, {Keeney}, {Stocke}, {Shull}, \&
  {Yao}}]{PG1553}
{Danforth}, C.~W., {Keeney}, B.~A., {Stocke}, J.~T., {Shull}, J.~M., \& {Yao},
  Y. 2010, \apj, 720, 976

\bibitem[{{D'Elia} {et~al.}(2013){D'Elia}, {Perri}, {Puccetti}, {Capalbi},
  {Giommi}, {Burrows}, {Campana}, {Tagliaferri}, {Cusumano}, {Evans},
  {Gehrels}, {Kennea}, {Moretti}, {Nousek}, {Osborne}, {Romano}, \&
  {Stratta}}]{deliaswift}
{D'Elia}, V., {Perri}, M., {Puccetti}, S., {et~al.} 2013, \aap, 551, A142

\bibitem[{Di~Mauro(2015)}]{IsoDifuseGammaBackGround}
Di~Mauro, M. 2015, in {5th International Fermi Symposium Nagoya, Japan, October
  20-24, 2014}

\bibitem[{{Di Mauro} {et~al.}(2014){Di Mauro}, {Donato}, {Lamanna}, {Sanchez},
  \& {Serpico}}]{DifuseGammaBLlacs}
{Di Mauro}, M., {Donato}, F., {Lamanna}, G., {Sanchez}, D.~A., \& {Serpico},
  P.~D. 2014, \apj, 786, 129

\bibitem[{{Dixon}(1970)}]{Radio7}
{Dixon}, R.~S. 1970, \apjs, 20, 1

\bibitem[{{Doert} \& {Errando}(2013)}]{UnassocFGL-AGN}
{Doert}, M. \& {Errando}, M. 2013, in Proceedings of the 33rd International
  Cosmic Ray Conference, Rio de Janeiro 2-9 July 2013

\bibitem[{{Eddington}(1913)}]{EddingtonBias}
{Eddington}, A.~S. 1913, \mnras, 73, 359

\bibitem[{Elvis {et~al.}(1992)Elvis, Plummer, Schachter, \&
  Fabbiano}]{Elvis1992}
Elvis, M., Plummer, D., Schachter, J., \& Fabbiano, G. 1992, ApJS, 80, 257

\bibitem[{Evans {et~al.}(2010)Evans, Primini, Glotfelty, Anderson, Bonaventura,
  Chen, Davis, Doe, Evans, Fabbiano, Galle, Gibbs, Grier, Hain, Hall, Harbo,
  He, Houck, Karovska, Kashyap, Lauer, McCollough, McDowell, Miller, Mitschang,
  Morgan, Mossman, Nichols, Nowak, Plummer, Refsdal, Rots, Siemiginowska,
  Sundheim, Tibbetts, {Van Stone}, Winkelman, \& Zografou}]{Evans2010}
Evans, I.~N., Primini, F.~A., Glotfelty, K.~J., {et~al.} 2010, ApJS, 189, 37

\bibitem[{Fornasa \& Sánchez-Conde(2015)}]{NatureIDGB}
Fornasa, M. \& Sánchez-Conde, M.~A. 2015, Phys. Rept., 598, 1

\bibitem[{{Franceschini} {et~al.}(2008){Franceschini}, {Rodighiero, G.}, \&
  {Vaccari, M.}}]{EBLabsorption}
{Franceschini}, A., {Rodighiero, G.}, \& {Vaccari, M.} 2008, A\&A, 487, 837

\bibitem[{Fujinaga {et~al.}(2015)Fujinaga, Niinuma, Kimura, Fujisawa, Oyama,
  Mizuno, Kono, Takemura, Sawada-Satoh, Akutagawa, Sugiyama, Motogi, \&
  Fukuzaki}]{Fujinaga07052015}
Fujinaga, Y., Niinuma, K., Kimura, A., {et~al.} 2015, Publications of the
  Astronomical Society of Japan

\bibitem[{{Furniss} {et~al.}(2013){Furniss}, {Williams}, {Danforth},
  {Fumagalli}, {Prochaska}, {Primack}, {Urry}, {Stocke}, {Filippenko}, \&
  {Neely}}]{pks1424p240HighZ}
{Furniss}, A., {Williams}, D.~A., {Danforth}, C., {et~al.} 2013, \apjl, 768,
  L31

\bibitem[{{Giommi} \& {Padovani}(2015)}]{EGBpaolo}
{Giommi}, P. \& {Padovani}, P. 2015, \mnras, 450, 2404

\bibitem[{{Gregory} {et~al.}(1996){Gregory}, {Scott}, {Douglas}, \&
  {Condon}}]{Radio2}
{Gregory}, P.~C., {Scott}, W.~K., {Douglas}, K., \& {Condon}, J.~J. 1996,
  \apjs, 103, 427

\bibitem[{{Harris} {et~al.}(1996){Harris}, {Forman}, {Gioa}, {Hale}, {Harnden},
  {Jones}, {Karakashian}, {Maccacaro}, {McSweeney}, {Primnini}, {Schwarz},
  {Tananbaum}, \& {Thurman}}]{IPC}
{Harris}, D.~E., {Forman}, W., {Gioa}, I.~M., {et~al.} 1996, VizieR Online Data
  Catalog, 9013, 0

\bibitem[{Harris {et~al.}(1993)Harris, Forman, Gioia, Hale, {Harnden, F. R.},
  Jones, Karakashian, Maccacaro, McSweeney, \& Primini}]{Harris1993}
Harris, D.~E., Forman, W., Gioia, I.~M., {et~al.} 1993, {The Einstein
  Observatory catalog of IPC X ray sources. Volume 1E: Documentation}

\bibitem[{{Inoue}(2014)}]{blazarbackground}
{Inoue}, Y. 2014, Fifth Fermi Symposium Proceedings, Nagoya, Japan.

\bibitem[{{Krawczynski} {et~al.}(2004){Krawczynski}, {Hughes}, {Horan},
  {Aharonian}, {Aller}, {Aller}, {Boltwood}, {Buckley}, {Coppi}, {Fossati},
  {G{\"o}tting}, {Holder}, {Horns}, {Kurtanidze}, {Marscher}, {Nikolashvili},
  {Remillard}, {Sadun}, \& {Schr{\"o}der}}]{variabilityTeV-WHSP}
{Krawczynski}, H., {Hughes}, S.~B., {Horan}, D., {et~al.} 2004, \apj, 601, 151

\bibitem[{{Malyshev} \& {Hogg}(2011)}]{PointSourcesBellowDetecLimit}
{Malyshev}, D. \& {Hogg}, D.~W. 2011, \apj, 738, 181

\bibitem[{{Masetti} {et~al.}(2013){Masetti}, {Sbarufatti}, {Parisi},
  {Jim{\'e}nez-Bail{\'o}n}, {Chavushyan}, {Vogt}, {Sguera}, {Stephen},
  {Palazzi}, {Bassani}, {Bazzano}, {Fiocchi}, {Galaz}, {Landi}, {Malizia},
  {Minniti}, {Morelli}, \& {Ubertini}}]{masetti2013}
{Masetti}, N., {Sbarufatti}, B., {Parisi}, P., {et~al.} 2013, \aap, 559, A58

\bibitem[{Massaro {et~al.}(2015)Massaro, Maselli, Leto, Marchegiani, Perri,
  Giommi, \& Piranomonte}]{5BZcat}
Massaro, E., Maselli, A., Leto, C., {et~al.} 2015, Ap\&SS, 357

\bibitem[{Massaro {et~al.}(2011)Massaro, D'Abrusco, Ajello, Grindlay, \&
  Smith}]{WiseBlazars}
Massaro, F., D'Abrusco, R., Ajello, M., Grindlay, J.~E., \& Smith, H.~A. 2011,
  The Astrophysical Journal Letters, 740, L48

\bibitem[{{Mattox} {et~al.}(1996){Mattox}, {Bertsch}, {Chiang}, {Dingus},
  {Digel}, {Esposito}, {Fierro}, {Hartman}, {Hunter}, {Kanbach}, {Kniffen},
  {Lin}, {Macomb}, {Mayer-Hasselwander}, {Michelson}, {von Montigny},
  {Mukherjee}, {Nolan}, {Ramanamurthy}, {Schneid}, {Sreekumar}, {Thompson}, \&
  {Willis}}]{TSmapsConfidenceRadius}
{Mattox}, J.~R., {Bertsch}, D.~L., {Chiang}, J., {et~al.} 1996, \apj, 461, 396

\bibitem[{{Nieppola} {et~al.}(2007){Nieppola}, {Tornikoski},
  {L{\"a}hteenm{\"a}ki}, {Valtaoja}, {Hakala}, {Hovatta}, {Kotiranta},
  {Nummila}, {Ojala}, {Parviainen}, {Ranta}, {Saloranta}, {Torniainen}, \&
  {Tr{\"o}ller}}]{Radio5}
{Nieppola}, E., {Tornikoski}, M., {L{\"a}hteenm{\"a}ki}, A., {et~al.} 2007,
  \aj, 133, 1947

\bibitem[{{Nolan} {et~al.}(2012){Nolan}, {Abdo}, {Ackermann}, {Ajello},
  {Allafort}, {Antolini}, {Atwood}, {Axelsson}, {Baldini}, {Ballet}, \&
  et~al.}]{2FGL}
{Nolan}, P.~L., {Abdo}, A.~A., {Ackermann}, M., {et~al.} 2012, \apjs, 199, 31

\bibitem[{{Padovani} \& {Giommi}(1995)}]{padgio95}
{Padovani}, P. \& {Giommi}, P. 1995, \apj, 444, 567

\bibitem[{{Padovani} {et~al.}(2016){Padovani}, {Resconi}, {Giommi}, {Arsioli},
  \& {Chang}}]{Neutrino-HSP}
{Padovani}, P., {Resconi}, E., {Giommi}, P., {Arsioli}, B., \& {Chang}, Y.~L.
  2016, \mnras, 457, 3582

\bibitem[{{Pita} {et~al.}(2014){Pita}, {Goldoni}, {Boisson}, {Lenain}, {Punch},
  {G{\'e}rard}, {Hammer}, {Kaper}, \& {Sol}}]{pita2013}
{Pita}, S., {Goldoni}, P., {Boisson}, C., {et~al.} 2014, \aap, 565, A12

\bibitem[{{Prokhorov} \& {Churazov}(2014)}]{CountGammaCluster}
{Prokhorov}, D.~A. \& {Churazov}, E.~M. 2014, \aap, 567, A93

\bibitem[{{Puccetti} {et~al.}(2011){Puccetti}, {Capalbi}, {Giommi}, {Perri},
  {Stratta}, {Angelini}, {Burrows}, {Campana}, {Chincarini}, {Cusumano},
  {Gehrels}, {Moretti}, {Nousek}, {Osborne}, \& {Tagliaferri}}]{SWIFT}
{Puccetti}, S., {Capalbi}, M., {Giommi}, P., {et~al.} 2011, \aap, 528, A122

\bibitem[{{Rieger} {et~al.}(2013){Rieger}, {de O{\~n}a-Wilhelmi}, \&
  {Aharonian}}]{TeVAstronomy}
{Rieger}, F.~M., {de O{\~n}a-Wilhelmi}, E., \& {Aharonian}, F.~A. 2013,
  Frontiers of Physics, 8, 714

\bibitem[{Rosen {et~al.}(2015)Rosen, Webb, Watson, Ballet, Barret, Braito,
  Carrera, Ceballos, Coriat, {Della Ceca}, Denkinson, Esquej, Farrell,
  Freyberg, Gris{\'{e}}, Guillout, Heil, Law-Green, Lamer, Lin, Martino,
  Michel, Motch, Gomez-Moran, Page, Page, Page, Pakull, Pye, Read, Rodriguez,
  Sakano, Saxton, Schwope, Scott, Sturm, Traulsen, Yershov, \&
  Zolotukhin}]{Rosen2015}
Rosen, S.~R., Webb, N.~A., Watson, M.~G., {et~al.} 2015, ArXiv e-prints,
  Submitt. to A{\&}A

\bibitem[{{Sahu} {et~al.}(2016){Sahu}, {Miranda}, \&
  {Rajpoot}}]{FlareMkr421-Hmodel}
{Sahu}, S., {Miranda}, L.~S., \& {Rajpoot}, S. 2016, European Physical Journal
  C, 76, 127

\bibitem[{Saxton {et~al.}(2008)Saxton, Read, Esquej, Freyberg, Altieri, \&
  Bermejo}]{Saxton2008}
Saxton, R.~D., Read, a.~M., Esquej, P., {et~al.} 2008, A{\&}A, 480, 611

\bibitem[{{Sbarufatti} {et~al.}(2005){Sbarufatti}, {Treves}, {Falomo}, {Heidt},
  {Kotilainen}, \& {Scarpa}}]{bll}
{Sbarufatti}, B., {Treves}, A., {Falomo}, R., {et~al.} 2005, \aj, 129, 559

\bibitem[{{Shaw} {et~al.}(2013{\natexlab{a}}){Shaw}, {Filippenko}, {Romani},
  {Cenko}, \& {Li}}]{shaw2}
{Shaw}, M.~S., {Filippenko}, A.~V., {Romani}, R.~W., {Cenko}, S.~B., \& {Li},
  W. 2013{\natexlab{a}}, \aj, 146, 127

\bibitem[{{Shaw} {et~al.}(2013{\natexlab{b}}){Shaw}, {Romani}, {Cotter},
  {Healey}, {Michelson}, {Readhead}, {Richards}, {Max-Moerbeck}, {King}, \&
  {Potter}}]{bllacz2}
{Shaw}, M.~S., {Romani}, R.~W., {Cotter}, G., {et~al.} 2013{\natexlab{b}},
  \apj, 764, 135

\bibitem[{{Vanden Berk} {et~al.}(2001){Vanden Berk}, {Richards}, {Bauer},
  {Strauss}, {Schneider}, {Heckman}, {York}, {Hall}, {Fan}, {Knapp},
  {Anderson}, {Annis}, {Bahcall}, {Bernardi}, {Briggs}, {Brinkmann}, {Brunner},
  {Burles}, {Carey}, {Castander}, {Connolly}, {Crocker}, {Csabai}, {Doi},
  {Finkbeiner}, {Friedman}, {Frieman}, {Fukugita}, {Gunn}, {Hennessy},
  {Ivezi{\'c}}, {Kent}, {Kunszt}, {Lamb}, {Leger}, {Long}, {Loveday}, {Lupton},
  {Meiksin}, {Merelli}, {Munn}, {Newberg}, {Newcomb}, {Nichol}, {Owen}, {Pier},
  {Pope}, {Rockosi}, {Schlegel}, {Siegmund}, {Smee}, {Snir}, {Stoughton},
  {Stubbs}, {SubbaRao}, {Szalay}, {Szokoly}, {Tremonti}, {Uomoto}, {Waddell},
  {Yanny}, \& {Zheng}}]{FSRQtemplate}
{Vanden Berk}, D.~E., {Richards}, G.~T., {Bauer}, A., {et~al.} 2001, \aj, 122,
  549

\bibitem[{{Voges} {et~al.}(1999){Voges}, {Aschenbach}, {Boller},
  {Br{\"a}uninger}, {Briel}, {Burkert}, {Dennerl}, {Englhauser}, {Gruber},
  {Haberl}, {Hartner}, {Hasinger}, {K{\"u}rster}, {Pfeffermann}, {Pietsch},
  {Predehl}, {Rosso}, {Schmitt}, {Tr{\"u}mper}, \& {Zimmermann}}]{ROSATBSC}
{Voges}, W., {Aschenbach}, B., {Boller}, T., {et~al.} 1999, \aap, 349, 389

\bibitem[{{Voges} {et~al.}(2000){Voges}, {Aschenbach}, {Boller}, {Brauninger},
  {Briel}, {Burkert}, {Dennerl}, {Englhauser}, {Gruber}, {Haberl}, {Hartner},
  {Hasinger}, {Pfeffermann}, {Pietsch}, {Predehl}, {Schmitt}, {Trumper}, \&
  {Zimmermann}}]{ROSATFSC}
{Voges}, W., {Aschenbach}, B., {Boller}, T., {et~al.} 2000, \iaucirc, 7432, 3

\bibitem[{{Warren} {et~al.}(2007){Warren}, {Cross}, {Dye}, {Hambly}, {Almaini},
  {Edge}, {Hewett}, {Hodgkin}, {Irwin}, {Jameson}, {Lawrence}, {Lucas},
  {Mortlock}, {Adamson}, {Bryant}, {Collins}, {Davis}, {Emerson}, {Evans},
  {Gonzales-Solares}, {Hirst}, {Kerr}, {Lewis}, {Mann}, {Rawlings}, {Read},
  {Riello}, {Sutorius}, \& {Varricatt}}]{UKIDSDR2}
{Warren}, S.~J., {Cross}, N.~J.~G., {Dye}, S., {et~al.} 2007, ArXiv
  Astrophysics e-prints: astro-ph/0703037

\bibitem[{{Watson} {et~al.}(2009){Watson}, {Schr{\"o}der}, {Fyfe}, {Page},
  {Lamer}, {Mateos}, {Pye}, {Sakano}, {Rosen}, {Ballet}, {Barcons}, {Barret},
  {Boller}, {Brunner}, {Brusa}, {Caccianiga}, {Carrera}, {Ceballos}, {Della
  Ceca}, {Denby}, {Denkinson}, {Dupuy}, {Farrell}, {Fraschetti}, {Freyberg},
  {Guillout}, {Hambaryan}, {Maccacaro}, {Mathiesen}, {McMahon}, {Michel},
  {Motch}, {Osborne}, {Page}, {Pakull}, {Pietsch}, {Saxton}, {Schwope},
  {Severgnini}, {Simpson}, {Sironi}, {Stewart}, {Stewart}, {Stobbart}, {Tedds},
  {Warwick}, {Webb}, {West}, {Worrall}, \& {Yuan}}]{XMM}
{Watson}, M.~G., {Schr{\"o}der}, A.~C., {Fyfe}, D., {et~al.} 2009, \aap, 493,
  339

\bibitem[{{White} \& {Becker}(1992)}]{Radio1}
{White}, R.~L. \& {Becker}, R.~H. 1992, \apjs, 79, 331

\bibitem[{{White} {et~al.}(1997){White}, {Becker}, {Helfand}, \&
  {Gregg}}]{Radio6First}
{White}, R.~L., {Becker}, R.~H., {Helfand}, D.~J., \& {Gregg}, M.~D. 1997,
  \apj, 475, 479

\bibitem[{{Wright} {et~al.}(2010){Wright}, {Eisenhardt}, {Mainzer}, {Ressler},
  {Cutri}, {Jarrett}, {Kirkpatrick}, {Padgett}, {McMillan}, {Skrutskie},
  {Stanford}, {Cohen}, {Walker}, {Mather}, {Leisawitz}, {Gautier}, {McLean},
  {Benford}, {Lonsdale}, {Blain}, {Mendez}, {Irace}, {Duval}, {Liu}, {Royer},
  {Heinrichsen}, {Howard}, {Shannon}, {Kendall}, {Walsh}, {Larsen}, {Cardon},
  {Schick}, {Schwalm}, {Abid}, {Fabinsky}, {Naes}, \& {Tsai}}]{WISE}
{Wright}, E.~L., {Eisenhardt}, P.~R.~M., {Mainzer}, A.~K., {et~al.} 2010, \aj,
  140, 1868

\end{thebibliography}
